\documentclass[twocolumn,longauth,traditabstract]{aa}

\usepackage{graphicx,amsmath}
\usepackage{epsf}
\usepackage{txfonts}
\usepackage[breaklinks, colorlinks, citecolor=blue]{hyperref}  
\usepackage{natbib}
\usepackage{upgreek}
\usepackage{url}

\bibpunct{(}{)}{;}{a}{}{,} 

\def\setsymbol#1#2{\expandafter\def\csname #1\endcsname{#2}}
\def\getsymbol#1{\csname #1\endcsname}

\def\Planck{{\it Planck\/}}

\def\allearlypapers{\nocite{planck2011-1.1, planck2011-1.3, planck2011-1.4, planck2011-1.5, planck2011-1.6, planck2011-1.7, planck2011-1.10, planck2011-1.10sup, planck2011-5.1a, planck2011-5.1b, planck2011-5.2a, planck2011-5.2b, planck2011-5.2c, planck2011-6.1, planck2011-6.2, planck2011-6.3a, planck2011-6.4a, planck2011-6.4b, planck2011-6.6, planck2011-7.0, planck2011-7.2, planck2011-7.3, planck2011-7.7a, planck2011-7.7b, planck2011-7.12, planck2011-7.13}}

\newbox\tablebox    \newdimen\tablewidth
\def\leaderfil{\leaders\hbox to 5pt{\hss.\hss}\hfil}
\def\tablenote#1 #2\par{\begingroup \parindent=0.8em
    \abovedisplayshortskip=0pt\belowdisplayshortskip=0pt
    \noindent
    $$\hss\vbox{\hsize\tablewidth \hangindent=\parindent \hangafter=1 \noindent
    \hbox to \parindent{\sup{\rm #1}\hss}\strut#2\strut\par}\hss$$
    \endgroup}

\def\L2{\ifmmode L_2\else $L_2$\fi}

\def\DeltaT{\ifmmode \Delta T\else $\Delta T$\fi}
\def\deltat{\ifmmode \Delta t\else $\Delta t$\fi}
\def\fknee{\ifmmode f_{\rm knee}\else $f_{\rm knee}$\fi}
\def\Fmax{\ifmmode F_{\rm max}\else $F_{\rm max}$\fi}
\def\solar{\ifmmode{\rm M}_{\mathord\odot}\else${\rm M}_{\mathord\odot}$\fi}

\def\inv{\ifmmode^{-1}\else$^{-1}$\fi}
\def\mo{\ifmmode^{-1}\else$^{-1}$\fi}
\def\sup#1{\ifmmode ^{\rm #1}\else $^{\rm #1}$\fi}
\def\expo#1{\ifmmode \times 10^{#1}\else $\times 10^{#1}$\fi}
\def\,{\thinspace}
\def\lsim{\mathrel{\raise .4ex\hbox{\rlap{$<$}\lower 1.2ex\hbox{$\sim$}}}}
\def\gsim{\mathrel{\raise .4ex\hbox{\rlap{$>$}\lower 1.2ex\hbox{$\sim$}}}}

\def\simprop{\mathrel{\raise .4ex\hbox{\rlap{$\propto$}\lower 1.2ex\hbox{$\sim$}}}}
\def\deg{\ifmmode^\circ\else$^\circ$\fi}
\def\pdeg{\ifmmode $\setbox0=\hbox{$^{\circ}$}\rlap{\hskip.11\wd0 .}$^{\circ}
          \else \setbox0=\hbox{$^{\circ}$}\rlap{\hskip.11\wd0 .}$^{\circ}$\fi}
\def\arcs{\ifmmode {^{\scriptstyle\prime\prime}}
          \else $^{\scriptstyle\prime\prime}$\fi}
\def\arcm{\ifmmode {^{\scriptstyle\prime}}
          \else $^{\scriptstyle\prime}$\fi}
\newdimen\sa  \newdimen\sb
\def\parcs{\sa=.07em \sb=.03em
     \ifmmode \hbox{\rlap{.}}^{\scriptstyle\prime\kern -\sb\prime}\hbox{\kern -\sa}
     \else \rlap{.}$^{\scriptstyle\prime\kern -\sb\prime}$\kern -\sa\fi}
\def\parcm{\sa=.08em \sb=.03em
     \ifmmode \hbox{\rlap{.}\kern\sa}^{\scriptstyle\prime}\hbox{\kern-\sb}
     \else \rlap{.}\kern\sa$^{\scriptstyle\prime}$\kern-\sb\fi}
\def\ra[#1 #2 #3.#4]{#1\sup{h}#2\sup{m}#3\sup{s}\llap.#4}
\def\dec[#1 #2 #3.#4]{#1\deg#2\arcm#3\arcs\llap.#4}
\def\deco[#1 #2 #3]{#1\deg#2\arcm#3\arcs}
\def\rra[#1 #2]{#1\sup{h}#2\sup{m}}

\def\dots{\relax\ifmmode \ldots\else $\ldots$\fi}
\def\WHzsr{\ifmmode $W\,Hz\mo\,sr\mo$\else W\,Hz\mo\,sr\mo\fi}
\def\mHz{\ifmmode $\,mHz$\else \,mHz\fi}
\def\GHz{\ifmmode $\,GHz$\else \,GHz\fi}
\def\mKs{\ifmmode $\,mK\,s$^{1/2}\else \,mK\,s$^{1/2}$\fi}
\def\muKs{\ifmmode \,\mu$K\,s$^{1/2}\else \,$\mu$K\,s$^{1/2}$\fi}
\def\muKRJs{\ifmmode \,\mu$K$_{\rm RJ}$\,s$^{1/2}\else \,$\mu$K$_{\rm RJ}$\,s$^{1/2}$\fi}
\def\muKHz{\ifmmode \,\mu$K\,Hz$^{-1/2}\else \,$\mu$K\,Hz$^{-1/2}$\fi}
\def\MJysr{\ifmmode \,$MJy\,sr\mo$\else \,MJy\,sr\mo\fi}
\def\MJysrmK{\ifmmode \,$MJy\,sr\mo$\,mK$_{\rm CMB}\mo\else \,MJy\,sr\mo\,mK$_{\rm CMB}\mo$\fi}
\def\microns{\ifmmode \,\mu$m$\else \,$\mu$m\fi}

\def\muK{\ifmmode \,\mu$K$\else \,$\mu$\hbox{K}\fi}
\def\microK{\ifmmode \,\mu$K$\else \,$\mu$\hbox{K}\fi}
\def\muW{\ifmmode \,\mu$W$\else \,$\mu$\hbox{W}\fi}
\def\kms{\ifmmode $\,km\,s$^{-1}\else \,km\,s$^{-1}$\fi}
\def\kmsMpc{\ifmmode $\,\kms\,Mpc\mo$\else \,\kms\,Mpc\mo\fi}
\setsymbol{LFI:center:frequency:70GHz:units}{70.3\,GHz}
\setsymbol{LFI:center:frequency:44GHz:units}{44.1\,GHz}
\setsymbol{LFI:center:frequency:30GHz:units}{28.5\,GHz}

\setsymbol{LFI:center:frequency:70GHz}{70.3}
\setsymbol{LFI:center:frequency:44GHz}{44.1}
\setsymbol{LFI:center:frequency:30GHz}{28.5}

\setsymbol{LFI:center:frequency:LFI18:Rad:M:units}{71.7\GHz}
\setsymbol{LFI:center:frequency:LFI19:Rad:M:units}{67.5\GHz}
\setsymbol{LFI:center:frequency:LFI20:Rad:M:units}{69.2\GHz}
\setsymbol{LFI:center:frequency:LFI21:Rad:M:units}{70.4\GHz}
\setsymbol{LFI:center:frequency:LFI22:Rad:M:units}{71.5\GHz}
\setsymbol{LFI:center:frequency:LFI23:Rad:M:units}{70.8\GHz}
\setsymbol{LFI:center:frequency:LFI24:Rad:M:units}{44.4\GHz}
\setsymbol{LFI:center:frequency:LFI25:Rad:M:units}{44.0\GHz}
\setsymbol{LFI:center:frequency:LFI26:Rad:M:units}{43.9\GHz}
\setsymbol{LFI:center:frequency:LFI27:Rad:M:units}{28.3\GHz}
\setsymbol{LFI:center:frequency:LFI28:Rad:M:units}{28.8\GHz}
\setsymbol{LFI:center:frequency:LFI18:Rad:S:units}{70.1\GHz}
\setsymbol{LFI:center:frequency:LFI19:Rad:S:units}{69.6\GHz}
\setsymbol{LFI:center:frequency:LFI20:Rad:S:units}{69.5\GHz}
\setsymbol{LFI:center:frequency:LFI21:Rad:S:units}{69.5\GHz}
\setsymbol{LFI:center:frequency:LFI22:Rad:S:units}{72.8\GHz}
\setsymbol{LFI:center:frequency:LFI23:Rad:S:units}{71.3\GHz}
\setsymbol{LFI:center:frequency:LFI24:Rad:S:units}{44.1\GHz}
\setsymbol{LFI:center:frequency:LFI25:Rad:S:units}{44.1\GHz}
\setsymbol{LFI:center:frequency:LFI26:Rad:S:units}{44.1\GHz}
\setsymbol{LFI:center:frequency:LFI27:Rad:S:units}{28.5\GHz}
\setsymbol{LFI:center:frequency:LFI28:Rad:S:units}{28.2\GHz}

\setsymbol{LFI:center:frequency:LFI18:Rad:M}{71.7}
\setsymbol{LFI:center:frequency:LFI19:Rad:M}{67.5}
\setsymbol{LFI:center:frequency:LFI20:Rad:M}{69.2}
\setsymbol{LFI:center:frequency:LFI21:Rad:M}{70.4}
\setsymbol{LFI:center:frequency:LFI22:Rad:M}{71.5}
\setsymbol{LFI:center:frequency:LFI23:Rad:M}{70.8}
\setsymbol{LFI:center:frequency:LFI24:Rad:M}{44.4}
\setsymbol{LFI:center:frequency:LFI25:Rad:M}{44.0}
\setsymbol{LFI:center:frequency:LFI26:Rad:M}{43.9}
\setsymbol{LFI:center:frequency:LFI27:Rad:M}{28.3}
\setsymbol{LFI:center:frequency:LFI28:Rad:M}{28.8}
\setsymbol{LFI:center:frequency:LFI18:Rad:S}{70.1}
\setsymbol{LFI:center:frequency:LFI19:Rad:S}{69.6}
\setsymbol{LFI:center:frequency:LFI20:Rad:S}{69.5}
\setsymbol{LFI:center:frequency:LFI21:Rad:S}{69.5}
\setsymbol{LFI:center:frequency:LFI22:Rad:S}{72.8}
\setsymbol{LFI:center:frequency:LFI23:Rad:S}{71.3}
\setsymbol{LFI:center:frequency:LFI24:Rad:S}{44.1}
\setsymbol{LFI:center:frequency:LFI25:Rad:S}{44.1}
\setsymbol{LFI:center:frequency:LFI26:Rad:S}{44.1}
\setsymbol{LFI:center:frequency:LFI27:Rad:S}{28.5}
\setsymbol{LFI:center:frequency:LFI28:Rad:S}{28.2}

\setsymbol{LFI:white:noise:sensitivity:70GHz:units}{152.6\muKs}
\setsymbol{LFI:white:noise:sensitivity:44GHz:units}{173.1\muKs}
\setsymbol{LFI:white:noise:sensitivity:30GHz:units}{146.8\muKs}

\setsymbol{LFI:white:noise:sensitivity:70GHz}{152.6}
\setsymbol{LFI:white:noise:sensitivity:44GHz}{173.1}
\setsymbol{LFI:white:noise:sensitivity:30GHz}{146.8}

\setsymbol{LFI:white:noise:sensitivity:LFI18:Rad:M:units}{512.0\muKs}
\setsymbol{LFI:white:noise:sensitivity:LFI19:Rad:M:units}{581.4\muKs}
\setsymbol{LFI:white:noise:sensitivity:LFI20:Rad:M:units}{590.8\muKs}
\setsymbol{LFI:white:noise:sensitivity:LFI21:Rad:M:units}{455.2\muKs}
\setsymbol{LFI:white:noise:sensitivity:LFI22:Rad:M:units}{492.0\muKs}
\setsymbol{LFI:white:noise:sensitivity:LFI23:Rad:M:units}{507.7\muKs}
\setsymbol{LFI:white:noise:sensitivity:LFI24:Rad:M:units}{462.2\muKs}
\setsymbol{LFI:white:noise:sensitivity:LFI25:Rad:M:units}{413.6\muKs}
\setsymbol{LFI:white:noise:sensitivity:LFI26:Rad:M:units}{478.6\muKs}
\setsymbol{LFI:white:noise:sensitivity:LFI27:Rad:M:units}{277.7\muKs}
\setsymbol{LFI:white:noise:sensitivity:LFI28:Rad:M:units}{312.3\muKs}
\setsymbol{LFI:white:noise:sensitivity:LFI18:Rad:S:units}{465.7\muKs}
\setsymbol{LFI:white:noise:sensitivity:LFI19:Rad:S:units}{555.6\muKs}
\setsymbol{LFI:white:noise:sensitivity:LFI20:Rad:S:units}{623.2\muKs}
\setsymbol{LFI:white:noise:sensitivity:LFI21:Rad:S:units}{564.1\muKs}
\setsymbol{LFI:white:noise:sensitivity:LFI22:Rad:S:units}{534.4\muKs}
\setsymbol{LFI:white:noise:sensitivity:LFI23:Rad:S:units}{542.4\muKs}
\setsymbol{LFI:white:noise:sensitivity:LFI24:Rad:S:units}{399.2\muKs}
\setsymbol{LFI:white:noise:sensitivity:LFI25:Rad:S:units}{392.6\muKs}
\setsymbol{LFI:white:noise:sensitivity:LFI26:Rad:S:units}{418.6\muKs}
\setsymbol{LFI:white:noise:sensitivity:LFI27:Rad:S:units}{302.9\muKs}
\setsymbol{LFI:white:noise:sensitivity:LFI28:Rad:S:units}{285.3\muKs}

\setsymbol{LFI:white:noise:sensitivity:LFI18:Rad:M}{512.0}
\setsymbol{LFI:white:noise:sensitivity:LFI19:Rad:M}{581.4}
\setsymbol{LFI:white:noise:sensitivity:LFI20:Rad:M}{590.8}
\setsymbol{LFI:white:noise:sensitivity:LFI21:Rad:M}{455.2}
\setsymbol{LFI:white:noise:sensitivity:LFI22:Rad:M}{492.0}
\setsymbol{LFI:white:noise:sensitivity:LFI23:Rad:M}{507.7}
\setsymbol{LFI:white:noise:sensitivity:LFI24:Rad:M}{462.2}
\setsymbol{LFI:white:noise:sensitivity:LFI25:Rad:M}{413.6}
\setsymbol{LFI:white:noise:sensitivity:LFI26:Rad:M}{478.6}
\setsymbol{LFI:white:noise:sensitivity:LFI27:Rad:M}{277.7}
\setsymbol{LFI:white:noise:sensitivity:LFI28:Rad:M}{312.3}
\setsymbol{LFI:white:noise:sensitivity:LFI18:Rad:S}{465.7}
\setsymbol{LFI:white:noise:sensitivity:LFI19:Rad:S}{555.6}
\setsymbol{LFI:white:noise:sensitivity:LFI20:Rad:S}{623.2}
\setsymbol{LFI:white:noise:sensitivity:LFI21:Rad:S}{564.1}
\setsymbol{LFI:white:noise:sensitivity:LFI22:Rad:S}{534.4}
\setsymbol{LFI:white:noise:sensitivity:LFI23:Rad:S}{542.4}
\setsymbol{LFI:white:noise:sensitivity:LFI24:Rad:S}{399.2}
\setsymbol{LFI:white:noise:sensitivity:LFI25:Rad:S}{392.6}
\setsymbol{LFI:white:noise:sensitivity:LFI26:Rad:S}{418.6}
\setsymbol{LFI:white:noise:sensitivity:LFI27:Rad:S}{302.9}
\setsymbol{LFI:white:noise:sensitivity:LFI28:Rad:S}{285.3}

\setsymbol{LFI:knee:frequency:70GHz:units}{29.5\mHz}
\setsymbol{LFI:knee:frequency:44GHz:units}{56.2\mHz}
\setsymbol{LFI:knee:frequency:30GHz:units}{113.7\mHz}

\setsymbol{LFI:knee:frequency:70GHz}{29.5}
\setsymbol{LFI:knee:frequency:44GHz}{56.2}
\setsymbol{LFI:knee:frequency:30GHz}{113.7}

\setsymbol{LFI:knee:frequency:LFI18:Rad:M:units}{16.3\mHz}
\setsymbol{LFI:knee:frequency:LFI19:Rad:M:units}{15.1\mHz}
\setsymbol{LFI:knee:frequency:LFI20:Rad:M:units}{18.7\mHz}
\setsymbol{LFI:knee:frequency:LFI21:Rad:M:units}{37.2\mHz}
\setsymbol{LFI:knee:frequency:LFI22:Rad:M:units}{12.7\mHz}
\setsymbol{LFI:knee:frequency:LFI23:Rad:M:units}{34.6\mHz}
\setsymbol{LFI:knee:frequency:LFI24:Rad:M:units}{46.2\mHz}
\setsymbol{LFI:knee:frequency:LFI25:Rad:M:units}{24.9\mHz}
\setsymbol{LFI:knee:frequency:LFI26:Rad:M:units}{67.6\mHz}
\setsymbol{LFI:knee:frequency:LFI27:Rad:M:units}{187.4\mHz}
\setsymbol{LFI:knee:frequency:LFI28:Rad:M:units}{122.2\mHz}
\setsymbol{LFI:knee:frequency:LFI18:Rad:S:units}{17.7\mHz}
\setsymbol{LFI:knee:frequency:LFI19:Rad:S:units}{22.0\mHz}
\setsymbol{LFI:knee:frequency:LFI20:Rad:S:units}{8.7\mHz}
\setsymbol{LFI:knee:frequency:LFI21:Rad:S:units}{25.9\mHz}
\setsymbol{LFI:knee:frequency:LFI22:Rad:S:units}{15.8\mHz}
\setsymbol{LFI:knee:frequency:LFI23:Rad:S:units}{129.8\mHz}
\setsymbol{LFI:knee:frequency:LFI24:Rad:S:units}{100.9\mHz}
\setsymbol{LFI:knee:frequency:LFI25:Rad:S:units}{38.9\mHz}
\setsymbol{LFI:knee:frequency:LFI26:Rad:S:units}{58.9\mHz}
\setsymbol{LFI:knee:frequency:LFI27:Rad:S:units}{104.4\mHz}
\setsymbol{LFI:knee:frequency:LFI28:Rad:S:units}{40.7\mHz}

\setsymbol{LFI:knee:frequency:LFI18:Rad:M}{16.3}
\setsymbol{LFI:knee:frequency:LFI19:Rad:M}{15.1}
\setsymbol{LFI:knee:frequency:LFI20:Rad:M}{18.7}
\setsymbol{LFI:knee:frequency:LFI21:Rad:M}{37.2}
\setsymbol{LFI:knee:frequency:LFI22:Rad:M}{12.7}
\setsymbol{LFI:knee:frequency:LFI23:Rad:M}{34.6}
\setsymbol{LFI:knee:frequency:LFI24:Rad:M}{46.2}
\setsymbol{LFI:knee:frequency:LFI25:Rad:M}{24.9}
\setsymbol{LFI:knee:frequency:LFI26:Rad:M}{67.6}
\setsymbol{LFI:knee:frequency:LFI27:Rad:M}{187.4}
\setsymbol{LFI:knee:frequency:LFI28:Rad:M}{122.2}
\setsymbol{LFI:knee:frequency:LFI18:Rad:S}{17.7}
\setsymbol{LFI:knee:frequency:LFI19:Rad:S}{22.0}
\setsymbol{LFI:knee:frequency:LFI20:Rad:S}{8.7}
\setsymbol{LFI:knee:frequency:LFI21:Rad:S}{25.9}
\setsymbol{LFI:knee:frequency:LFI22:Rad:S}{15.8}
\setsymbol{LFI:knee:frequency:LFI23:Rad:S}{129.8}
\setsymbol{LFI:knee:frequency:LFI24:Rad:S}{100.9}
\setsymbol{LFI:knee:frequency:LFI25:Rad:S}{38.9}
\setsymbol{LFI:knee:frequency:LFI26:Rad:S}{58.9}
\setsymbol{LFI:knee:frequency:LFI27:Rad:S}{104.4}
\setsymbol{LFI:knee:frequency:LFI28:Rad:S}{40.7}

\setsymbol{LFI:slope:70GHz:units}{$-1.03$\mHz}
\setsymbol{LFI:slope:44GHz:units}{$-0.89$\mHz}
\setsymbol{LFI:slope:30GHz:units}{$-0.87$\mHz}

\setsymbol{LFI:slope:70GHz}{$-1.03$}
\setsymbol{LFI:slope:44GHz}{$-0.89$}
\setsymbol{LFI:slope:30GHz}{$-0.87$}

\setsymbol{LFI:slope:LFI18:Rad:M:units}{$-1.04$\mHz}
\setsymbol{LFI:slope:LFI19:Rad:M:units}{$-1.09$\mHz}
\setsymbol{LFI:slope:LFI20:Rad:M:units}{$-0.69$\mHz}
\setsymbol{LFI:slope:LFI21:Rad:M:units}{$-1.56$\mHz}
\setsymbol{LFI:slope:LFI22:Rad:M:units}{$-1.01$\mHz}
\setsymbol{LFI:slope:LFI23:Rad:M:units}{$-0.96$\mHz}
\setsymbol{LFI:slope:LFI24:Rad:M:units}{$-0.83$\mHz}
\setsymbol{LFI:slope:LFI25:Rad:M:units}{$-0.91$\mHz}
\setsymbol{LFI:slope:LFI26:Rad:M:units}{$-0.95$\mHz}
\setsymbol{LFI:slope:LFI27:Rad:M:units}{$-0.87$\mHz}
\setsymbol{LFI:slope:LFI28:Rad:M:units}{$-0.88$\mHz}
\setsymbol{LFI:slope:LFI18:Rad:S:units}{$-1.15$\mHz}
\setsymbol{LFI:slope:LFI19:Rad:S:units}{$-1.00$\mHz}
\setsymbol{LFI:slope:LFI20:Rad:S:units}{$-0.95$\mHz}
\setsymbol{LFI:slope:LFI21:Rad:S:units}{$-0.92$\mHz}
\setsymbol{LFI:slope:LFI22:Rad:S:units}{$-1.01$\mHz}
\setsymbol{LFI:slope:LFI23:Rad:S:units}{$-0.95$\mHz}
\setsymbol{LFI:slope:LFI24:Rad:S:units}{$-0.73$\mHz}
\setsymbol{LFI:slope:LFI25:Rad:S:units}{$-1.16$\mHz}
\setsymbol{LFI:slope:LFI26:Rad:S:units}{$-0.79$\mHz}
\setsymbol{LFI:slope:LFI27:Rad:S:units}{$-0.82$\mHz}
\setsymbol{LFI:slope:LFI28:Rad:S:units}{$-0.91$\mHz}

\setsymbol{LFI:slope:LFI18:Rad:M}{$-1.04$}
\setsymbol{LFI:slope:LFI19:Rad:M}{$-1.09$}
\setsymbol{LFI:slope:LFI20:Rad:M}{$-0.69$}
\setsymbol{LFI:slope:LFI21:Rad:M}{$-1.56$}
\setsymbol{LFI:slope:LFI22:Rad:M}{$-1.01$}
\setsymbol{LFI:slope:LFI23:Rad:M}{$-0.96$}
\setsymbol{LFI:slope:LFI24:Rad:M}{$-0.83$}
\setsymbol{LFI:slope:LFI25:Rad:M}{$-0.91$}
\setsymbol{LFI:slope:LFI26:Rad:M}{$-0.95$}
\setsymbol{LFI:slope:LFI27:Rad:M}{$-0.87$}
\setsymbol{LFI:slope:LFI28:Rad:M}{$-0.88$}
\setsymbol{LFI:slope:LFI18:Rad:S}{$-1.15$}
\setsymbol{LFI:slope:LFI19:Rad:S}{$-1.00$}
\setsymbol{LFI:slope:LFI20:Rad:S}{$-0.95$}
\setsymbol{LFI:slope:LFI21:Rad:S}{$-0.92$}
\setsymbol{LFI:slope:LFI22:Rad:S}{$-1.01$}
\setsymbol{LFI:slope:LFI23:Rad:S}{$-0.95$}
\setsymbol{LFI:slope:LFI24:Rad:S}{$-0.73$}
\setsymbol{LFI:slope:LFI25:Rad:S}{$-1.16$}
\setsymbol{LFI:slope:LFI26:Rad:S}{$-0.79$}
\setsymbol{LFI:slope:LFI27:Rad:S}{$-0.82$}
\setsymbol{LFI:slope:LFI28:Rad:S}{$-0.91$}

\setsymbol{LFI:FWHM:70GHz:units}{13\parcm01}
\setsymbol{LFI:FWHM:44GHz:units}{27\parcm92}
\setsymbol{LFI:FWHM:30GHz:units}{32\parcm65}

\setsymbol{LFI:FWHM:70GHz}{13.01}
\setsymbol{LFI:FWHM:44GHz}{27.92}
\setsymbol{LFI:FWHM:30GHz}{32.65}

\setsymbol{LFI:FWHM:LFI18:units}{13\parcm39}
\setsymbol{LFI:FWHM:LFI19:units}{13\parcm01}
\setsymbol{LFI:FWHM:LFI20:units}{12\parcm75}
\setsymbol{LFI:FWHM:LFI21:units}{12\parcm74}
\setsymbol{LFI:FWHM:LFI22:units}{12\parcm87}
\setsymbol{LFI:FWHM:LFI23:units}{13\parcm27}
\setsymbol{LFI:FWHM:LFI24:units}{22\parcm98}
\setsymbol{LFI:FWHM:LFI25:units}{30\parcm46}
\setsymbol{LFI:FWHM:LFI26:units}{30\parcm31}
\setsymbol{LFI:FWHM:LFI27:units}{32\parcm65}
\setsymbol{LFI:FWHM:LFI28:units}{32\parcm66}

\setsymbol{LFI:FWHM:LFI18}{13.39}
\setsymbol{LFI:FWHM:LFI19}{13.01}
\setsymbol{LFI:FWHM:LFI20}{12.75}
\setsymbol{LFI:FWHM:LFI21}{12.74}
\setsymbol{LFI:FWHM:LFI22}{12.87}
\setsymbol{LFI:FWHM:LFI23}{13.27}
\setsymbol{LFI:FWHM:LFI24}{22.98}
\setsymbol{LFI:FWHM:LFI25}{30.46}
\setsymbol{LFI:FWHM:LFI26}{30.31}
\setsymbol{LFI:FWHM:LFI27}{32.65}
\setsymbol{LFI:FWHM:LFI28}{32.66}

\setsymbol{LFI:FWHM:uncertainty:LFI18:units}{0.170\arcm}
\setsymbol{LFI:FWHM:uncertainty:LFI19:units}{0.174\arcm}
\setsymbol{LFI:FWHM:uncertainty:LFI20:units}{0.170\arcm}
\setsymbol{LFI:FWHM:uncertainty:LFI21:units}{0.156\arcm}
\setsymbol{LFI:FWHM:uncertainty:LFI22:units}{0.164\arcm}
\setsymbol{LFI:FWHM:uncertainty:LFI23:units}{0.171\arcm}
\setsymbol{LFI:FWHM:uncertainty:LFI24:units}{0.652\arcm}
\setsymbol{LFI:FWHM:uncertainty:LFI25:units}{1.075\arcm}
\setsymbol{LFI:FWHM:uncertainty:LFI26:units}{1.131\arcm}
\setsymbol{LFI:FWHM:uncertainty:LFI27:units}{1.266\arcm}
\setsymbol{LFI:FWHM:uncertainty:LFI28:units}{1.287\arcm}

\setsymbol{LFI:FWHM:uncertainty:LFI18}{0.170}
\setsymbol{LFI:FWHM:uncertainty:LFI19}{0.174}
\setsymbol{LFI:FWHM:uncertainty:LFI20}{0.170}
\setsymbol{LFI:FWHM:uncertainty:LFI21}{0.156}
\setsymbol{LFI:FWHM:uncertainty:LFI22}{0.164}
\setsymbol{LFI:FWHM:uncertainty:LFI23}{0.171}
\setsymbol{LFI:FWHM:uncertainty:LFI24}{0.652}
\setsymbol{LFI:FWHM:uncertainty:LFI25}{1.075}
\setsymbol{LFI:FWHM:uncertainty:LFI26}{1.131}
\setsymbol{LFI:FWHM:uncertainty:LFI27}{1.266}
\setsymbol{LFI:FWHM:uncertainty:LFI28}{1.287}

\setsymbol{HFI:center:frequency:100GHz:units}{100\,GHz}
\setsymbol{HFI:center:frequency:143GHz:units}{143\,GHz}
\setsymbol{HFI:center:frequency:217GHz:units}{217\,GHz}
\setsymbol{HFI:center:frequency:353GHz:units}{353\,GHz}
\setsymbol{HFI:center:frequency:545GHz:units}{545\,GHz}
\setsymbol{HFI:center:frequency:857GHz:units}{857\,GHz}

\setsymbol{HFI:center:frequency:100GHz}{100}
\setsymbol{HFI:center:frequency:143GHz}{143}
\setsymbol{HFI:center:frequency:217GHz}{217}
\setsymbol{HFI:center:frequency:353GHz}{353}
\setsymbol{HFI:center:frequency:545GHz}{545}
\setsymbol{HFI:center:frequency:857GHz}{857}

\setsymbol{HFI:Ndetectors:100GHz}{8}
\setsymbol{HFI:Ndetectors:143GHz}{11}
\setsymbol{HFI:Ndetectors:217GHz}{12}
\setsymbol{HFI:Ndetectors:353GHz}{12}
\setsymbol{HFI:Ndetectors:545GHz}{3}
\setsymbol{HFI:Ndetectors:857GHz}{4}

\setsymbol{HFI:FWHM:Maps:100GHz:units}{9\parcm88}
\setsymbol{HFI:FWHM:Maps:143GHz:units}{7\parcm18}
\setsymbol{HFI:FWHM:Maps:217GHz:units}{4\parcm87}
\setsymbol{HFI:FWHM:Maps:353GHz:units}{4\parcm65}
\setsymbol{HFI:FWHM:Maps:545GHz:units}{4\parcm72}
\setsymbol{HFI:FWHM:Maps:857GHz:units}{4\parcm39}
\setsymbol{HFI:FWHM:Maps:100GHz}{9.88}
\setsymbol{HFI:FWHM:Maps:143GHz}{7.18}
\setsymbol{HFI:FWHM:Maps:217GHz}{4.87}
\setsymbol{HFI:FWHM:Maps:353GHz}{4.65}
\setsymbol{HFI:FWHM:Maps:545GHz}{4.72}
\setsymbol{HFI:FWHM:Maps:857GHz}{4.39}

\setsymbol{HFI:beam:ellipticity:Maps:100GHz}{1.15}
\setsymbol{HFI:beam:ellipticity:Maps:143GHz}{1.01}
\setsymbol{HFI:beam:ellipticity:Maps:217GHz}{1.06}
\setsymbol{HFI:beam:ellipticity:Maps:353GHz}{1.05}
\setsymbol{HFI:beam:ellipticity:Maps:545GHz}{1.14}
\setsymbol{HFI:beam:ellipticity:Maps:857GHz}{1.19}

\setsymbol{HFI:FWHM:Mars:100GHz:units}{9\parcm37}
\setsymbol{HFI:FWHM:Mars:143GHz:units}{7\parcm04}
\setsymbol{HFI:FWHM:Mars:217GHz:units}{4\parcm68}
\setsymbol{HFI:FWHM:Mars:353GHz:units}{4\parcm43}
\setsymbol{HFI:FWHM:Mars:545GHz:units}{3\parcm80}
\setsymbol{HFI:FWHM:Mars:857GHz:units}{3\parcm67}

\setsymbol{HFI:FWHM:Mars:100GHz}{9.37}
\setsymbol{HFI:FWHM:Mars:143GHz}{7.04}
\setsymbol{HFI:FWHM:Mars:217GHz}{4.68}
\setsymbol{HFI:FWHM:Mars:353GHz}{4.43}
\setsymbol{HFI:FWHM:Mars:545GHz}{3.80}
\setsymbol{HFI:FWHM:Mars:857GHz}{3.67}

\setsymbol{HFI:beam:ellipticity:Mars:100GHz}{1.18}
\setsymbol{HFI:beam:ellipticity:Mars:143GHz}{1.03}
\setsymbol{HFI:beam:ellipticity:Mars:217GHz}{1.14}
\setsymbol{HFI:beam:ellipticity:Mars:353GHz}{1.09}
\setsymbol{HFI:beam:ellipticity:Mars:545GHz}{1.25}
\setsymbol{HFI:beam:ellipticity:Mars:857GHz}{1.03}

\setsymbol{HFI:CMB:relative:calibration:100GHz}{$\lsim 1\%$}
\setsymbol{HFI:CMB:relative:calibration:143GHz}{$\lsim 1\%$}
\setsymbol{HFI:CMB:relative:calibration:217GHz}{$\lsim 1\%$}
\setsymbol{HFI:CMB:relative:calibration:353GHz}{$\lsim 1\%$}
\setsymbol{HFI:CMB:relative:calibration:545GHz}{}
\setsymbol{HFI:CMB:relative:calibration:857GHz}{}

\setsymbol{HFI:CMB:absolute:calibration:100GHz}{$\lsim 2\%$}
\setsymbol{HFI:CMB:absolute:calibration:143GHz}{$\lsim 2\%$}
\setsymbol{HFI:CMB:absolute:calibration:217GHz}{$\lsim 2\%$}
\setsymbol{HFI:CMB:absolute:calibration:353GHz}{$\lsim 2\%$}
\setsymbol{HFI:CMB:absolute:calibration:545GHz}{}
\setsymbol{HFI:CMB:absolute:calibration:857GHz}{}

\setsymbol{HFI:FIRAS:gain:calibration:accuracy:statistical:100GHz}{}
\setsymbol{HFI:FIRAS:gain:calibration:accuracy:statistical:143GHz}{}
\setsymbol{HFI:FIRAS:gain:calibration:accuracy:statistical:217GHz}{}
\setsymbol{HFI:FIRAS:gain:calibration:accuracy:statistical:353GHz}{2.5\%}
\setsymbol{HFI:FIRAS:gain:calibration:accuracy:statistical:545GHz}{1\%}
\setsymbol{HFI:FIRAS:gain:calibration:accuracy:statistical:857GHz}{0.5\%}

\setsymbol{HFI:FIRAS:gain:calibration:accuracy:systematic:100GHz}{}
\setsymbol{HFI:FIRAS:gain:calibration:accuracy:systematic:143GHz}{}
\setsymbol{HFI:FIRAS:gain:calibration:accuracy:systematic:217GHz}{}
\setsymbol{HFI:FIRAS:gain:calibration:accuracy:systematic:353GHz}{}
\setsymbol{HFI:FIRAS:gain:calibration:accuracy:systematic:545GHz}{7\%}
\setsymbol{HFI:FIRAS:gain:calibration:accuracy:systematic:857GHz}{7\%}

\setsymbol{HFI:FIRAS:zero:point:accuracy:100GHz:units}{0.8\MJysr}
\setsymbol{HFI:FIRAS:zero:point:accuracy:143GHz:units}{}
\setsymbol{HFI:FIRAS:zero:point:accuracy:217GHz:units}{}
\setsymbol{HFI:FIRAS:zero:point:accuracy:353GHz:units}{1.4\MJysr}
\setsymbol{HFI:FIRAS:zero:point:accuracy:545GHz:units}{2.2\MJysr}
\setsymbol{HFI:FIRAS:zero:point:accuracy:857GHz:units}{1.7\MJysr}

\setsymbol{HFI:FIRAS:zero:point:accuracy:100GHz}{0.8}
\setsymbol{HFI:FIRAS:zero:point:accuracy:143GHz}{}
\setsymbol{HFI:FIRAS:zero:point:accuracy:217GHz}{}
\setsymbol{HFI:FIRAS:zero:point:accuracy:353GHz}{1.4}
\setsymbol{HFI:FIRAS:zero:point:accuracy:545GHz}{2.2}
\setsymbol{HFI:FIRAS:zero:point:accuracy:857GHz}{1.7}

\setsymbol{HFI:unit:conversion:100GHz:units}{0.2415\MJysrmK}
\setsymbol{HFI:unit:conversion:143GHz:units}{0.3694\MJysrmK}
\setsymbol{HFI:unit:conversion:217GHz:units}{0.4811\MJysrmK}
\setsymbol{HFI:unit:conversion:353GHz:units}{0.2883\MJysrmK}
\setsymbol{HFI:unit:conversion:545GHz:units}{0.05826\MJysrmK}
\setsymbol{HFI:unit:conversion:857GHz:units}{0.002238\MJysrmK}

\setsymbol{HFI:unit:conversion:100GHz}{0.2415}
\setsymbol{HFI:unit:conversion:143GHz}{0.3694}
\setsymbol{HFI:unit:conversion:217GHz}{0.4811}
\setsymbol{HFI:unit:conversion:353GHz}{0.2883}
\setsymbol{HFI:unit:conversion:545GHz}{0.05826}
\setsymbol{HFI:unit:conversion:857GHz}{0.002238}

\setsymbol{HFI:colour:correction:alpha=-2:V1.01:100GHz}{0.9893}
\setsymbol{HFI:colour:correction:alpha=-2:V1.01:143GHz}{0.9759}
\setsymbol{HFI:colour:correction:alpha=-2:V1.01:217GHz}{1.0007}
\setsymbol{HFI:colour:correction:alpha=-2:V1.01:353GHz}{1.0028}
\setsymbol{HFI:colour:correction:alpha=-2:V1.01:545GHz}{1.0019}
\setsymbol{HFI:colour:correction:alpha=-2:V1.01:857GHz}{0.9889}

\setsymbol{HFI:colour:correction:alpha=0:V1.01:100GHz}{1.0008}
\setsymbol{HFI:colour:correction:alpha=0:V1.01:143GHz}{1.0148}
\setsymbol{HFI:colour:correction:alpha=0:V1.01:217GHz}{0.9909}
\setsymbol{HFI:colour:correction:alpha=0:V1.01:353GHz}{0.9888}
\setsymbol{HFI:colour:correction:alpha=0:V1.01:545GHz}{0.9878}
\setsymbol{HFI:colour:correction:alpha=0:V1.01:857GHz}{1.0014}

\newcommand{\herschel}{{\it Herschel }}  %
\newcommand{\cobe}{{\it COBE }}  %
\newcommand{\dirbe}{{\it DIRBE }}  %
\newcommand{\archeops}{{\it Archeops }}  %
\newcommand{\wmap}{{\it WMAP }}  %
\newcommand{\iras}{{\it IRAS }}  %
\newcommand{\iris}{{\it IRAS-IRIS }}  %
\newcommand{\hfi}{{\it HFI }}  %
\newcommand{\lfi}{{\it LFI }}  %
\newcommand{\firas}{{\it FIRAS }}  %
\newcommand{\spitzer}{{\it Spitzer }}  %
\newcommand{\parkes}{{\it Parkes }}  %
\newcommand{\atca}{{\it ATCA }}  %
\newcommand{\spire}{{\it SPIRE }}  %

\newcommand{\Td}{T_{\rm D}} 
\newcommand{\NH}{N_{\rm H}} 
\newcommand{\InuTOT}{I_{\nu}^{\rm tot}} %
\newcommand{\InuNOCMB}{I_{\nu}^{\rm noCMB}} %
\newcommand{\InuSUB}{I_{\nu}^{\rm sub}} %
\newcommand{\lcor}{l_{\rm cor}} %

\newcommand{\mic}{\,{\rm \upmu m} } 
\newcommand{\NHUNIT}{\rm \,Hcm^{-2}} 
\newcommand{\Kkms}{\rm \,Kkms^{-1}}
\newcommand{\sigII}{\rm \sigma_{II}} 
\newcommand{\sigabs}{\rm \sigma_{abs}} 
\newcommand{\radeux}{\rm \alpha_{2000}} 
\newcommand{\decdeux}{\rm \delta_{2000}} 

\newcommand{\bII}{$\rm b_{II}$}

\newcommand{\betalmc}{1.5} 
\newcommand{\betasmc}{1.2} 

\newcommand{\halpha}{\rm H_\alpha}
\newcommand{\whi}{W_{\rm HI}}
\newcommand{\wco}{W_{\rm CO}}
\newcommand{\lambref}{\lambda_{\rm ref}}

\newcommand{\ralmc}{05h15m30s}
\newcommand{\declmc}{-68\degr30'10"}
\newcommand{\Rlmc}{4.09}

\newcommand{\rasmc}{01h04m16s}
\newcommand{\decsmc}{-72\degr51'36"}
\newcommand{\Rsmc}{2.38}

\newcommand{\IRASfourfreq}{2997.92}
\newcommand{\HFIonefreq}{\getsymbol{HFI:center:frequency:857GHz}}
\newcommand{\HFItwofreq}{\getsymbol{HFI:center:frequency:545GHz}}
\newcommand{\HFIthreefreq}{\getsymbol{HFI:center:frequency:353GHz}}
\newcommand{\HFIfourfreq}{\getsymbol{HFI:center:frequency:217GHz}}
\newcommand{\HFIfivefreq}{\getsymbol{HFI:center:frequency:143GHz}}
\newcommand{\HFIsixfreq}{\getsymbol{HFI:center:frequency:100GHz}}
\newcommand{\LFIonefreq}{\getsymbol{LFI:center:frequency:70GHz}}
\newcommand{\LFItwofreq}{\getsymbol{LFI:center:frequency:44GHz}}
\newcommand{\LFIthreefreq}{\getsymbol{LFI:center:frequency:30GHz}}
\newcommand{\WMAPonefreq}{93.69}
\newcommand{\WMAPtwofreq}{61.18}
\newcommand{\WMAPthreefreq}{41.07}
\newcommand{\WMAPfourfreq}{32.94}
\newcommand{\WMAPfivefreq}{23.06}
\newcommand{\IRASfourreso}{4.30}
\newcommand{\HFIonereso}{\getsymbol{HFI:FWHM:Mars:857GHz}}
\newcommand{\HFItworeso}{\getsymbol{HFI:FWHM:Mars:545GHz}}
\newcommand{\HFIthreereso}{\getsymbol{HFI:FWHM:Mars:353GHz}}
\newcommand{\HFIfourreso}{\getsymbol{HFI:FWHM:Mars:217GHz}}
\newcommand{\HFIfivereso}{\getsymbol{HFI:FWHM:Mars:143GHz}}
\newcommand{\HFIsixreso}{\getsymbol{HFI:FWHM:Mars:100GHz}}
\newcommand{\LFIonereso}{\getsymbol{LFI:FWHM:70GHz}}
\newcommand{\LFItworeso}{\getsymbol{LFI:FWHM:44GHz}}
\newcommand{\LFIthreereso}{\getsymbol{LFI:FWHM:30GHz}}
\newcommand{\WMAPonereso}{13.2}
\newcommand{\WMAPtworeso}{21.0}
\newcommand{\WMAPthreereso}{30.6}
\newcommand{\WMAPfourreso}{39.6}
\newcommand{\WMAPfivereso}{52.8}
\newcommand{\IRASfourabserr}{13.6}
\newcommand{\HFIoneabserr}{\getsymbol{HFI:FIRAS:gain:calibration:accuracy:systematic:857GHz}}
\newcommand{\HFItwoabserr}{\getsymbol{HFI:FIRAS:gain:calibration:accuracy:systematic:545GHz}}
\newcommand{\HFIthreeabserr}{\getsymbol{HFI:CMB:absolute:calibration:353GHz}}
\newcommand{\HFIfourabserr}{\getsymbol{HFI:CMB:absolute:calibration:217GHz}}

\newcommand{\HFIsixabserr}{\getsymbol{HFI:CMB:absolute:calibration:100GHz}}
\newcommand{\LFIoneabserr}{5.0}
\newcommand{\LFItwoabserr}{5.0}
\newcommand{\LFIthreeabserr}{5.0}
\newcommand{\WMAPoneabserr}{1.0}
\newcommand{\WMAPtwoabserr}{1.0}
\newcommand{\WMAPthreeabserr}{1.0}
\newcommand{\WMAPfourabserr}{1.0}
\newcommand{\WMAPfiveabserr}{1.0}

\newcommand{\Healpix}{\rm HEALPix}

\newcommand{\medwHIforeLMC}{\rm 297}   
\newcommand{\medwHIforeSMC}{\rm 172}   
\newcommand{\medHFIoneforeLMC}{\rm 3.0}  
\newcommand{\medHFIoneforeSMC}{\rm 1.7}  

\newcommand{\raNsixsix}{\rm 00h59m27.40s}
\newcommand{\decNsixsix}{\rm -72\degr10'11"}
\newcommand{\raNheightthree}{\rm 01h14m21.0s}
\newcommand{\decNheightthree}{\rm -73\degr17'12"}
\newcommand{\raNheightone}{\rm 01h09m13.6s}
\newcommand{\decNheightone}{\rm -73\degr11'41"}
\newcommand{\raNheightheight}{\rm 01h24m08.1s}
\newcommand{\decNheightheight}{\rm -73\degr08'55"}
\newcommand{\raSfivefour}{\rm 00h50m25.9s}
\newcommand{\decSfivefour}{\rm -72\degr53'10"}
\newcommand{\raNsevensix}{\rm 01h04m01.2s}
\newcommand{\decNsevensix}{\rm -72\degr01'52"}

\begin{document}

\author{\small
Planck Collaboration:
P.~A.~R.~Ade\inst{69}
\and
N.~Aghanim\inst{45}
\and
M.~Arnaud\inst{55}
\and
M.~Ashdown\inst{53, 75}
\and
J.~Aumont\inst{45}
\and
C.~Baccigalupi\inst{67}
\and
A.~Balbi\inst{28}
\and
A.~J.~Banday\inst{73, 6, 60}
\and
R.~B.~Barreiro\inst{50}
\and
J.~G.~Bartlett\inst{3, 51}
% ...
}
\institute{\small
Aalto University Mets\"{a}hovi Radio Observatory, Mets\"{a}hovintie 114, FIN-02540 Kylm\"{a}l\"{a}, Finland\\
\and
Agenzia Spaziale Italiana Science Data Center, c/o ESRIN, via Galileo Galilei, Frascati, Italy\\
\and
Astroparticule et Cosmologie, CNRS (UMR7164), Universit\'{e} Denis Diderot Paris 7, B\^{a}timent Condorcet, 10 rue A. Domon et L\'{e}onie Duquet, Paris, France\\
\and
Atacama Large Millimeter/submillimeter Array, ALMA Santiago Central Offices Alonso de Cordova 3107, Vitacura, Casilla 763 0355, Santiago, Chile\\
\and
CITA, University of Toronto, 60 St. George St., Toronto, ON M5S 3H8, Canada\\
\and
CNRS, IRAP, 9 Av. colonel Roche, BP 44346, F-31028 Toulouse cedex 4, France\\
\and
California Institute of Technology, Pasadena, California, U.S.A.\\
\and
DAMTP, Centre for Mathematical Sciences, Wilberforce Road, Cambridge CB3 0WA, U.K.\\
\and
DSM/Irfu/SPP, CEA-Saclay, F-91191 Gif-sur-Yvette Cedex, France\\
\and
DTU Space, National Space Institute, Juliane Mariesvej 30, Copenhagen, Denmark\\
\and
Departamento de F\'{\i}sica, Universidad de Oviedo, Avda. Calvo Sotelo s/n, Oviedo, Spain\\
\and
Department of Astronomy and Astrophysics, University of Toronto, 50 Saint George Street, Toronto, Ontario, Canada\\
\and
Department of Astronomy and Earth Sciences, Tokyo Gakugei University, Koganei, Tokyo 184-8501, Japan\\
\and
Department of Physical Science, Graduate School of Science, Osaka Prefecture University, 1-1 Gakuen-cho, Naka-ku, Sakai, Osaka 599-8531, Japan\\
\and
Department of Physics \& Astronomy, University of British Columbia, 6224 Agricultural Road, Vancouver, British Columbia, Canada\\
\and
Department of Physics, Gustaf H\"{a}llstr\"{o}min katu 2a, University of Helsinki, Helsinki, Finland\\
\and
Department of Physics, Nagoya University, Chikusa-ku, Nagoya, 464-8602, Japan\\
\and
Department of Physics, Princeton University, Princeton, New Jersey, U.S.A.\\
\and
Department of Physics, Purdue University, 525 Northwestern Avenue, West Lafayette, Indiana, U.S.A.\\
\and
Department of Physics, University of California, Berkeley, California, U.S.A.\\
\and
Department of Physics, University of California, One Shields Avenue, Davis, California, U.S.A.\\
\and
Department of Physics, University of California, Santa Barbara, California, U.S.A.\\
\and
Department of Physics, University of Illinois at Urbana-Champaign, 1110 West Green Street, Urbana, Illinois, U.S.A.\\
\and
Dipartimento di Fisica G. Galilei, Universit\`{a} degli Studi di Padova, via Marzolo 8, 35131 Padova, Italy\\
\and
Dipartimento di Fisica, Universit\`{a} La Sapienza, P. le A. Moro 2, Roma, Italy\\
\and
Dipartimento di Fisica, Universit\`{a} degli Studi di Milano, Via Celoria, 16, Milano, Italy\\
\and
Dipartimento di Fisica, Universit\`{a} degli Studi di Trieste, via A. Valerio 2, Trieste, Italy\\
\and
Dipartimento di Fisica, Universit\`{a} di Roma Tor Vergata, Via della Ricerca Scientifica, 1, Roma, Italy\\
\and
Discovery Center, Niels Bohr Institute, Blegdamsvej 17, Copenhagen, Denmark\\
\and
Dpto. Astrof\'{i}sica, Universidad de La Laguna (ULL), E-38206 La Laguna, Tenerife, Spain\\
\and
European Southern Observatory, ESO Vitacura, Alonso de Cordova 3107, Vitacura, Casilla 19001, Santiago, Chile\\
\and
European Space Agency, ESAC, Planck Science Office, Camino bajo del Castillo, s/n, Urbanizaci\'{o}n Villafranca del Castillo, Villanueva de la Ca\~{n}ada, Madrid, Spain\\
\and
European Space Agency, ESTEC, Keplerlaan 1, 2201 AZ Noordwijk, The Netherlands\\
\and
Helsinki Institute of Physics, Gustaf H\"{a}llstr\"{o}min katu 2, University of Helsinki, Helsinki, Finland\\
\and
INAF - Osservatorio Astrofisico di Catania, Via S. Sofia 78, Catania, Italy\\
\and
INAF - Osservatorio Astronomico di Padova, Vicolo dell'Osservatorio 5, Padova, Italy\\
\and
INAF - Osservatorio Astronomico di Roma, via di Frascati 33, Monte Porzio Catone, Italy\\
\and
INAF - Osservatorio Astronomico di Trieste, Via G.B. Tiepolo 11, Trieste, Italy\\
\and
INAF/IASF Bologna, Via Gobetti 101, Bologna, Italy\\
\and
INAF/IASF Milano, Via E. Bassini 15, Milano, Italy\\
\and
INRIA, Laboratoire de Recherche en Informatique, Universit\'{e} Paris-Sud 11, B\^{a}timent 490, 91405 Orsay Cedex, France\\
\and
IPAG: Institut de Plan\'{e}tologie et d'Astrophysique de Grenoble, Universit\'{e} Joseph Fourier, Grenoble 1 / CNRS-INSU, UMR 5274, Grenoble, F-38041, France\\
\and
Imperial College London, Astrophysics group, Blackett Laboratory, Prince Consort Road, London, SW7 2AZ, U.K.\\
\and
Infrared Processing and Analysis Center, California Institute of Technology, Pasadena, CA 91125, U.S.A.\\
\and
Institut d'Astrophysique Spatiale, CNRS (UMR8617) Universit\'{e} Paris-Sud 11, B\^{a}timent 121, Orsay, France\\
\and
Institut d'Astrophysique de Paris, CNRS UMR7095, Universit\'{e} Pierre \& Marie Curie, 98 bis boulevard Arago, Paris, France\\
\and
Institute of Astronomy and Astrophysics, Academia Sinica, Taipei, Taiwan\\
\and
Institute of Theoretical Astrophysics, University of Oslo, Blindern, Oslo, Norway\\
\and
Instituto de Astrof\'{\i}sica de Canarias, C/V\'{\i}a L\'{a}ctea s/n, La Laguna, Tenerife, Spain\\
\and
Instituto de F\'{\i}sica de Cantabria (CSIC-Universidad de Cantabria), Avda. de los Castros s/n, Santander, Spain\\
\and
Jet Propulsion Laboratory, California Institute of Technology, 4800 Oak Grove Drive, Pasadena, California, U.S.A.\\
\and
Jodrell Bank Centre for Astrophysics, Alan Turing Building, School of Physics and Astronomy, The University of Manchester, Oxford Road, Manchester, M13 9PL, U.K.\\
\and
Kavli Institute for Cosmology Cambridge, Madingley Road, Cambridge, CB3 0HA, U.K.\\
\and
LERMA, CNRS, Observatoire de Paris, 61 Avenue de l'Observatoire, Paris, France\\
\and
Laboratoire AIM, IRFU/Service d'Astrophysique - CEA/DSM - CNRS - Universit\'{e} Paris Diderot, B\^{a}t. 709, CEA-Saclay, F-91191 Gif-sur-Yvette Cedex, France\\
\and
Laboratoire Traitement et Communication de l'Information, CNRS (UMR 5141) and T\'{e}l\'{e}com ParisTech, 46 rue Barrault F-75634 Paris Cedex 13, France\\
\and
Laboratoire de Physique Subatomique et de Cosmologie, CNRS, Universit\'{e} Joseph Fourier Grenoble I, 53 rue des Martyrs, Grenoble, France\\
\and
Laboratoire de l'Acc\'{e}l\'{e}rateur Lin\'{e}aire, Universit\'{e} Paris-Sud 11, CNRS/IN2P3, Orsay, France\\
\and
Lawrence Berkeley National Laboratory, Berkeley, California, U.S.A.\\
\and
Max-Planck-Institut f\"{u}r Astrophysik, Karl-Schwarzschild-Str. 1, 85741 Garching, Germany\\
\and
MilliLab, VTT Technical Research Centre of Finland, Tietotie 3, Espoo, Finland\\
\and
National University of Ireland, Department of Experimental Physics, Maynooth, Co. Kildare, Ireland\\
\and
Niels Bohr Institute, Blegdamsvej 17, Copenhagen, Denmark\\
\and
Observational Cosmology, Mail Stop 367-17, California Institute of Technology, Pasadena, CA, 91125, U.S.A.\\
\and
Observatoire Astronomique de Strasbourg, CNRS, UMR7550, F-67000 Strasbourg, France\\
\and
Optical Science Laboratory, University College London, Gower Street, London, U.K.\\
\and
SISSA, Astrophysics Sector, via Bonomea 265, 34136, Trieste, Italy\\
\and
SUPA, Institute for Astronomy, University of Edinburgh, Royal Observatory, Blackford Hill, Edinburgh EH9 3HJ, U.K.\\
\and
School of Physics and Astronomy, Cardiff University, Queens Buildings, The Parade, Cardiff, CF24 3AA, U.K.\\
\and
Space Sciences Laboratory, University of California, Berkeley, California, U.S.A.\\
\and
Spitzer Science Center, 1200 E. California Blvd., Pasadena, California, U.S.A.\\
\and
Stanford University, Dept of Physics, Varian Physics Bldg, 382 Via Pueblo Mall, Stanford, California, U.S.A.\\
\and
Universit\'{e} de Toulouse, UPS-OMP, IRAP, F-31028 Toulouse cedex 4, France\\
\and
Universities Space Research Association, Stratospheric Observatory for Infrared Astronomy, MS 211-3, Moffett Field, CA 94035, U.S.A.\\
\and
University of Cambridge, Cavendish Laboratory, Astrophysics group, J J Thomson Avenue, Cambridge, U.K.\\
% ...
}

\allearlypapers

\title{\textit{Planck} Early Results: Origin of the submm excess dust emission in the Magellanic Clouds}

\titlerunning{Origin of the millimetre excess in the LMC and SMC}
\authorrunning{{\Planck} collaboration}


\abstract{
The integrated Spectral Energy Distributions (SED) of the Large
Magellanic Cloud (LMC) and Small Magellanic Cloud (SMC) appear
significantly flatter than expected from dust models based on their
far-infrared and radio emission.
The origin of this millimetre excess is still unexplained, and is here 
investigated using the {\Planck}
data.  The integrated SED of the two galaxies before subtraction of
the foreground (Milky Way) and background (CMB fluctuations) emission
are in good agreement with previous determinations, confirming the
presence of the millimetre excess.

The background CMB contribution is subtracted using an Internal Linear
Combination (ILC) method performed locally around the galaxies. The
foreground emission from the Milky Way is subtracted as a Galactic
\ion{H}{i} template and the dust emissivity is derived in a region
surrounding the two galaxies and dominated by Milky Way emission.
After subtraction, the remaining emission of both galaxies correlates
closely with the atomic and molecular gas emission of the LMC and SMC. The millimetre
excess in the LMC can be explained by CMB fluctuations, but a
significant excess is still present in the SMC SED.

The {\Planck} and \iris data at $100 \mic$ are combined to produce
thermal dust temperature and optical depth maps of the two
galaxies. The LMC temperature map shows the presence of a warm inner
arm already found with the {\spitzer} data, but also shows the
existence of a previously unidentified cold outer arm.  Several cold
regions are found along this arm, some of which are associated with
known molecular clouds.  The dust optical depth maps are used to
constrain the thermal dust emissivity power law index ($\beta$). The
average spectral index is found to be consistent with
$\beta=${\betalmc} and $\beta=${\betasmc} below $500\mic$ for the LMC
and SMC respectively, significantly flatter than the values observed
in the Milky Way. Furthermore, there is evidence in the SMC for a
further flattening of the SED in the sub-mm, unlike for the LMC where
the SED remains consistent with $\beta=${\betalmc}.  The spatial
distribution of the millimetre dust excess in the SMC follows the gas
and thermal dust distribution.

Different models are explored in order to fit the dust emission in the
SMC. It is concluded that the millimetre excess is unlikely to be
caused by very cold dust emission and that it could be due to a
combination of spinning dust emission and thermal dust emission by
more amorphous dust grains than those present in our Galaxy.
}

\keywords{
ISM: general, dust, extinction, clouds --
Galaxies: ISM --
Infrared: ISM --
Submillimeter: ISM
               }

\maketitle
%

\section{Introduction}

Star formation and the exchange and evolution of materials
between the stars and the interstallar medium (ISM) are continuous 
processes that drive the evolution
of galaxies. As stars evolve, die and renew the life cycle of dust
and gas, the amount of dust and its distribution in a galaxy has
important consequences for its subsequent star formation. Thus knowing
the dust content throughout cosmic history would provide clues to the
star formation history of the universe as the metallicities evolve.
One of the puzzling results that has emerged from the studies of 
dust SEDs in the early universe, is finding very high dust masses in
high redshift galaxies. For example, recent \herschel observations of
submillimetre galaxies (SMGs) find excessively high dust masses,
and high dust-to-gas mass ratios (D/G) \citep{Santini2010}. These
findings contradict the low metallicities measured in the gas.  Large
dust masses have also been measured for a range of low metallicity
local universe dwarf galaxies \citep{Dumke2003,
Galliano2003, Galliano2005, Bendo2006, Galametz2009, Grossi2010,
O'Halloran2010}. 
Our understanding of how dust masses are estimated and what
observational contraints are necessary is called into question by 
the discovery that the dust masses measured in these galaxies appear
significantly higher than expected from their metal content. 

Excessively large dust masses only seem to be found in low
metallicity galaxies to date. The evidence is found in the behavior
of the submillimetre (submm) emission, beyond about $400-500\mic$.
From a study of a broad range of metal-rich and metal-poor galaxies
observed with wide wavelength coverage that included submm
observations beyond $500\mic$, a submm excess, beyond the normal dust
SED models, was found only for low metallicity systems
\citep{Galametz2010}.  Such submm excess is particularly evident in
the Magellanic Clouds \citep{Israel2010,Bot2010}. The origin of such
excess is still uncertain but several suggestions have been put
forward:
 
\begin{itemize}

\item{The submm excess has been modeled as a cold dust component with
a submm emissivity index ($\beta$) of $\beta$=1, which suggests a low dust
temperature of $\sim$10\,K \citep{Galliano2003, Galliano2005,
Galametz2009, Galametz2010}.  There are few observational constraints
in the submm wavelength window, hence this cold component is a rather
ad hoc solution. This description often results in discrepantly large
D/G ratios being found for the low metallicity galaxies if the
estimates of the total gas reservoir are well known.}

\item{\cite{Meny2007} show that emission by amorphous grains is
expected to produce excess emission in the submm. As a result,
the spectral shape of emission by amorphous grains cannot be
reproduced by a single emissivity index over the whole Far-InfraRed (FIR) to millimetre (mm) range,
and is expected to flatten at long wavelengths. The intensity of the
excess is also predicted to depend strongly on the temperature of the
dust grains.  Modifying the dust optical properties to incorporate the
effects of the disordered structure of the amorphous grains, showed
that the effective submm emissivity index decreases (thus flattening
the submm Rayleigh-Jeans slope) as the dust temperature
increases. Similarly, an inverse correlation between the dust
temperature and the spectral index has been observed in data from the
FIR to the submm \citep[e.g.][]{Dupac2003a, Paradis2010}.
However, these authors as well as \cite{Shetty2009b} advise caution in
the interpretation of the observed inverse $T$--$\beta$ relationship,
which is affected by the natural correlation between these two
parameters, and requires careful treatment of observational
uncertainties.  \cite{Shetty2009a} also studied the effect on the
observed $T$--$\beta$ relation of temperature variations along the
line-of-sight.  \cite{Paradis2009b} using the {\dirbe}, {\archeops}
and {\wmap} data showed that the FIR-submm SEDs of galactic molecular
clouds and their neutral surroundings indeed show a flattening beyond
$\lambda\simeq 500 \mic$.}

\item{Spinning dust emitted in the ionised gas of galaxies has been
suggested as the explanation for the radio emission often seen in
galaxies \citep{Ferrara1994}. This idea was further explored by
\cite{Anderson1993} and \cite{Draine1998a,Draine1998b} who
characterized the 10 to 100\,GHz anomalous foreground emission
component as spinning dust from very small dust particles such as
Polycyclic Aromatic Hydrocarbons (PAHs).  Recent improvements to the
model suggest the peak frequency, which could occur in the submm,
depends on many parameters, including the radiation field intensity,
the dust size distribution, the dipole moment distribution, the
physical parameters of the gas phases, etc. \citep{Hoang2010,
Silsbee2010, Ysard2010a, Ysard2010b}.  Spinning dust has been a
preferred explanation for the submm excess observed on the global
scale in the Magellanic Clouds \citep{Israel2010,Bot2010}.}

\item{
\cite{Serra2008} have suggested hydrogenated amorphous carbon as the
most likely carbonaceous grain material instead of graphite. The
advantage of using amorphous carbon, particularly in the cases where a
submm excess is found, is that amorphous carbons are more emissive and
result in a flatter submm slope and less dust mass. This has been used
with new \herschel observations to model some low metallicity galaxies
\citep{Galametz2010, O'Halloran2008, Meixner2010}.}

\item{The enhancement of hot, very small, stochastically heated grains
with a low dust emissivity, has also been suggested to characterize
the submm emission when there is a submm excess present
\citep{Lisenfeld2001, Zhu2009}.}

\end{itemize}

Some of the possible causes of the submm excess could be ruled out or
constrained if the gas mass estimates were better established.
However, estimates of the total gas reservoir in low metallicity
environments have been uncertain, particularly the molecular gas
component. Observations of CO in low metallicity galaxies have been a
challenge, and the dearth of detected CO has often been interpreted as
meaning there is very little mass in molecular gas present in galaxies
which are otherwise actively forming massive stars \citep{Leroy2009}.
However, relying on CO alone to trace the H$_{2}$ mass could
potentially miss a large reservoir of molecular gas residing outside
the CO-emitting region, as indicated by the FIR fine structure
observations in the Magellanic Clouds and other low metallicity
galaxies \citep[e.g.][]{Poglitsch1995, Madden1997, Madden2000}.  The
presence of a hidden gas mass in low metallicity galaxies is also
suggested by dust observations of the Magellanic Clouds
\citep[e.g.][]{Bernard2008, Dobashi2008, Leroy2007,
Roman-Duval2010,Bot2010b}.

Most of these studies which find a submm excess in low-metallicity
galaxies are derived from models on global galaxy scales and use data
with a wavelength coverage that is limited to $\lambda<500\mic$ with
\herschel or to $\lambda<870\mic$ for ground-based observations.
The \cite{Galametz2010} study demonstrated that dust masses can differ
by an order of magnitude depending on the availability of submm
constraints on the SED modelling.  These require wider wavelength
coverage of the submm and millimetre (mm) region and better spatial
information, to map out the submm excess and determine the local
physical conditions.

The closest low-metallicity galaxies are our neighboring Large
Magellanic Cloud (LMC) and Small Magellanic Cloud (SMC), which have
metallicities of 50\% and 20\% solar metallicity
(Z$_{\odot}$) respectively.
They are ideal laboratories for studying the ISM and
star formation in different low metallicity environments.  Recent
studies with {\spitzer} \citep[SAGE:][]{Meixner2006, Bolatto2007} and
{\herschel} \citep[HERITAGE:][]{Meixner2010} have mapped out the the
temperature and dust mass distribution \citep{Bernard2008, Gordon2009,
Gordon2010, Leroy2007}.  Recent {\herschel} studies of the LMC already
point to a possible submm band excess and an excessive dust mass,
which may be suggesting the presence of molecular gas not probed by
CO. This could be explained by the presence of amorphous carbon grains
\citep{Roman-Duval2010, Meixner2010, Gordon2010}.  It was difficult to
be conclusive about the cause of the excess seen in the \herschel
observation because of the lack of longer wavelength coverage.  Global
excess mm and submm emission in the LMC and SMC was indisputably shown
by \cite{Bot2010} and \cite{Israel2010} using broader wavelength
coverage that included the submm to cm observations of the Top-Hat
balloon experiment, {\wmap} and {\cobe} and their results favour
spinning dust as the explanation for the submm and mm excess, but
require additional spatial coverage to be firmly conclusive. {\Planck}
\footnote{\Planck\ (http://www.esa.int/\Planck ) is a project of the
European Space Agency (ESA) with instruments provided by two
scientific consortia funded by ESA member states (in particular the
lead countries: France and Italy) with contributions from NASA (USA),
and telescope reflectors provided in a collaboration between ESA and a
scientific consortium led and funded by Denmark.}  observations have
both wide wavelength coverage and the spatial resolution to locate the submm
emission within the LMC and SMC, which will help determine the origin
of the submm excess.

\section{Observations} 
\label{sec_obs}
\subsection{Planck data}
\label{sec_planckdata}

The {\Planck} first mission results are presented in
\cite{planck2011-1.1} and the in-flight performances of the two focal
plane instruments \hfi (High Frequency Instrument) and \lfi (Low
Frequency Instrument) are given in \cite{planck2011-1.5} and
\cite{planck2011-1.4} respectively. The data processing and
calibration of the {\hfi} and {\lfi} data used here is described in
\cite{planck2011-1.7} and \cite{planck2011-5.1a} respectively.

\begin{figure*}[ht]
\begin{center}
\includegraphics[width=9.1cm,angle=180]{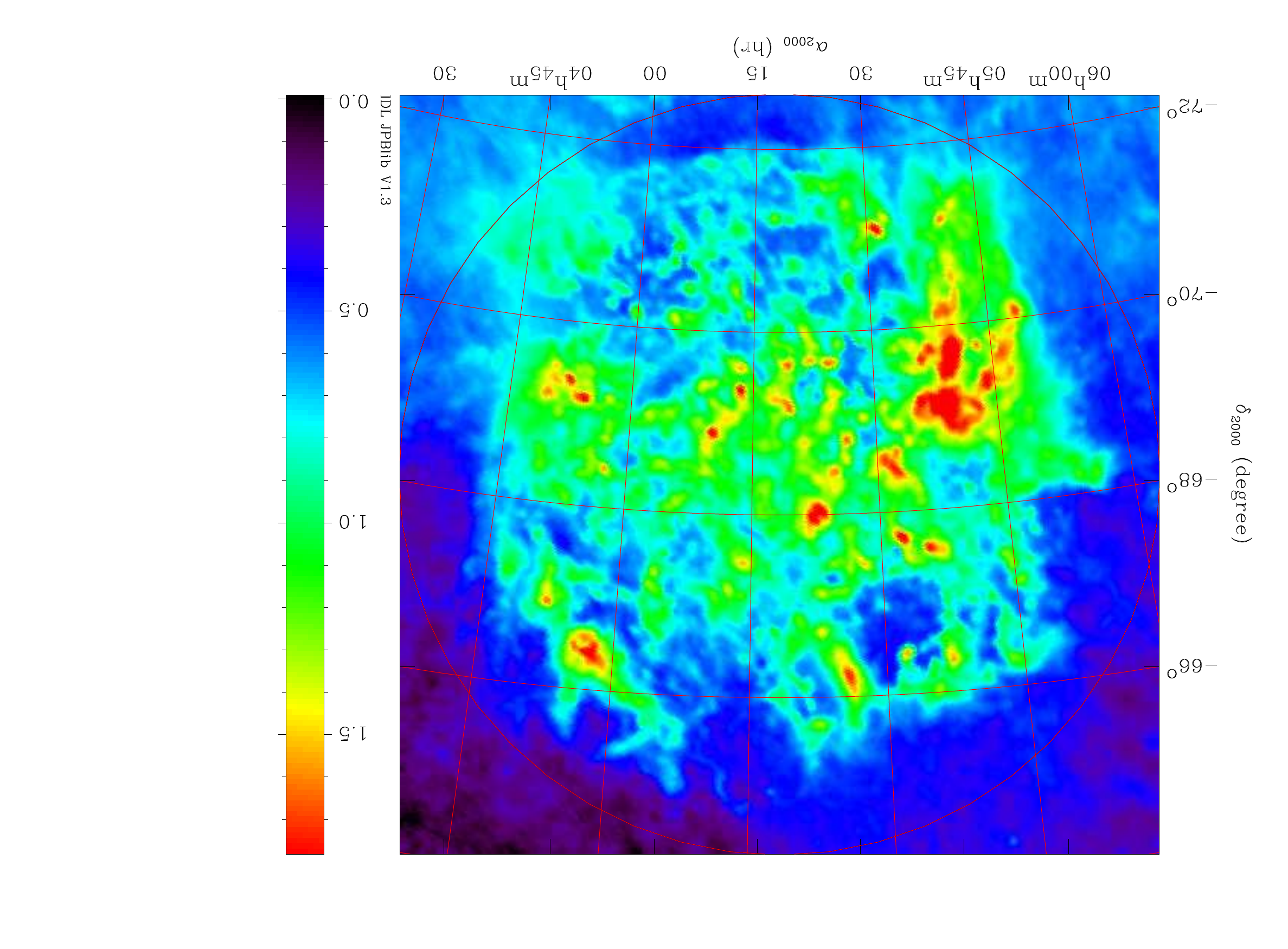}
\includegraphics[width=9.1cm,angle=180]{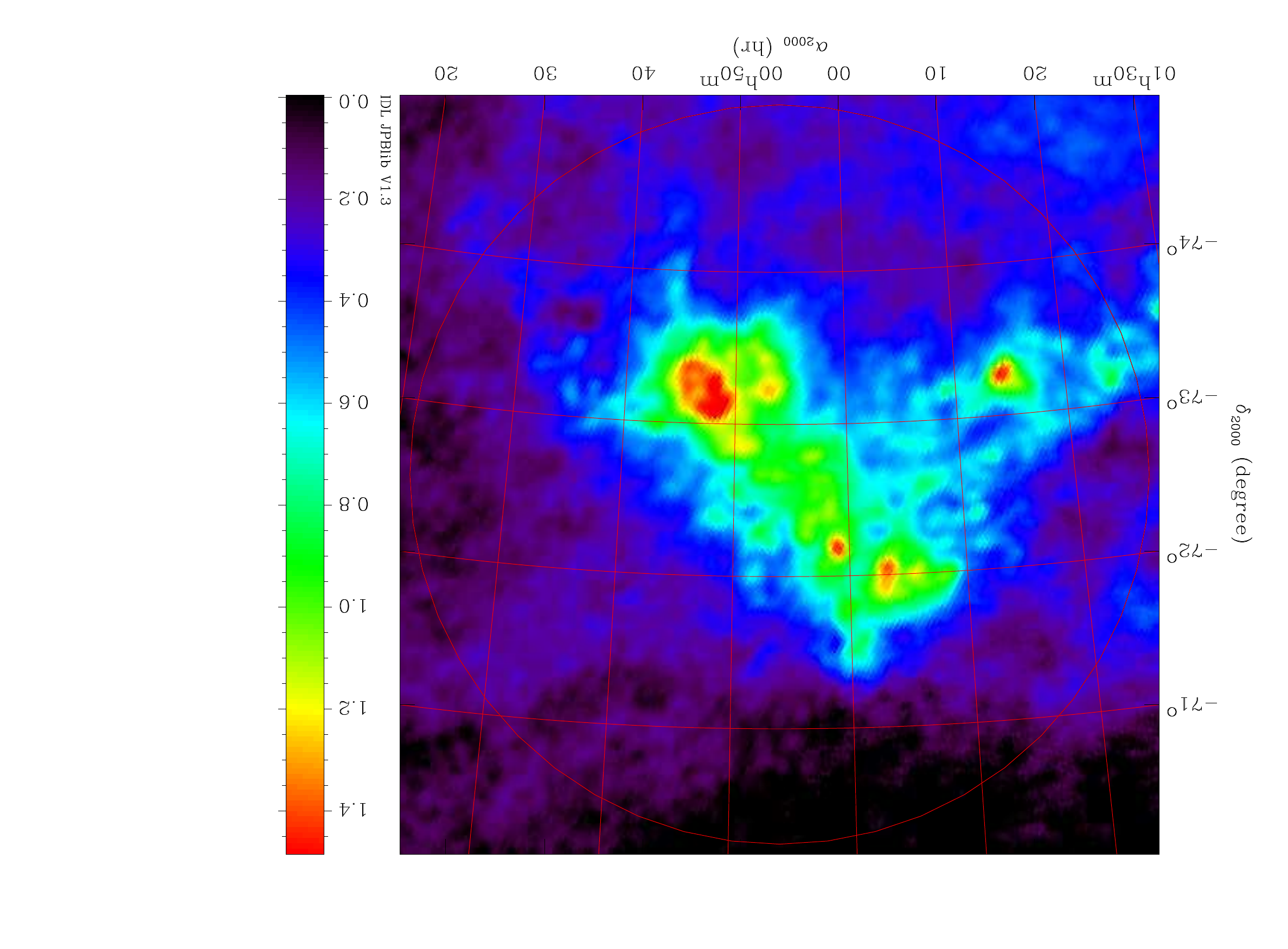}
\includegraphics[width=9.1cm,angle=180]{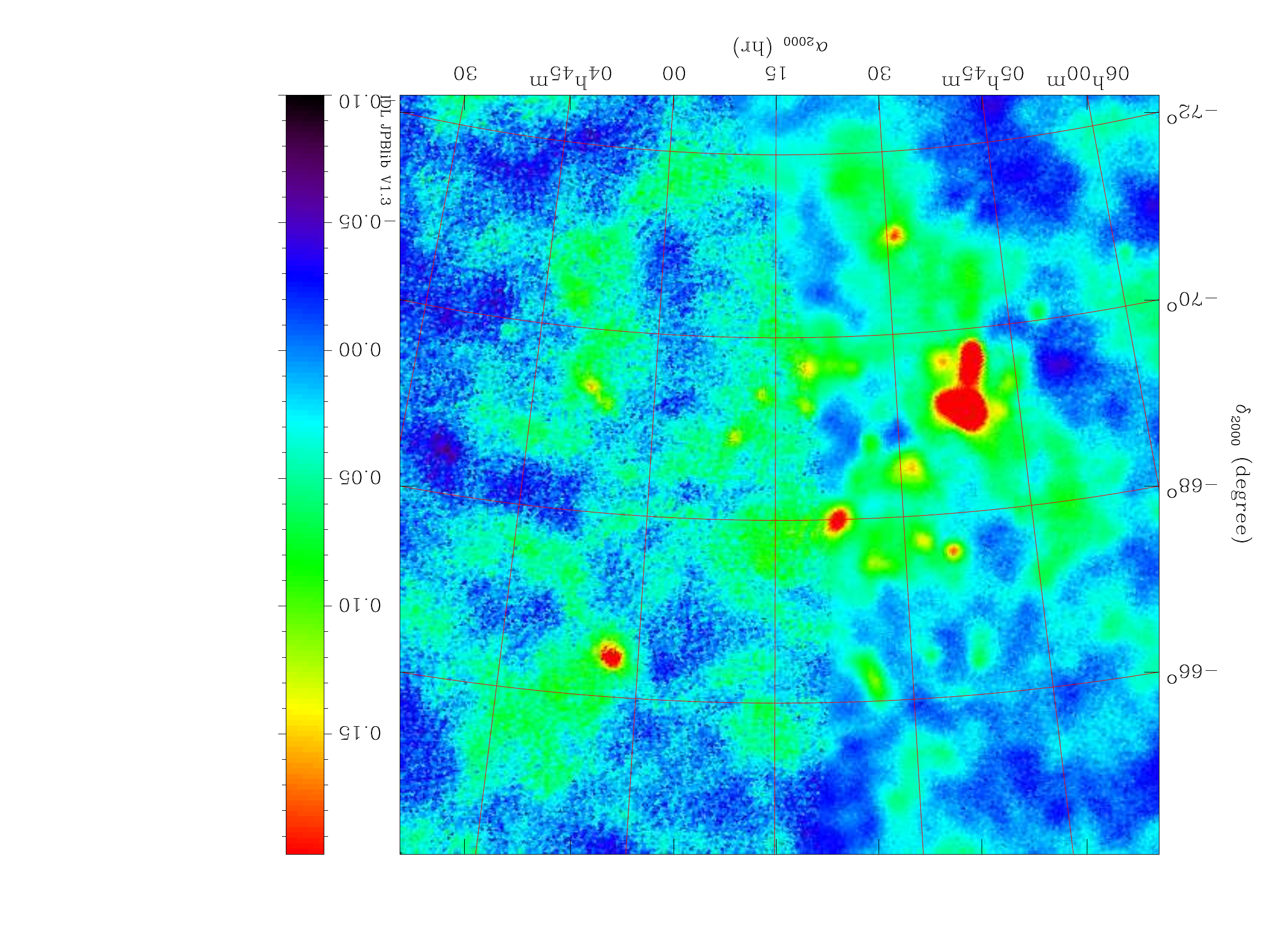}
\includegraphics[width=9.1cm,angle=180]{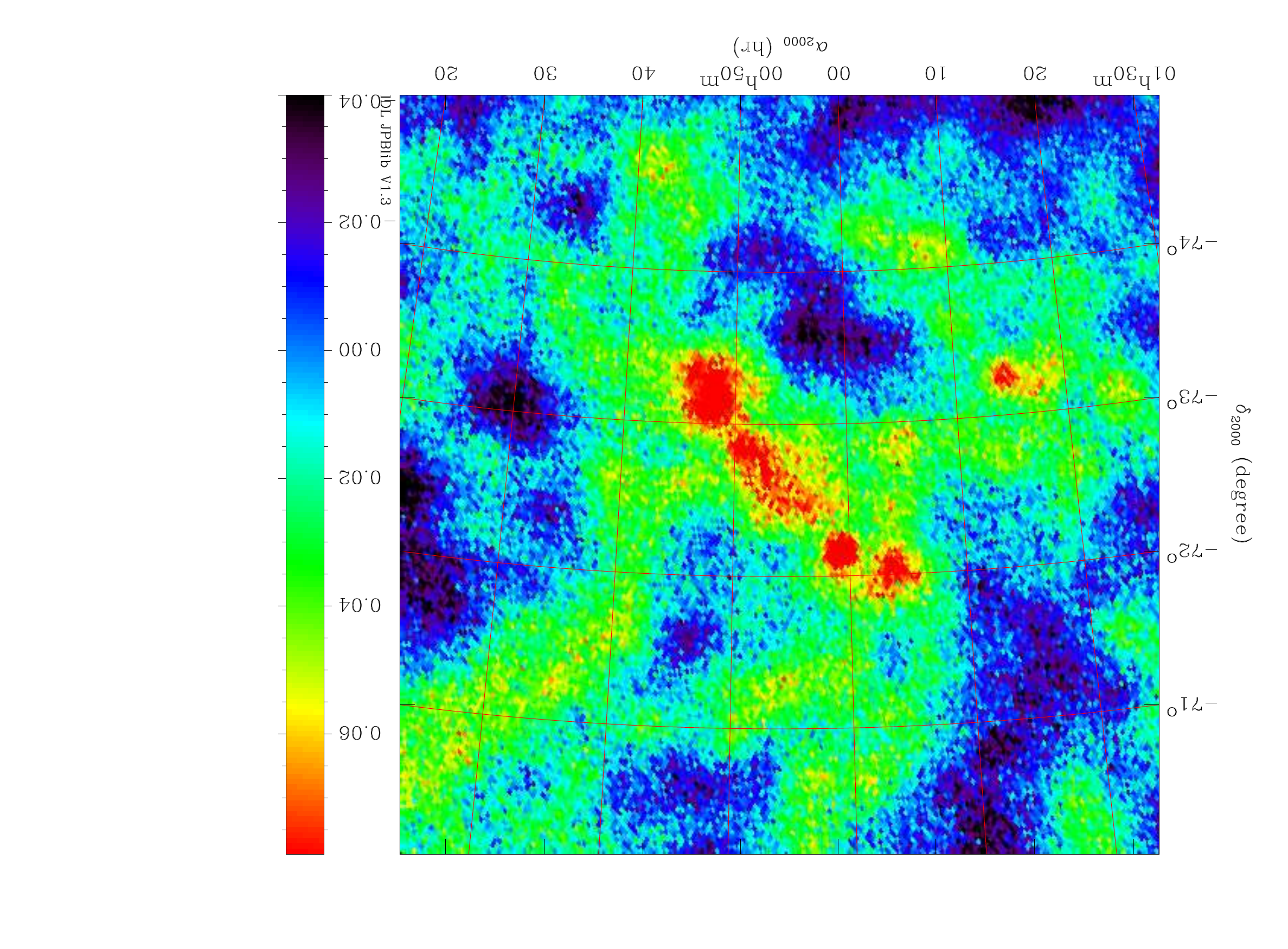}
\includegraphics[width=9.1cm,angle=180]{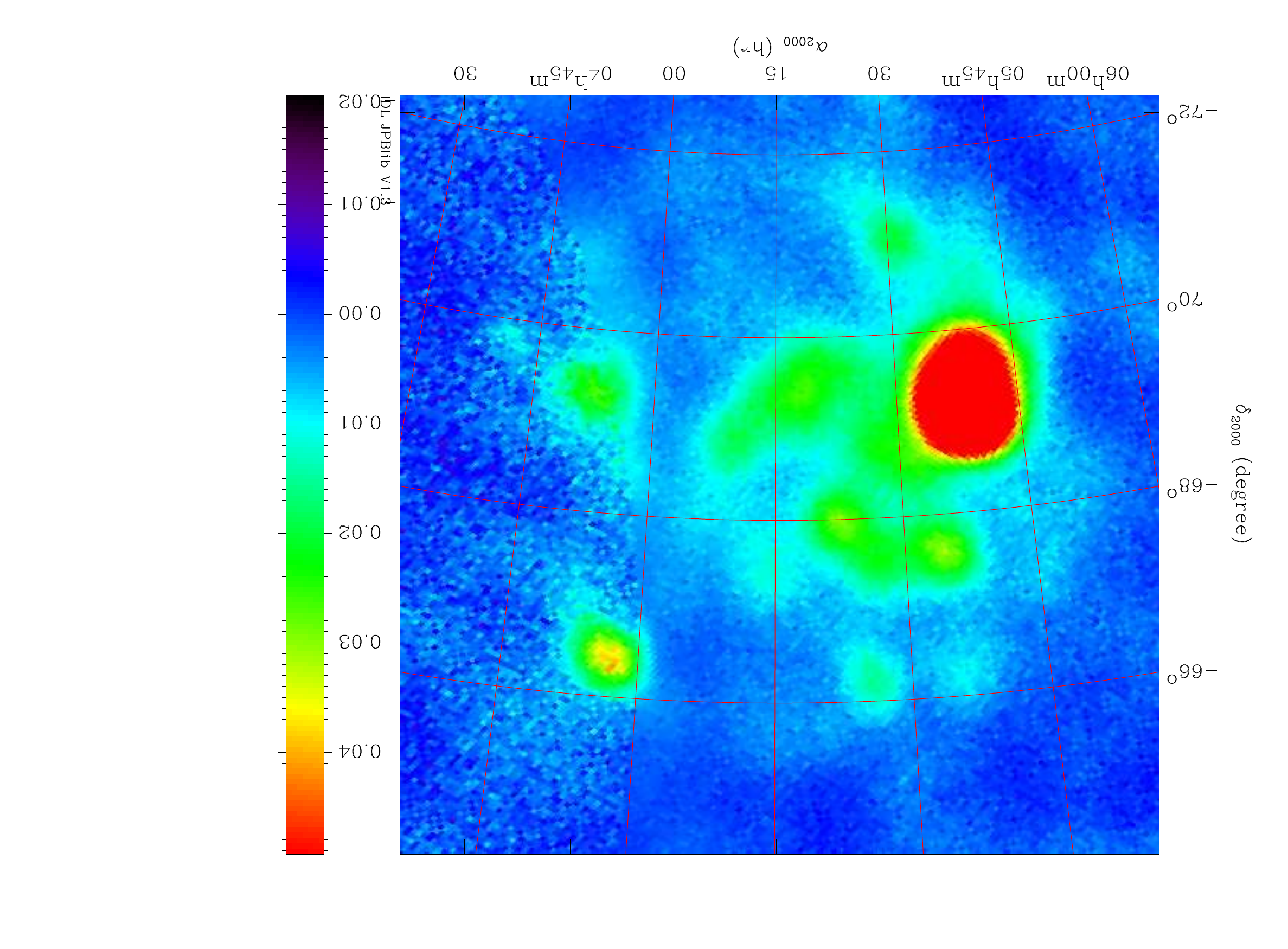}
\includegraphics[width=9.1cm,angle=180]{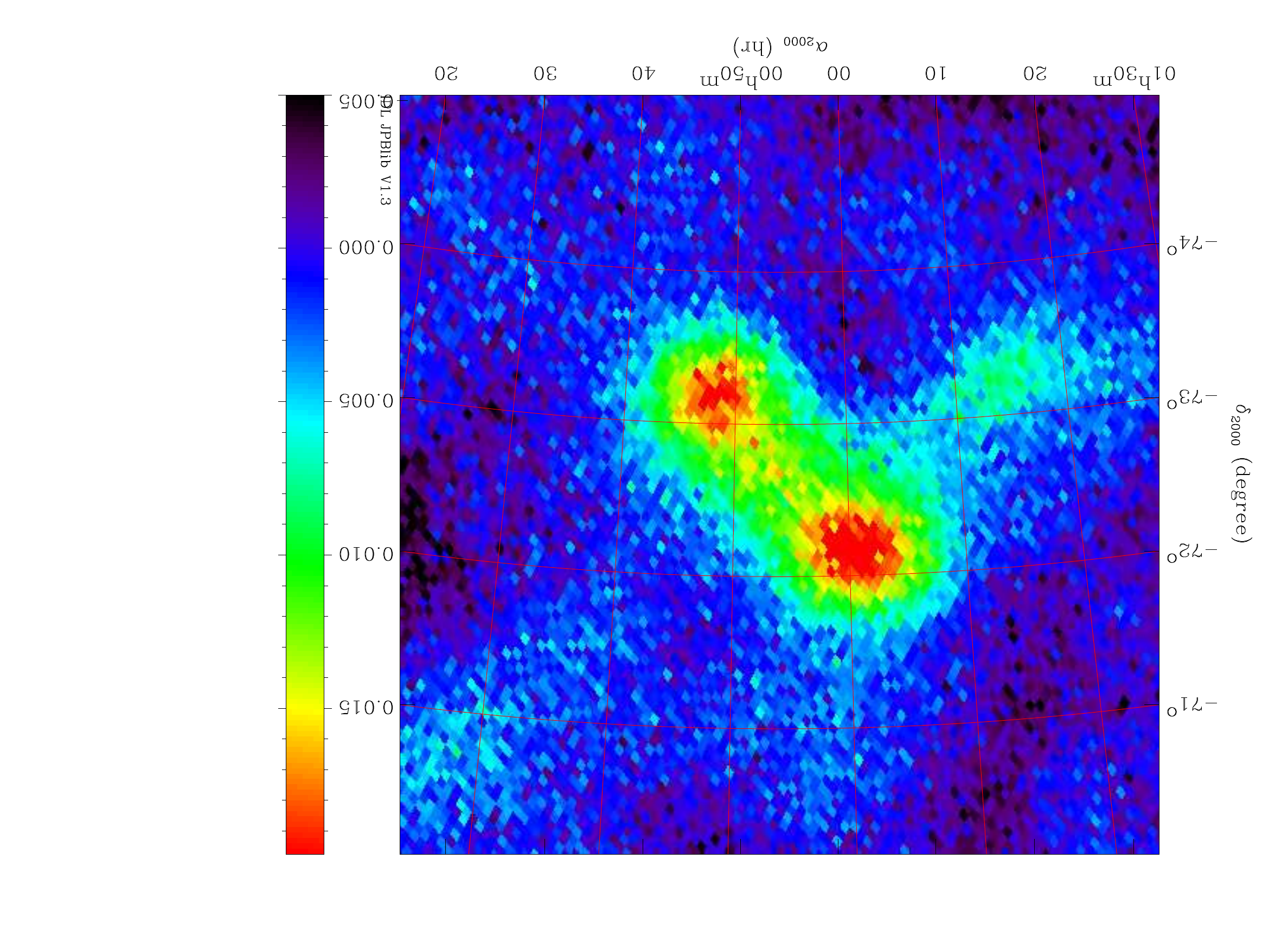}
\caption{\label{fig:Intensitymaps}
{\Planck} total Intensity data for the LMC (left) and SMC (right) at {\HFIonefreq}
(top), {\HFIsixfreq} (middle) and {\LFIthreefreq}\,\GHz (bottom) at full
resolution. The top panels are shown in log scale. The circle in the
top panels shows
the region used to extract average SEDs.
}
\end{center}
\end{figure*}

Figure\,\ref{fig:Intensitymaps} shows the total intensity maps
observed toward the LMC and SMC at a few {\hfi} and {\lfi}
frequencies.  Both galaxies are well detected at
high frequencies.  Around {\HFIsixfreq}\,GHz, their emission can barely
be distinguished from CMB fluctuations. At lower frequencies, the
contrast between the galaxies' emission and the CMB fluctuations
becomes larger again. Note that the apparent variation of the noise
level at a constant right-ascension value across the LMC is real and
is due to the LMC being positioned at the edge of the {\Planck} deep
field.

The {\Planck} DR2 data have had the CMB fluctuations removed in a way
which is inappropriate for detailed examination of some foreground sources
like the LMC and SMC.
We therefore use the original data before CMB subtraction (version DX4
of the {\Planck}-{\hfi} and {\Planck}-{\lfi} data) and perform our own
subtraction of the CMB fluctuations, based on a local ILC (see
Sec.\,\ref{sec:cmb_sub}).

We use the internal variance ($\sigII^2$) provided with the {\Planck}
data, which represents the white noise on the intensity.  The LMC and
SMC have been observed many times, as the {\Planck} scanning strategy
\citep{planck2011-1.1} covers the region close to the ecliptic pole
repeatedly. In particular, the LMC is located at the boundary of the
{\Planck} deep field. Its eastern half has a much higher signal to
noise ratio than its western half.  We assume the absolute
uncertainties from calibration given in \cite{planck2011-1.7} and
\cite{planck2011-5.1a} for {\hfi} and {\lfi} respectively and
summarized in Table\,\ref{tab:ancillary}.

\begin{table}
\begin{center}
\caption{\label{tab:ancillary} Characteristics of the FIR/submm data used in this study.}
\begin{tabular}{llllll}
\hline
\hline
Data   & $\rm \lambref$ & $\rm \nu_{ref}$ & $\theta$ & $\sigII$ & $\sigabs$ \\
       & [$\mic$]   & [\GHz]          & [arcmin]        & [$\rm MJy\,sr$$^{-1}$] & [\%] \\
\hline
 \iras &    100    &   \IRASfourfreq   &  \IRASfourreso &      0.06$^\dag$$^\ddag$ &   \IRASfourabserr$^\ddag$ \\
  \hfi &    349.82 &    \HFIonefreq    &  \HFIonereso   &      0.12$^\flat$       &    \HFIoneabserr \\
  \hfi &    550.08 &    \HFItwofreq    &  \HFItworeso   &      0.12$^\flat$       &    \HFItwoabserr \\
  \hfi &    849.27 &    \HFIthreefreq  &  \HFIthreereso &      0.08$^\flat$       &    \HFIthreeabserr \\
  \hfi &   1381.5 &    \HFIfourfreq   &  \HFIfourreso  &      0.08$^\flat$       &    \HFIfourabserr \\
  \hfi &   2096.4 &    \HFIfivefreq   &  \HFIfivereso  &      0.08$^\flat$       &    \HFIfourabserr \\
  \hfi &   2997.9 &    \HFIsixfreq    &  \HFIsixreso   &      0.07$^\flat$       &    \HFIsixabserr \\
  \lfi &   4285.7 &    \LFIonefreq    &  \LFIonereso   &        --              &    \LFIoneabserr \\
  \lfi &   6818.2 &    \LFItwofreq    &  \LFItworeso   &        --              &    \LFItwoabserr \\
  \lfi &  10000 &    \LFIthreefreq  &  \LFIthreereso &        --              &    \LFIthreeabserr \\
 \wmap &   3200 &    \WMAPonefreq   &  \WMAPonereso  &      1.76$^\sharp$     &    \WMAPoneabserr \\
 \wmap &   4900 &    \WMAPtwofreq   &  \WMAPtworeso  &      0.36$^\sharp$      &    \WMAPtwoabserr \\
 \wmap &   7300 &    \WMAPthreefreq &  \WMAPthreereso &     0.11$^\sharp$      &   \WMAPthreeabserr \\
 \wmap &   9100 &    \WMAPfourfreq  &  \WMAPfourreso &      0.05$^\sharp$      &    \WMAPfourabserr \\
 \wmap &  13000 &    \WMAPfivefreq  &  \WMAPfivereso &      0.02$^\sharp$      &    \WMAPfiveabserr \\
\hline
\hline
\end{tabular}
\end{center}
$^\dag$ Assumed to be for the average \iras coverage. $\sigII$ computed by rescaling this value to actual coverage\\
$^\ddag$ From \cite{Miville2005}\\
$^\flat$ $1\sigma$ average value in one beam scaled from \cite{planck2011-1.7}. We actually use the internal variance maps for $\sigII$\\
$^\sharp$ From \cite{Bennett2003}\\
\end{table}

\subsection{Ancillary data}
\label{sec_ancillarydata}

\subsubsection{\ion{H}{i} LMC/SMC data}
\label{sec_HILMCSMCdata}

To trace the atomic gas in the Magellanic Clouds, we used \ion{H}{i}
maps in the 21cm line, obtained by combining data from the Australia
Telescope Compact Array ({\atca}) and the {\parkes} single dish telescope.
For the LMC, this data was obtained by \citet{Kim2003} and
\citet{Staveley2003} and covers $11.1\degr \times 12.4\degr$ on the
sky. The spatial resolution is $1\arcmin$ corresponding to a physical
resolution of about $14.5$ pc at the distance of the LMC.  For the SMC,
the data was obtained by \citet{Staveley-Smith:1997fk} and
\citet{Stanimirovic1999}. It covers a $4.5\degr\times 4.5\degr$ region. H{\sc i}
observations of the SMC tail ($7\degr\times 6\degr$) were obtained by
\cite{Muller:2003eu} and combined to the one in the direction of the
SMC.  The spatial resolution is $98\arcsec$, corresponding to 30pc at
the distance of the SMC.

\subsubsection{\ion{H}{i} Galactic data}
\label{sec_HIMWdata}

The LMC and SMC are located at galactic latitudes \bII=$-34\degr$ and
\bII=$-44\degr$ respectively. They can therefore suffer from
significant contamination by Galactic foreground emission, which has
to be removed from the IR data.  In order to account for the Galactic
foreground emission, we used Galactic \ion{H}{i} column density maps.

For the LMC, the \ion{H}{i} foreground map was constructed by
\cite{Staveley2003} by integrating the {\parkes} \ion{H}{i} data in
the velocity range from $\rm -64<v_{hel}< 100\,\kms$, which excludes
all LMC and SMC associated gas ($\rm v > 100\,\kms$) but includes
essentially all Galactic emission. The spatial resolution is
14\arcmin.  For the SMC, a map of combined {\atca} and {\parkes} data
was build by integrating the Galactic velocities by E. Muller (private
communication). The resolution is $98\arcsec$.

These maps show that the Galactic foreground across the LMC is as
strong as $\NH=1.3\times10^{21}\NHUNIT$, with significant variation
across the LMC, in particular a wide filamentary structure oriented
southwest to northeast. The Galactic foreground across the SMC is
weaker with values $\NH\simeq3.2\times10^{20}\NHUNIT$ but the
associated FIR-submm emission is still non-negligible since the SMC
emission is also weaker than that of the LMC, due to its lower dust
and gas content.

These foreground maps are used to
subtract the foreground IR emission from the IR maps, using the
emissivity SED described in Sect.\,\ref{sec_MWsubtraction}.  

\subsubsection{CO data}
\label{sec_COLMCdata}

The $^{12}$CO($J$=1$\rightarrow$0) molecular data used in this work was obtained
using the NANTEN telescope, a 4-m radio telescope of Nagoya University
at Las Campanas Observatory, Chile \citep[see][]{Fukui2008}.  The
observed region covers about 30 square degrees where CO clouds were
detected in the NANTEN first survey
\citep[e.g.][]{Fukui1999,Mizuno2001,Fukui2008}.
The observed grid spacing was $2\arcmin$, corresponding
to about 30 and 35\,pc at the distance of the LMC and SMC, while the
half-power beam width was $2.6\arcmin$ at 115\,GHz.

We used the CO maps
of the
Magellanic Clouds to construct an integrated intensity map ($\wco$),
integrating over the full $\rm v_{lsr}$ range of the data ($\rm 100 <
v_{lsr} < 400$\,\kms for about 80 \% of the data, while the remaining
20 \% had a velocity range of about $\rm 100$\,\kms covering the
\ion{H}{i} emitting regions.)

\subsubsection{$\halpha$ data}

In order to estimate the free-free contribution to the millimetre
fluxes, we used the continuum-subtracted $\halpha$ maps from the
Southern H-Alpha Sky Survey Atlas \citep[SHASSA,][]{Gaustad2001}
centered on the Magellanic Clouds.

\subsubsection{FIR-Submm data}
\label{sec_FIRdata}

We used the following FIR-submm ancillary data.

\begin{itemize}
\item{ \iris (Improved Reprocessing of the \iras Survey)
$100\mic$ in order to constrain the dust
temperature. The characteristics of this data, including the noise
properties were taken from \cite{Miville2005} and are summarized in
Table\,\ref{tab:ancillary}}
\item{{\wmap} 7yr data. The
characteristics of this data, including the noise properties were taken
from \cite{Bennett2003} and \cite{Jarosik2010} and are summarized in
Table\,\ref{tab:ancillary}}
\end{itemize}

\subsection{Additional Data processing}

\subsubsection{Common angular resolution and pixelisation}

For data already available in the {\Healpix} \citep{Gorski2005} format
(e.g.,$\wmap$), we obtained the data from the Lambda web site
(http://lambda.gsfc.nasa.gov/).  For data not originally presented in
the {\Healpix} format, the ancillary data were brought to the
{\Healpix} pixelisation, using a method where the surface of the
intersection between each {\Healpix} and FITS pixel of the original data
was used as a weight to regrid the data.
The {\Healpix} resolution was chosen so as to match the Shannon
sampling of the original data at resolution $\theta$, with a
{\Healpix} resolution set so that the pixel size is $\rm <\theta/2.4$.
The ancillary data and the description of their processing will be
presented in \cite{Paradis2011}.

All ancillary data were then smoothed to an appropriate resolution by
convolution with a Gaussian smoothing function with appropriate FWHM
using the smoothing {\Healpix} function, and were brought to a pixel
size matching the Shannon sampling of the final resolution.

\subsubsection{CMB subtraction}
\label{sec:cmb_sub}

\begin{figure}[ht]
\begin{center}
\includegraphics[height=6cm, angle=0]{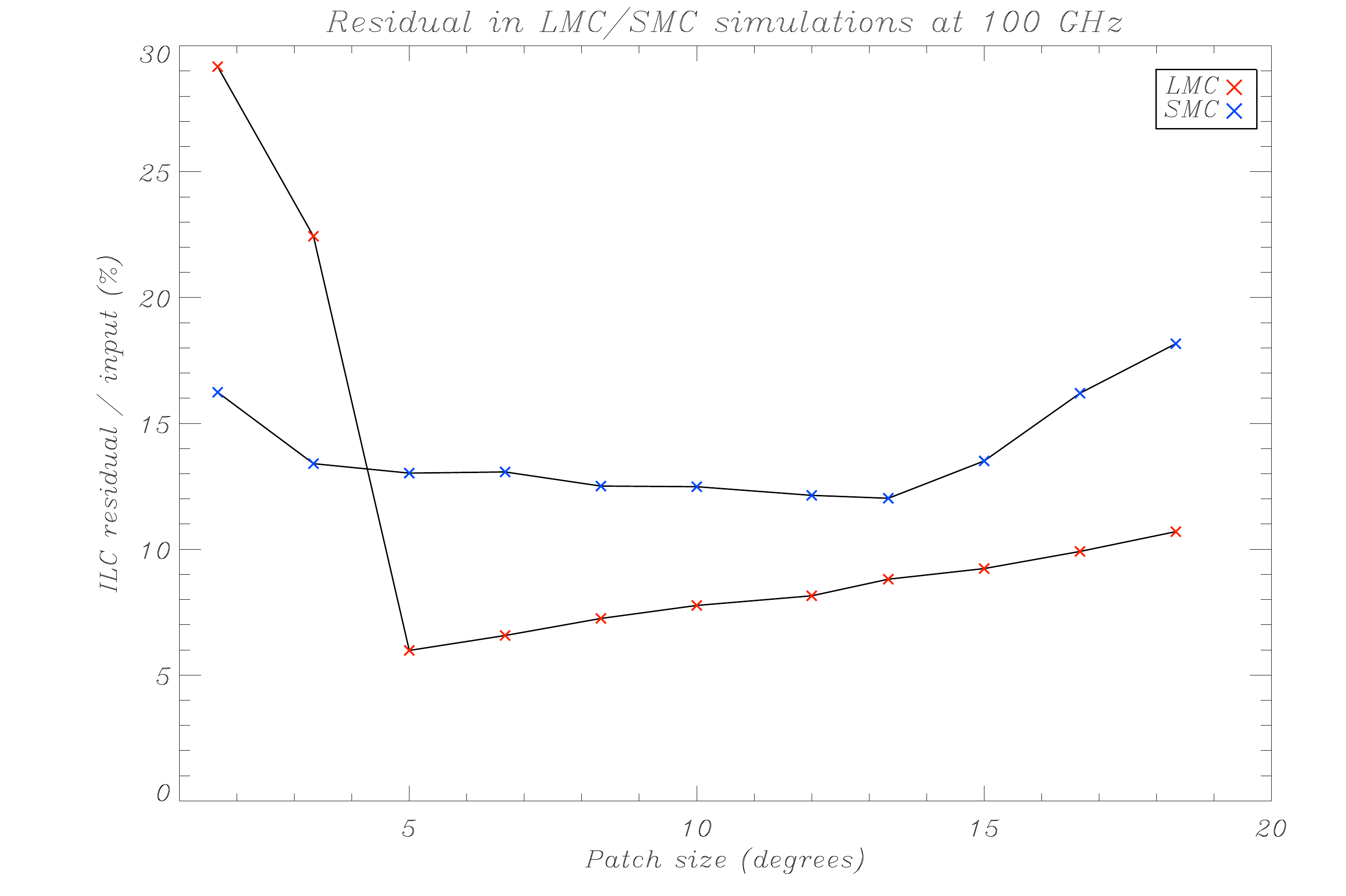}
\caption{
Error due to the ILC CMB subtraction derived from Monte-Carlo
simulations for both LMC and SMC at 100\,GHz, as a function of the
patch size. The values shown are the difference between the recovered
and the input CMB divided by the simulated LMC and SMC, both
integrated in a $4^\circ$ ring.}
\label{simu_residuals}
\end{center}
\end{figure}

\begin{figure*}[ht]
\begin{center}
\includegraphics[width=9.1cm,angle=180]{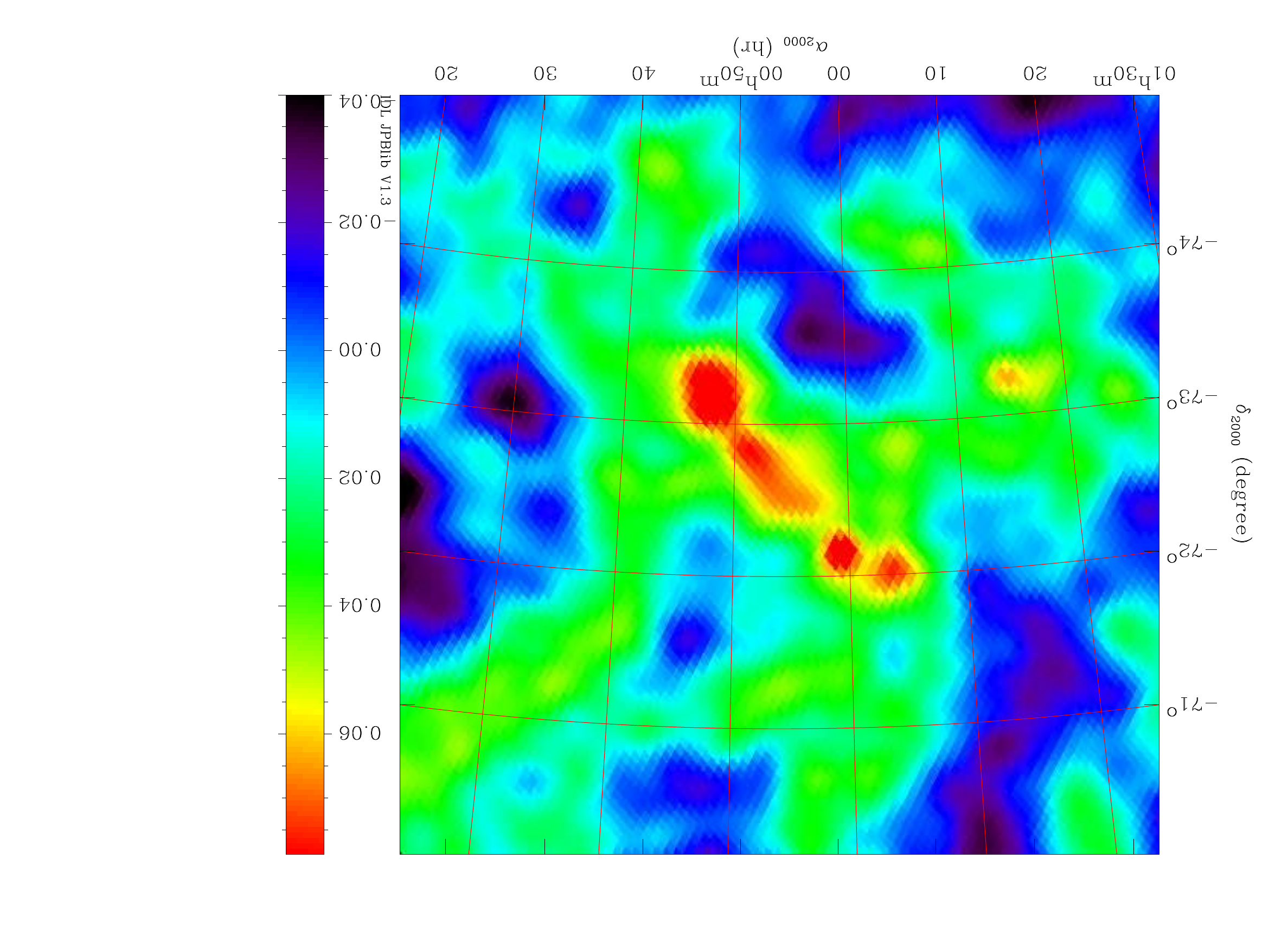}
\includegraphics[width=9.1cm,angle=180]{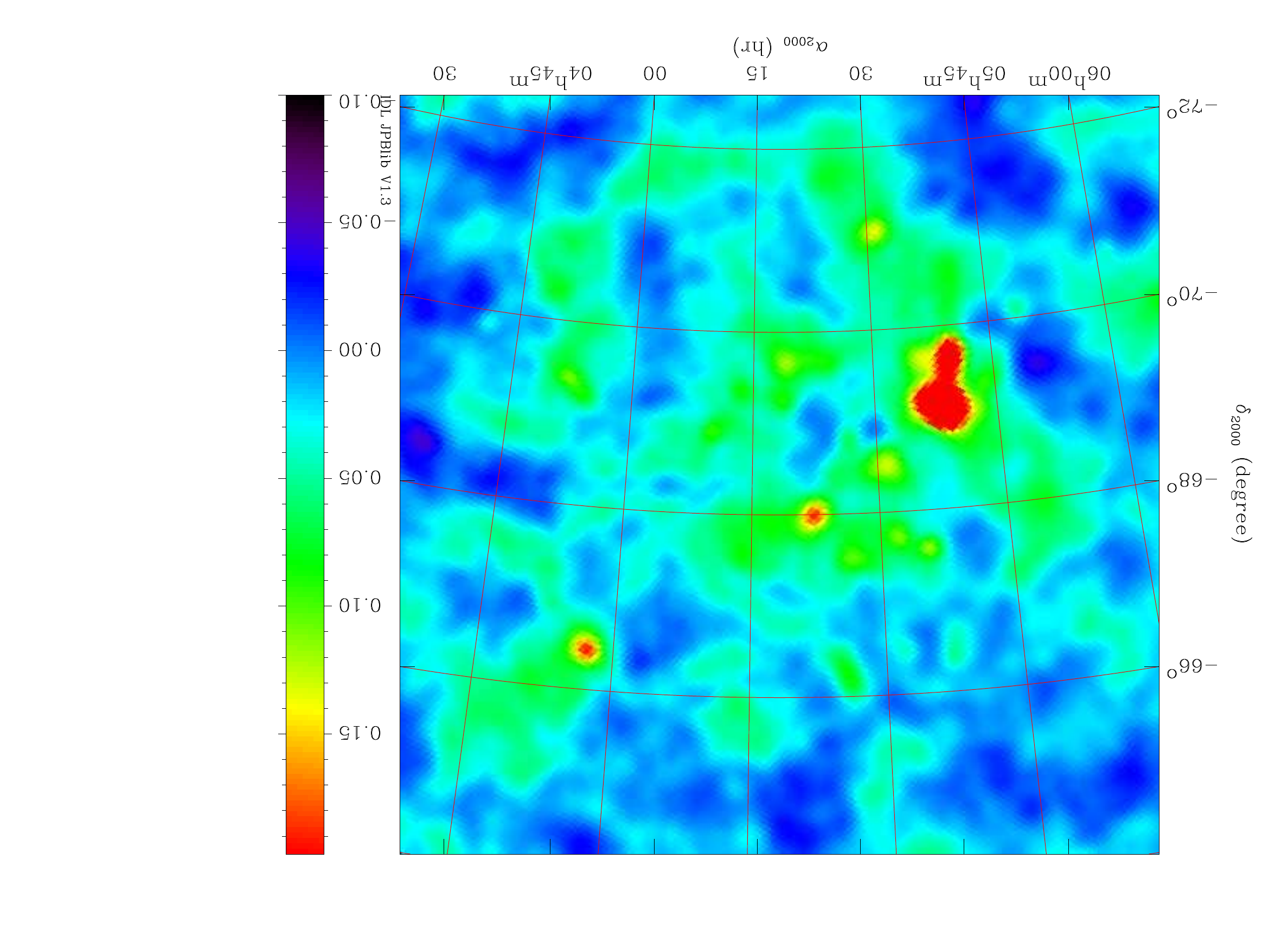}
\includegraphics[width=9.1cm,angle=180]{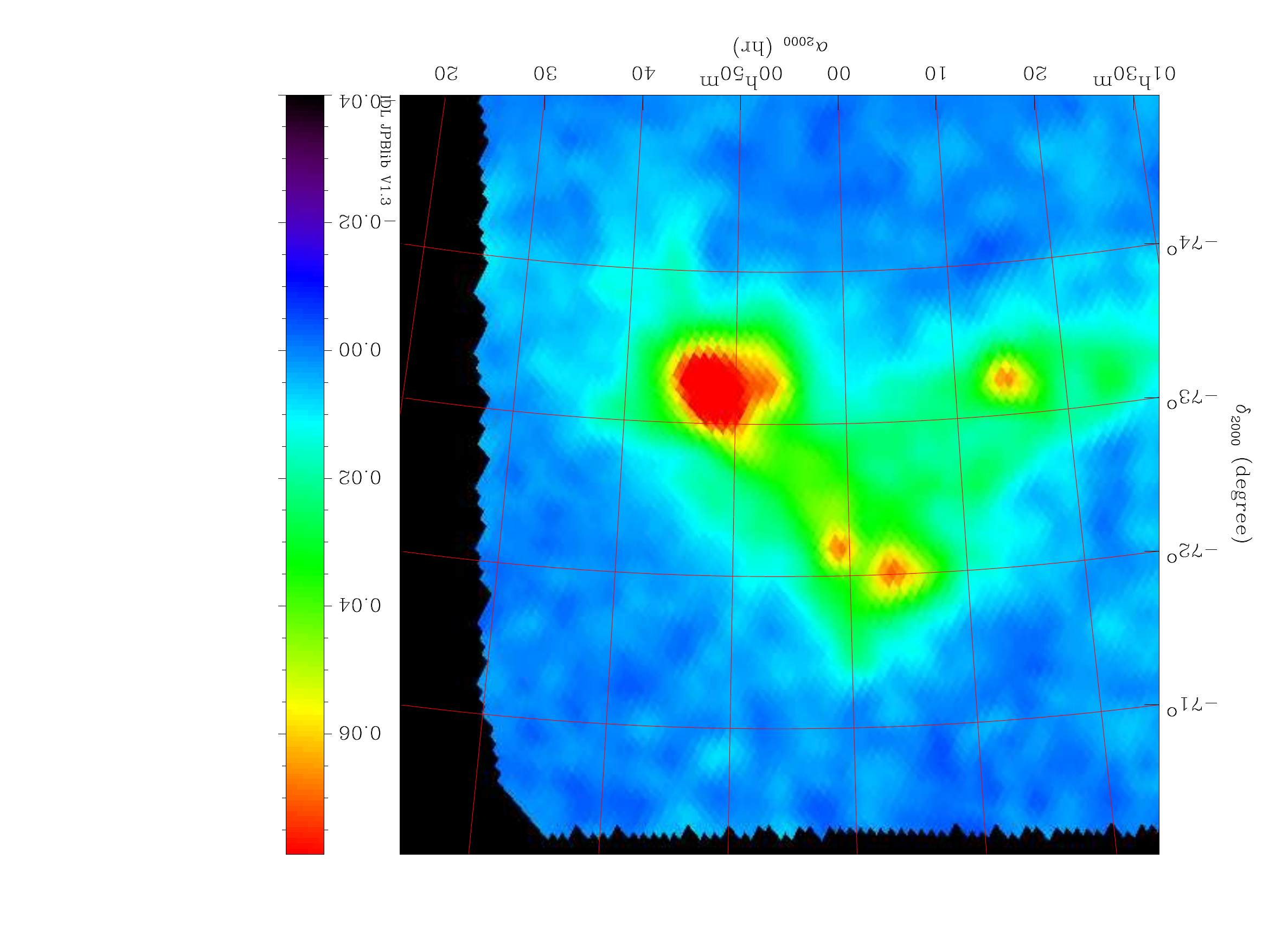}
\includegraphics[width=9.1cm,angle=180]{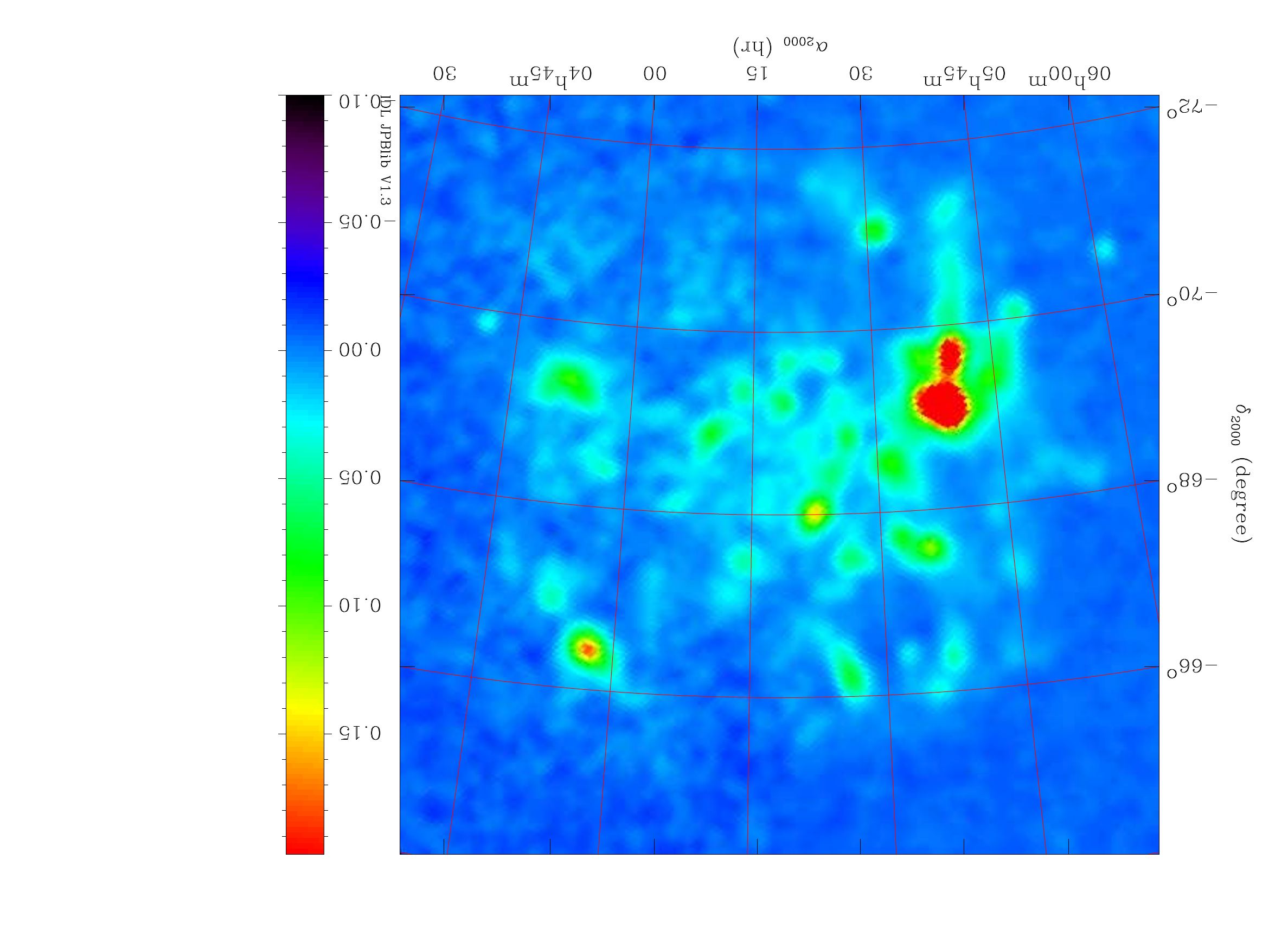}
\caption{\label{fig:before_after_cmbsub}
SMC (left) and LMC (right) total intensity maps before (top) and after (bottom) CMB
subtraction in the {\LFIonefreq}\,{\,GHz} band at the {\LFIonereso}
arcmin resolution.}
\end{center}
\end{figure*}

The upper panel in Fig.\,\ref{fig:before_after_cmbsub} shows the total intensity maps
observed toward the LMC and SMC before CMB subtraction in the {\lfi}
{\LFIonefreq}\,{\,GHz} chanel.  The amplitude of CMB fluctuations is
of the order of that of the diffuse emission of the galaxies and it
is clear that, in order to study the integrated SED of the LMC and SMC
galaxies, an efficient CMB subtraction has to be performed.

The standard CMB--subtracted maps produced by the Data
Processing Center (DPC) \citep{planck2011-1.7} were not used in this
analysis. They were processed by a Needlet Internal Linear Combination
\citep{planck2011-1.7} (NILC) that left a significant amount of
foreground emission in the CMB estimate towards the LMC and the SMC.

For this reason we have subtracted an estimate of the CMB optimized
locally for the LMC and SMC regions, as described below. This CMB
component was reconstructed through a classical Internal Linear
Combination (ILC) by means of Lagrange multipliers
\citep{Eriksen2004}. $12\degr \times 12\degr$ patches around the LMC
and SMC were extracted from the \hfi CMB frequency channels maps
(\HFIsixfreq, \HFIfivefreq, \HFIfourfreq and \HFIthreefreq\,\GHz)
reduced to a common resolution (10 arcmin), in units of K$_{\rm
CMB}$. The CMB component obtained on these patches clearly contains
less LMC and SMC residual than the standard \hfi DPC CMB and therefore
will less affect the SED determination.  The lower panel in
Fig.\,\ref{fig:before_after_cmbsub} shows the maps after CMB
subtraction in the {\lfi} {\LFIonefreq}\,{\GHz} chanel.

We performed Monte-Carlo simulations in order to estimate the error
induced on the SED by our CMB removal. The LMC and SMC were simulated
as a sum of two correlated components. The first component spatial
template is the \iris $100\mic$ map. The second component spatial
template is the \HFItwofreq\,{\GHz} LMC or SMC {\Planck} map.  Their
correlation coefficients are 83\% for the LMC and 92\% for the SMC. We
normalized their fluxes inside a $4\degr$ ring to the value of a
typical LMC or SMC dust component and a typical millimetre excess at
each \hfi CMB frequency, respectively. 200 independent realizations
of a {\wmap} 7yr best fit CMB \citep{Komatsu2010} and of nominal
inhomogeneous \hfi white noise were added to the synthetic LMC and SMC
on patches of varying sizes. For each of these simulations, for each
size, our ILC is performed and compared to the input CMB. We estimate
the error due to the CMB subtraction in both the LMC and the SMC by
comparing the residual in the CMB map to the emission of the two
components at a given frequency. Results at \HFIsixfreq\,{\GHz} are
displayed in Fig.\,\ref{simu_residuals}. For both the LMC and the SMC,
the error increases at small and large patch sizes and an optimal
patch size with respect to the CMB subtraction can be found at
$5\degr$ for the LMC and $13\degr$ for the SMC. This behavior is due
to the fact that the narrower the patch, the lower the contribution of
the CMB to the total variance which is to be minimized in the ILC. On the other
hand, when their sizes increase, patches include foreground emission
which is uncorrelated with the galaxies and has a different spectrum, and
thus both small and large patches contribute to the total variance which must be minimized.

For the $12\degr \times 12\degr$ patches used in the following
analysis, we estimate the error due to CMB subtraction as $\rm 10.38\,\upmu
K_{CMB}$ and $\rm 28.2\,\upmu K_{CMB}$ for the LMC and SMC respectively. In
terms of the fraction of the total galaxy brightness, these correspond
to 8.1, 6.2, 2.3 and 0.3 \% for the LMC and 12.1, 10.7, 6.4 and 1.4 \%
for the SMC at \HFIsixfreq, \HFIfivefreq, \HFIfourfreq and \HFIthreefreq\,\GHz,
respectively.

\subsubsection{Galactic foreground subtraction}
\label{sec_MWsubtraction}

\begin{figure}[ht]
\begin{center}
\includegraphics[height=6cm,angle=0]{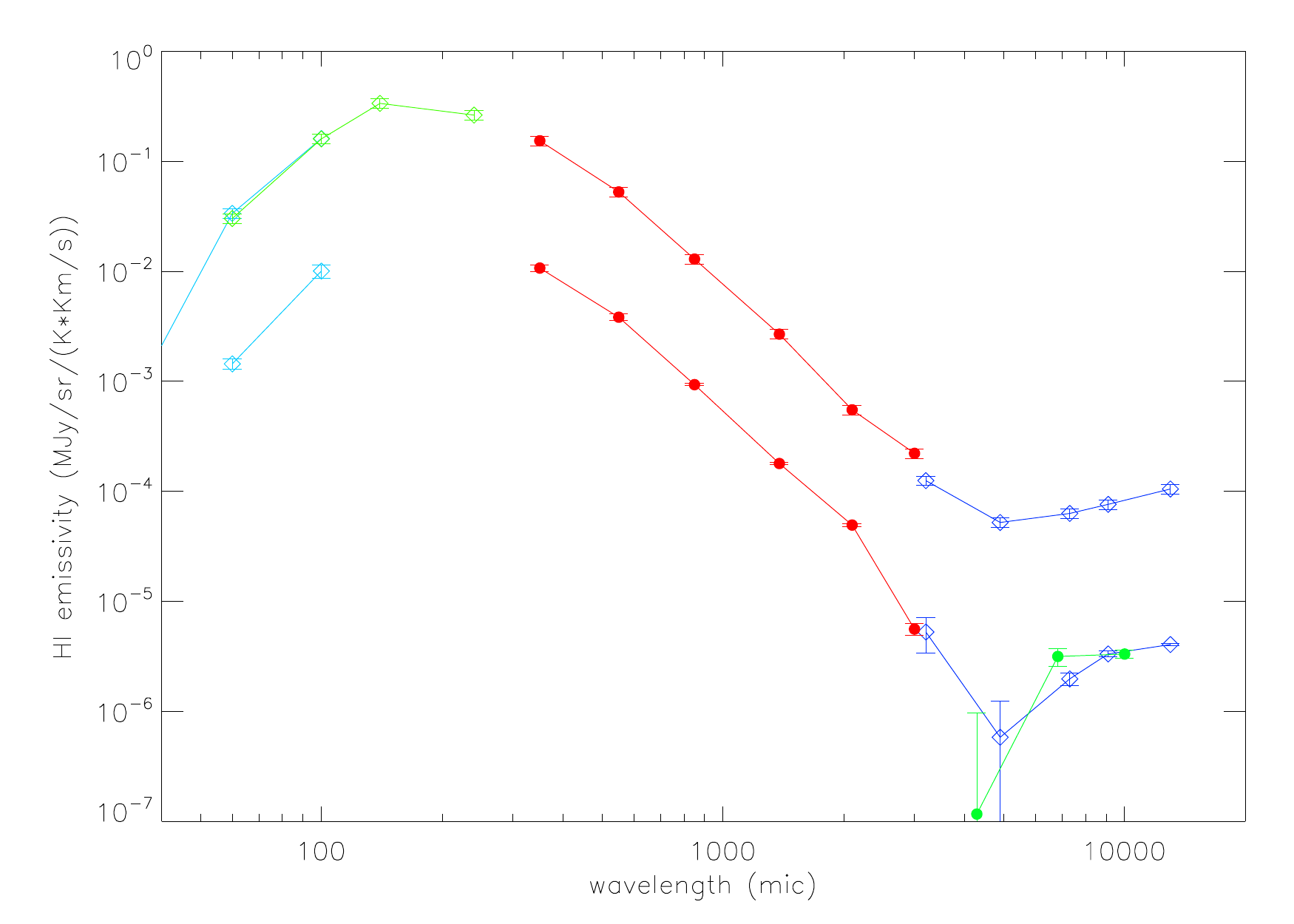}
\caption{
Average foreground SED in the direction of the LMC and SMC (lower
curve) including $\iris$ (light blue), {\Planck}-{\hfi} (red),
{\Planck}-{\lfi} (green) and $\wmap$ data (dark blue), compared to the
SED of the high latitude low column-density MW SED (upper curve) derived in
\cite{planck2011-7.0}, which has been scaled up by a factor of 10 for
clarity.
\label{fig_foregroud_sed}}
\end{center}
\end{figure}

The Milky Way (MW) emission is non--negligible compared to the
emission of the LMC and SMC. We remove this contribution at all wavelengths
using the MW \ion{H}{i} template described in Sec.\,\ref{sec_HIMWdata}. We
first computed the correlation between all FIR-submm data with this
template, in the region where both the MW \ion{H}{i} template and the CMB
estimate are available, but excluding a circular region centered on
each galaxy (centre coordinates taken as
$\rm \radeux=\ralmc$,
$\rm \decdeux=\declmc$ and
$\rm \radeux=\rasmc$,
$\rm \decdeux=\decsmc$
for the LMC and SMC respectively) with radius $\rm \Rlmc\degr$ and $\rm \Rsmc\degr$
for the LMC and SMC respectively.  The spectral distribution of this
correlation factor, taken to represent the SED of the MW foreground is
shown in Fig.\,\ref{fig_foregroud_sed}. The SED is compared to that of
the high galactic latitude reference region used in
\cite{planck2011-7.0} in Fig.\,\ref{fig_foregroud_sed}. It can seen
that the two SEDs are similar, although the MW foreground towards the
LMC and SMC appears slightly colder and has a relatively stronger
non-thermal component in the millimetre. We subtracted the MW
foreground from all data using this SED multiplied by the MW \ion{H}{i}
template over the full map extent.
Note that the median foreground \ion{H}{i} integrated intensities over
the LMC and SMC are $\rm \simeq \medwHIforeLMC\,K\kms$ and $\rm \simeq
\medwHIforeSMC\,K\kms$, which correspond to a brightnesses of $\rm
\medHFIoneforeLMC\,MJy\,sr$$^{-1}$ and $\rm \medHFIoneforeSMC\,MJy\,sr$$^{-1}$
respectively at \HFIonefreq\,\GHz. This is of the same order as the
average brightness of the galaxies at that frequency (see
Table\,\ref{tab:LMC_SED}). However, most of the
MW emission is canceled when subtracting a local background around the
galaxies and the differential correction due to the spatial structure
of the MW foreground then accounts for about $\rm 0.4\%$ and $\rm 21\%$ of the
LMC and the SMC brightness respectively.  

\section{Integrated SEDs}
\label{sec_seds}

\begin{figure*}[]
\begin{center}
\includegraphics[height=6cm,angle=0]{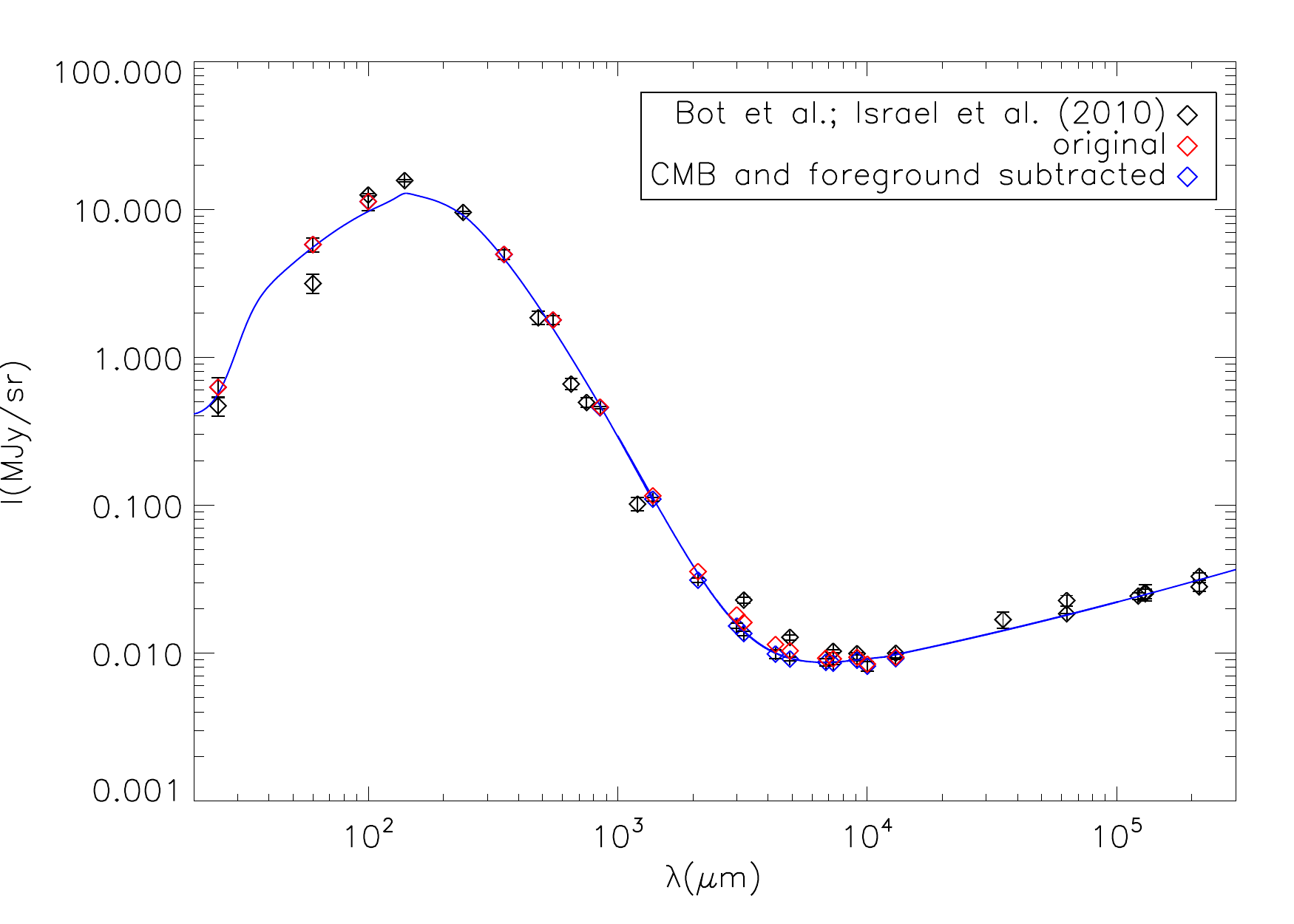}
\includegraphics[height=6cm,angle=0]{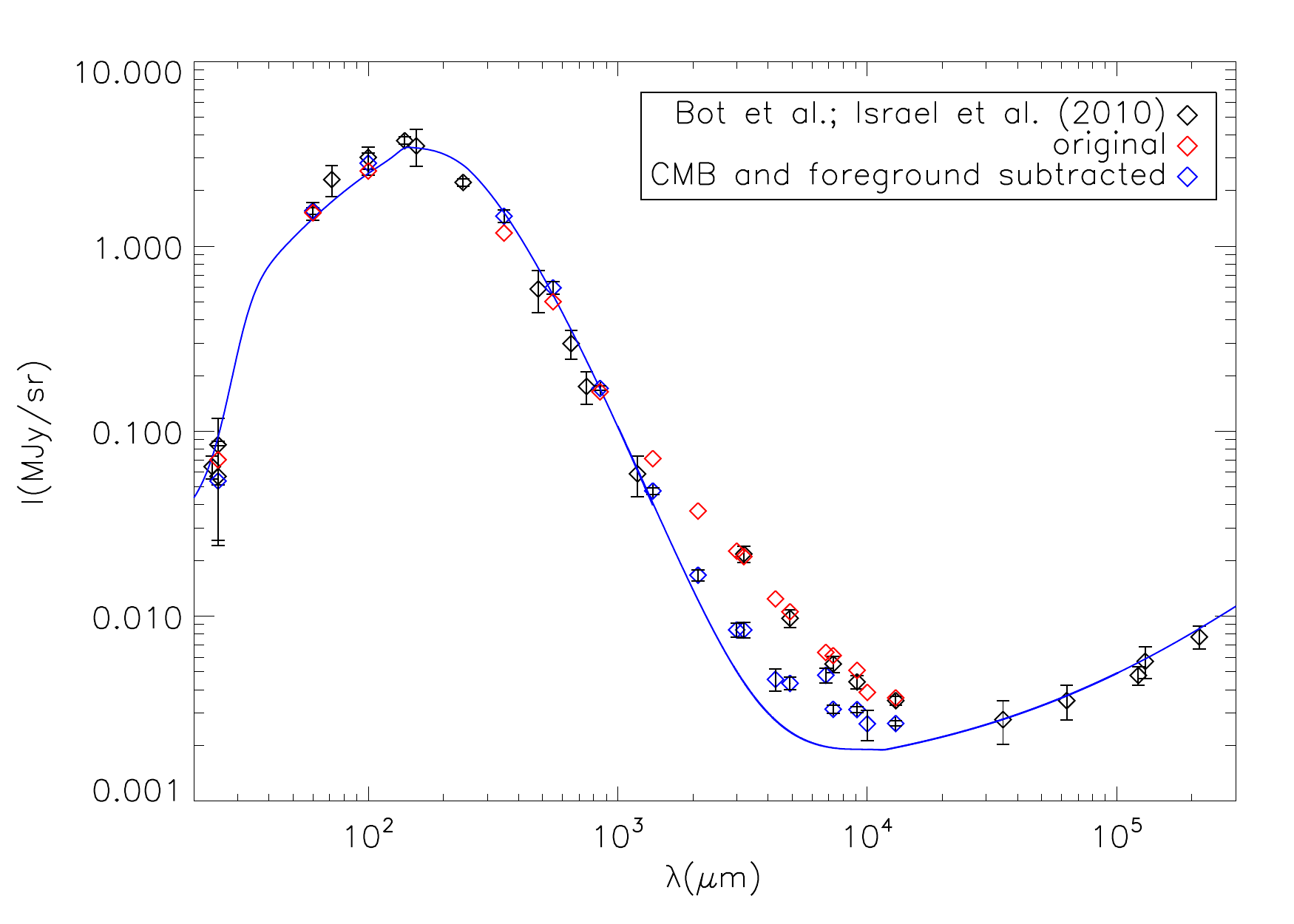}
\caption{
Integrated SEDs of the LMC (left) and SMC (right) before and after CMB
subtraction.  The black points and model are taken from
\cite{Bot2010}. The red symbols show the SEDs derived from the
$\dirbe$, $\iras$, and $\wmap$ data before CMB subtraction. The blue
symbols show the same after CMB subtraction.
\label{fig_beforeaftersed}}
\end{center}
\end{figure*}

The integrated SED of the LMC and SMC before and after CMB subtraction
are shown in Fig.\,\ref{fig_beforeaftersed} (red and blue diamonds,
respectively). They were computed by averaging values in the circular
area around each galaxy defined in Sec.\,\ref{sec_MWsubtraction}, and
subtracting an estimate of the sky off the source taken in an annulus with radius $\rm 1\deg$. They
are compared to the data taken from \cite{Israel2010} and \cite{Bot2010}, which
were integrated over the same region.  It can be seen that the flux
observed in the HFI and LFI bands is consistent with that obtained
with previous studies in this wavelength range before subtraction
of the CMB fluctuations. A model of thermal dust, free-free and
synchrotron emission is fitted to the data as follows. As in \cite{Bot2010}, the
thermal dust emission is adjusted according to the \cite{Draine:2007lr} dust
model. The free-free emission is deduced from the $\halpha$ integrated
flux, using the expression from \cite{Hunt2004}, assuming an
electronic temperature $\rm T_e=10^4$ K, the ratio of ionized helium
to hydrogen $\rm n_{He^+}/n_{H}^+=0.087$, and no extinction. The
synchrotron emission was fitted to the radio data from the literature
as in \cite{Israel2010}. The combined dust, free-free and synchrotron
emission is shown by the blue line.  It can be seen from
Fig.\,\ref{fig_beforeaftersed} that the SED after subtraction of the
CMB fluctuations is in fact compatible with no millimeter emission
excess for the LMC, for this particular model.  However, the
CMB-subtracted SED still shows a significant excess for the SMC, with
most data points above $\rm \lambda\simeq1\,mm$ being in excess over
the model by more than $5\sigma$, leading to an overall significance of the
excess of about $50\sigma$.

The CMB subtraction removes part of the millimetre excess in both
galaxies.  This shows that CMB fluctuations behind the LMC and the SMC
average out to a small but positive contribution when integrated over
the extent of the galaxies. We note that an excess emission remains in
the SMC independently of the dust model used and the assumptions made
on the free-free or synchrotron emission. The shape and intensity of
the millimetre excess can change accordingly, but we could not find a
solution where the SMC SED is explained purely by thermal dust
emission, free-free and synchrotron.  We emphasize also that the dust
model, used to reproduce the dust emission up to the sub-millimetre,
assumes components heated by a radiation field 10 times lower than the
solar neighborhood radiation field. Compared to other nearby galaxies,
this result is rather extreme \citep{Draine2007,Bot2010}.

\begin{figure}[tmb]
\begin{center}
\includegraphics[height=6cm,angle=0]{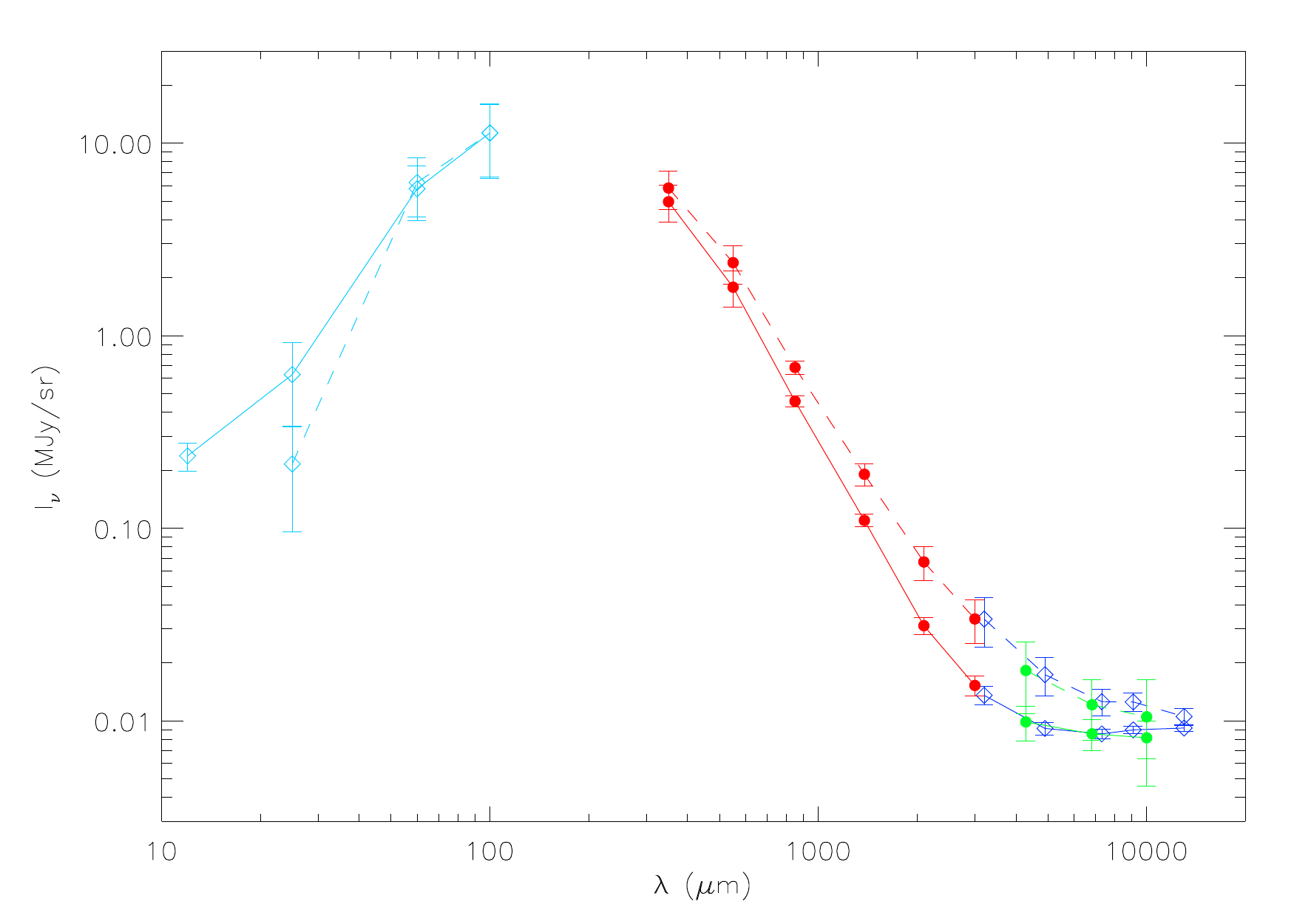}
\caption{\label{fig_avgsed}
Integrated SEDs of the LMC (solid) and SMC (dashed) after CMB and
galactic foreground subtraction, including data from {\Planck}-{\hfi}
(red), {\Planck}-{\lfi} (green), $\iris$ (light blue) and $\wmap$
(dark blue). The uncertainties shown are $\pm 3\sigma$.  The SMC SED
was scaled by a factor 4 providing normalization at $100\mic$.
}
\end{center}
\end{figure}

The integrated SED of the LMC and SMC after CMB and foreground
subtraction are compared in Fig.\,\ref{fig_avgsed}. These SED were
computed in the same integration region as those in
Fig.\,\ref{fig_beforeaftersed}.  The comparison in
Fig.\,\ref{fig_avgsed} shows that the SMC SED is flatter than the LMC
one in the submm, while the SEDs when normalized in the FIR, reconcile
above 10 mm, where the emission is presumably dominated by free-free
and/or spinning dust (see Sec.\,\ref{sec_discussion}).

The SEDs at various stages of the background and foreground
subtraction are given in Table\,\ref{tab:LMC_SED}
for the LMC and SMC. The uncertainties given include the
contribution from the data variance combined for the integration
region, the data variance combined for the background region, and
the absolute calibration uncertainties, using the values given in
Table\,\ref{tab:ancillary}. They also include the noise resulting from
the background (CMB) subtraction and from the foreground MW
subtraction and the free-free removal. 
All uncertainty contributions were added quadratically.

\begin{table*}[tmb]
\caption[ ]{\label{tab:LMC_SED}
LMC (column 2-4) and SMC (column 5-7) SEDs averaged in a circular region
for each galaxy. The integration region is
centered on $\radeux=\ralmc$, $\decdeux=\declmc$, with radius $\rm R_{LMC}=\Rlmc\degr$ faor the LMC
and centered on $\radeux=\rasmc$, $\decdeux=\decsmc$,
 with radius $\rm R_{SMC}=\Rsmc\degr$ for the SMC.  A common
background was subtracted in a $1\degr$ annulus around this region.
Brightness values are in $\rm MJy\,sr$$^{-1}$ in the $\rm \nu
I_\nu=cste$ flux convention. The last two lines give the average
$\halpha$ emission (in Rayleigh), Galactic HI emission (in $\Kkms$) in
the same area. The table lists the total SED ($\InuTOT$),
the CMB subtracted SED ($\InuNOCMB$) and the CMB, MW
foreground and free-free subtracted SED ($\InuSUB$) and
their associated 1$\sigma$ uncertainties.
}
\begin{flushleft}
\begin{tabular}{l|lll|lll}
\hline
\hline
 $\lambda$ & $\InuTOT$     & $\InuNOCMB$   & $\InuSUB$     &    $\InuTOT$     & $\InuNOCMB$   & $\InuSUB$ \cr
 $[\mic]$ & [$\rm MJy\,sr$$^{-1}$] & [$\rm MJy\,sr$$^{-1}$] & [$\rm MJy\,sr$$^{-1}$] &    [$\rm MJy\,sr$$^{-1}$] & [$\rm MJy\,sr$$^{-1}$] & [$\rm MJy\,sr$$^{-1}$] \cr
\hline
\multicolumn{7}{l}{\iras:} \\
\hline
12      & (2.36$\pm$0.12)$\times10^{-1}$ & (2.36$\pm$0.12)$\times10^{-1}$ & (2.34$\pm$0.13)$\times10^{-1}$  & (1.37$\pm$0.11)$\times10^{-2}$ & (1.37$\pm$0.11)$\times10^{-2}$ &  (-2.79$\pm$0.27)$\times10^{-2}$ \\  
25      & (6.28$\pm$0.95)$\times10^{-1}$ & (6.28$\pm$0.95)$\times10^{-1}$ & (6.25$\pm$0.96)$\times10^{-1}$  & (7.01$\pm$1.10)$\times10^{-2}$ & (7.01$\pm$1.10)$\times10^{-2}$ & (5.30$\pm$0.98)$\times10^{-2}$ \\
60      & 5.79$\pm$0.60             & 5.79$\pm$0.60              & 5.78$\pm$0.61              & 1.52$\pm$0.16 & 1.52$\pm$0.16 & 1.56$\pm$0.18\\
100     & (1.13$\pm$0.15)$\times10^{1}$  & (1.13$\pm$0.15)$\times10^{1}$  & (1.13$\pm$0.15)$\times10^{1}$  & 2.56$\pm$0.35 & 2.56$\pm$0.35 & 2.82$\pm$0.39 \\
\hline
\multicolumn{7}{l}{\Planck:} \\
\hline
349.82   & 4.97$\pm$0.35                  & 4.97$\pm$0.35                 & 4.96$\pm$0.35                 & 1.18$\pm$0.08                  & 1.18$\pm$0.08 & 1.46$\pm$0.11\\
550.08   & 1.79$\pm$0.13                  & 1.79$\pm$0.13                 & 1.78$\pm$0.13                 & (5.02$\pm$0.37)$\times10^{-1}$ & (4.99$\pm$0.37)$\times10^{-1}$ & (5.96$\pm$0.45)$\times10^{-1}$\\
849.27   & (4.61$\pm$0.10)$\times10^{-1}$ & (4.58$\pm$0.10)$\times10^{-1}$ & (4.51$\pm$0.10)$\times10^{-1}$ & (1.63$\pm$0.04)$\times10^{-1}$ & (1.47$\pm$0.04)$\times10^{-1}$ & (1.69$\pm$0.05)$\times10^{-1}$\\
1381.5   & (1.16$\pm$0.02)$\times10^{-1}$ & (1.10$\pm$0.03)$\times10^{-1}$ & (1.04$\pm$0.03)$\times10^{-1}$ & (7.12$\pm$0.18)$\times10^{-2}$ & (4.30$\pm$0.19)$\times10^{-2}$ & (4.63$\pm$0.20)$\times10^{-2}$\\
2096.4   & (3.56$\pm$0.08)$\times10^{-2}$ & (3.13$\pm$0.10)$\times10^{-2}$ & (2.51$\pm$0.09)$\times10^{-2}$ & (3.71$\pm$0.10)$\times10^{-2}$ & (1.54$\pm$0.11)$\times10^{-2}$ & (1.54$\pm$0.11)$\times10^{-2}$\\
2997.9   & (1.81$\pm$0.05)$\times10^{-2}$ & (1.53$\pm$0.06)$\times10^{-2}$ & (8.96$\pm$0.48)$\times10^{-3}$ & (2.25$\pm$0.07)$\times10^{-2}$ & (8.29$\pm$0.71)$\times10^{-3}$ & (7.09$\pm$0.69)$\times10^{-3}$\\
4285.7   & (1.14$\pm$0.07)$\times10^{-2}$ & (9.87$\pm$0.68)$\times10^{-3}$ & (3.34$\pm$0.36)$\times10^{-3}$ & (1.24$\pm$0.09)$\times10^{-2}$ & (4.55$\pm$0.61)$\times10^{-3}$ & (3.16$\pm$0.54)$\times10^{-3}$ \\
6818.2   & (9.24$\pm$0.52)$\times10^{-3}$ & (8.57$\pm$0.51)$\times10^{-3}$ & (1.72$\pm$0.17)$\times10^{-3}$ & (6.37$\pm$0.49)$\times10^{-3}$ & (2.94$\pm$0.34)$\times10^{-3}$ & (1.57$\pm$0.28)$\times10^{-3}$\\
10000   & (8.43$\pm$0.61)$\times10^{-3}$ & (8.17$\pm$0.60)$\times10^{-3}$ & (1.05$\pm$0.25)$\times10^{-3}$  & (3.87$\pm$0.55)$\times10^{-3}$ & (2.53$\pm$0.49)$\times10^{-3}$ & (1.10$\pm$0.41)$\times10^{-3}$\\
\hline
\multicolumn{7}{l}{\wmap:} \\
\hline
3200   & (1.61$\pm$0.04)$\times10^{-2}$ & (1.36$\pm$0.05)$\times10^{-2}$ & (7.23$\pm$0.43)$\times10^{-3}$ & (2.10$\pm$0.08)$\times10^{-2}$ & (8.28$\pm$0.80)$\times10^{-3}$ & (7.07$\pm$0.79)$\times10^{-3}$\\
4900   & (1.04$\pm$0.02)$\times10^{-2}$ & (9.13$\pm$0.23)$\times10^{-3}$ & (2.51$\pm$0.17)$\times10^{-3}$ & (1.06$\pm$0.03)$\times10^{-2}$ & (4.32$\pm$0.33)$\times10^{-3}$ & (2.92$\pm$0.31)$\times10^{-3}$\\
7300   & (9.15$\pm$0.13)$\times10^{-3}$ & (8.56$\pm$0.15)$\times10^{-3}$ & (1.66$\pm$0.09)$\times10^{-3}$ & (6.12$\pm$0.15)$\times10^{-3}$ & (3.09$\pm$0.16)$\times10^{-3}$ & (1.67$\pm$0.15)$\times10^{-3}$\\
9100   & (9.38$\pm$0.12)$\times10^{-3}$ & (8.98$\pm$0.13)$\times10^{-3}$ & (1.93$\pm$0.06)$\times10^{-3}$ & (5.10$\pm$0.11)$\times10^{-3}$ & (3.04$\pm$0.11)$\times10^{-3}$ & (1.63$\pm$0.10)$\times10^{-3}$\\
13000   & (9.38$\pm$0.11)$\times10^{-3}$ & (9.19$\pm$0.12)$\times10^{-3}$ & (1.88$\pm$0.05)$\times10^{-3}$ & (3.62$\pm$0.08)$\times10^{-3}$ & (2.52$\pm$0.09)$\times10^{-3}$ & (1.07$\pm$0.07)$\times10^{-3}$\\
\hline
\multicolumn{7}{l}{$\halpha$ (R):} \\
\hline
--   & (1.93$\pm$0.19)$\times10^{1}$  &  & & 7.00$\pm$0.72 & & \\
\hline
\multicolumn{7}{l}{$\whi$$^{\rm MW}$ ($\Kkms$):} \\
\hline
--   & 1.01$\pm$0.13 & & & ($-$2.57$\pm$0.26$)\times10^{+1}$ &  & \\
\hline
\end{tabular}
\end{flushleft}
\end{table*}

\section{Dust temperature and emissivity}
\label{sec_temperature}

\subsection{Temperature determination}

\begin{table}
\caption[ ]{\label{tab:Tbeta_values}
Dust temperature, $\beta$ values and fit reduced $\chi^2$ for various
methods experimented to derive the temperature maps.  The values
listed are median values in the SED integration region for each
galaxy.}
\begin{flushleft}
\begin{tabular}{llll}
\hline
\hline
Method & $\Td\pm\Delta\Td$ & $\beta\pm\Delta\beta$ & $\chi^2$ \\
       &       [K]         &                       &  \\
\hline
\multicolumn{4}{l}{LMC:}\\
\hline
 free  $\beta$ &     21.0$\pm$1.9 & 1.48$\pm$0.25  &      1.91\\
 fixed $\beta$ &     20.7$\pm$1.7 &      1.5$^\dag$    &      1.73\\
 fixed $\beta$ &     19.2$\pm$1.6 &      1.8    &      1.63\\
 fixed $\beta$ &     18.3$\pm$1.6 &      2.0    &      1.57\\
      dustem   &     17.7$\pm$1.6 &      --     &      1.60\\
\hline
\multicolumn{4}{l}{SMC:}\\
\hline
 free  $\beta$ &     22.3$\pm$2.3 &     1.21$\pm$0.27 &      2.28\\
 fixed $\beta$ &     21.6$\pm$1.9 &      1.2$^\dag$   &      1.90\\
 fixed $\beta$ &     18.8$\pm$1.7 &      1.8   &      9.94 \\
 fixed $\beta$ &     17.9$\pm$1.6 &      2.0   &     14.0 \\
        dustem &     17.3$\pm$1.6 &      --    &     12.66 \\
\hline
\hline
\end{tabular}
\end{flushleft}
$^\dag$ fixed $\beta$ model used in this paper.
\end{table}

As shown by several previous studies
\citep[e.g.][]{Reach1995,Finkbeiner99,Paradis2009b,planck2011-7.0,planck2011-7.12},
the dust emissivity spectrum in our Galaxy cannot be represented by a
single dust emissivity index $\beta$ over the full FIR-submm
domain. The data available indicate that $\beta$ is usually stronger
(steeper emissivity) in the FIR and lower (flatter emissivity) in the
submm, with a transition around $500\mic$
\citep[see][]{Paradis2009b,planck2011-7.12}.  This is likely to be the
case also for the LMC and the SMC.  The dust temperature derived will
depend on the assumptions made about $\beta$, since these two
parameters are somewhat degenerate in $\chi^2$ space.  Note that some
of the past attempts at constraining the equilibrium temperature of
the large grains in the LMC used the \iras $60\mic$ emission. Emission
at $60\mic$ is highly contaminated by out--of--equilibrium emission
from Very Small Grans (VSGs) and this is even more the case in the
Magellanic Clouds, due to the presence of the $70\mic$ excess
\citep{Bot2004,Bernard2008}. Combining the \iras $60\mic$ and
$100\mic$ data therefore strongly over-estimates the temperature and
accordingly under-estimates the abundances of all types of dust
particles. A good sampling at frequencies dominated by Big Grain (BG)
emission became possible using the combination of the \iras and
{\spitzer} data
\citep{Leroy2007,Bolatto2007,Bernard2008,Sandstrom2010}.

As dust temperature is best derived from the FIR data, we limit the
range of frequencies used in the determination to the FIR, which
limits the impact of potential changes of the dust emissivity index
$\beta$ with frequency.  In the determination of the dust temperature
($\Td$), we used the {\iris}\,$100\mic$ map and the two highest \hfi
frequencies at \HFIonefreq \,and \HFItwofreq\,\GHz.  Temperature maps
were derived at the common resolution of the 3 bands used (5') and at
lower resolution for further analysis.  In each case, the emission was
computed in the photometric channels of the instruments used ($\iras$,
{\Planck}, and $\wmap$), including the color corrections, using the
actual transmission profiles for each instrument, and following the
flux convention description given in the respective explanatory
supplements.

In order to derive the thermal dust temperature, we use the same
strategy as described in \cite{planck2011-7.0}. To minimise
computation time, the predictions of the model were tabulated for a large
set of parameters ($\Td$, $\beta$).
For each map pixel, the $\chi^2$ was computed for each entry
of the table and the shape of the $\chi^2$ distribution around the
minimum value was used to derive the uncertainty on the free
parameters.  This includes the effect of the data variance
$\sigII^2$ and the absolute uncertainties.

We explored several options for deriving maps of the apparent dust
temperature, which are summarized in Table\,\ref{tab:Tbeta_values}.

We first fitted each pixel of the maps with a modified black body of
the form $\rm I_{\nu} \propto \nu^\beta B_\nu (\Td)$ in the above
spectral range (method referred to as ``free $\beta$'' in
Table\,\ref{tab:Tbeta_values}).  The median values of $\Td$ and $\beta$
derived using this method are given in the first line of
Table\,\ref{tab:Tbeta_values}. This led to median $\beta$ values of
$\rm \beta_{LMC} \simeq \betalmc$ and $\rm \beta_{SMC} \simeq
\betasmc$ for the LMC and SMC respectively. Note that the value for
the LMC is consistent with that derived using a combination of the
\iras and \herschel data by \cite{Gordon2010}.
However, inspection of the corresponding maps shows correlated
variations of the two parameters which are also correlated with the
noise level in the maps.  This suggests that the spurious values
probably originate from the correlation in parameter
space and the presence of noise in the data, particularly in low
brightness regions of the maps.

\begin{figure*}[ht]
\begin{center}
\includegraphics[width=9cm,angle=180]{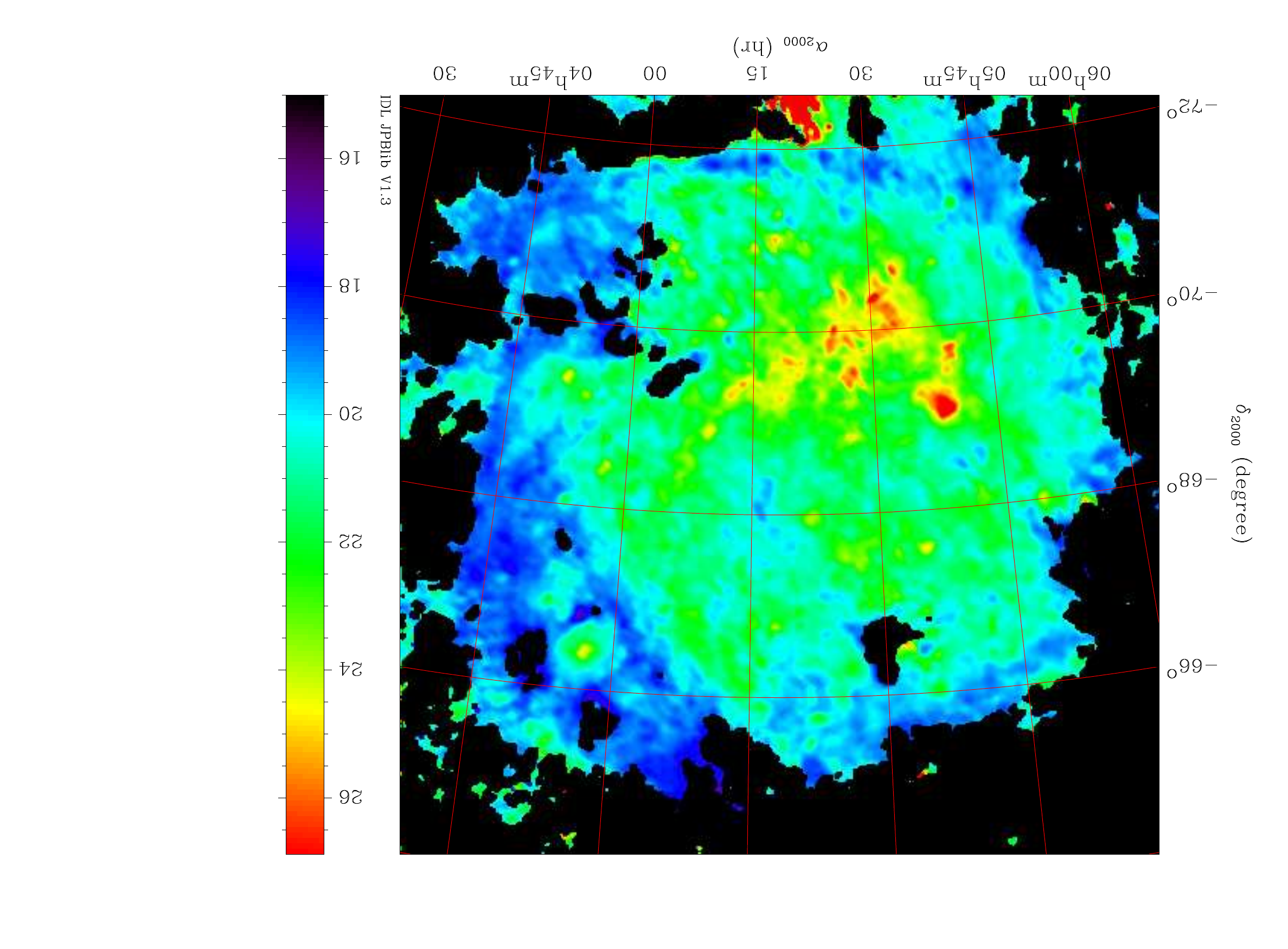}
\includegraphics[width=9cm,angle=180]{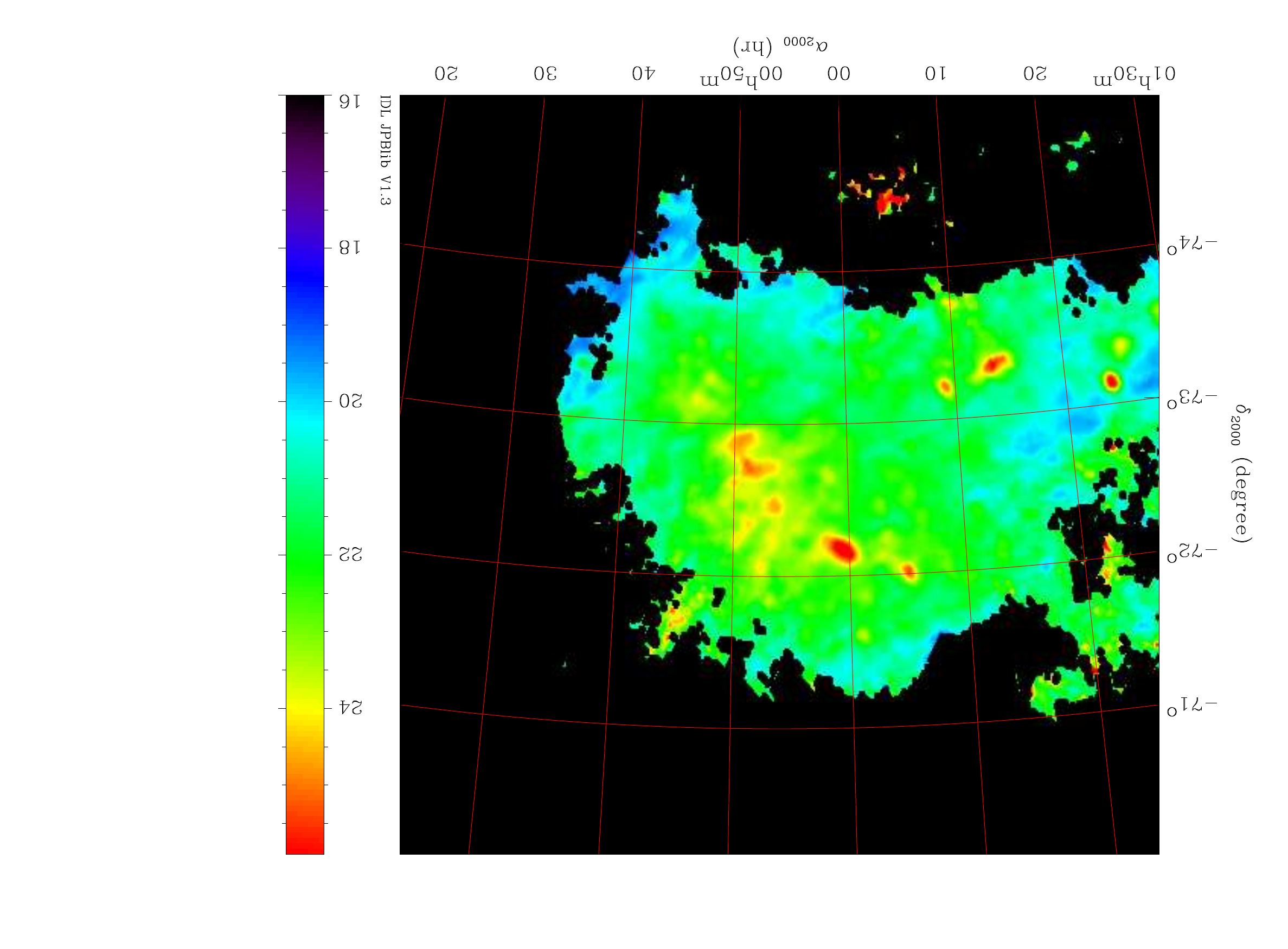}
\includegraphics[width=9cm,angle=180]{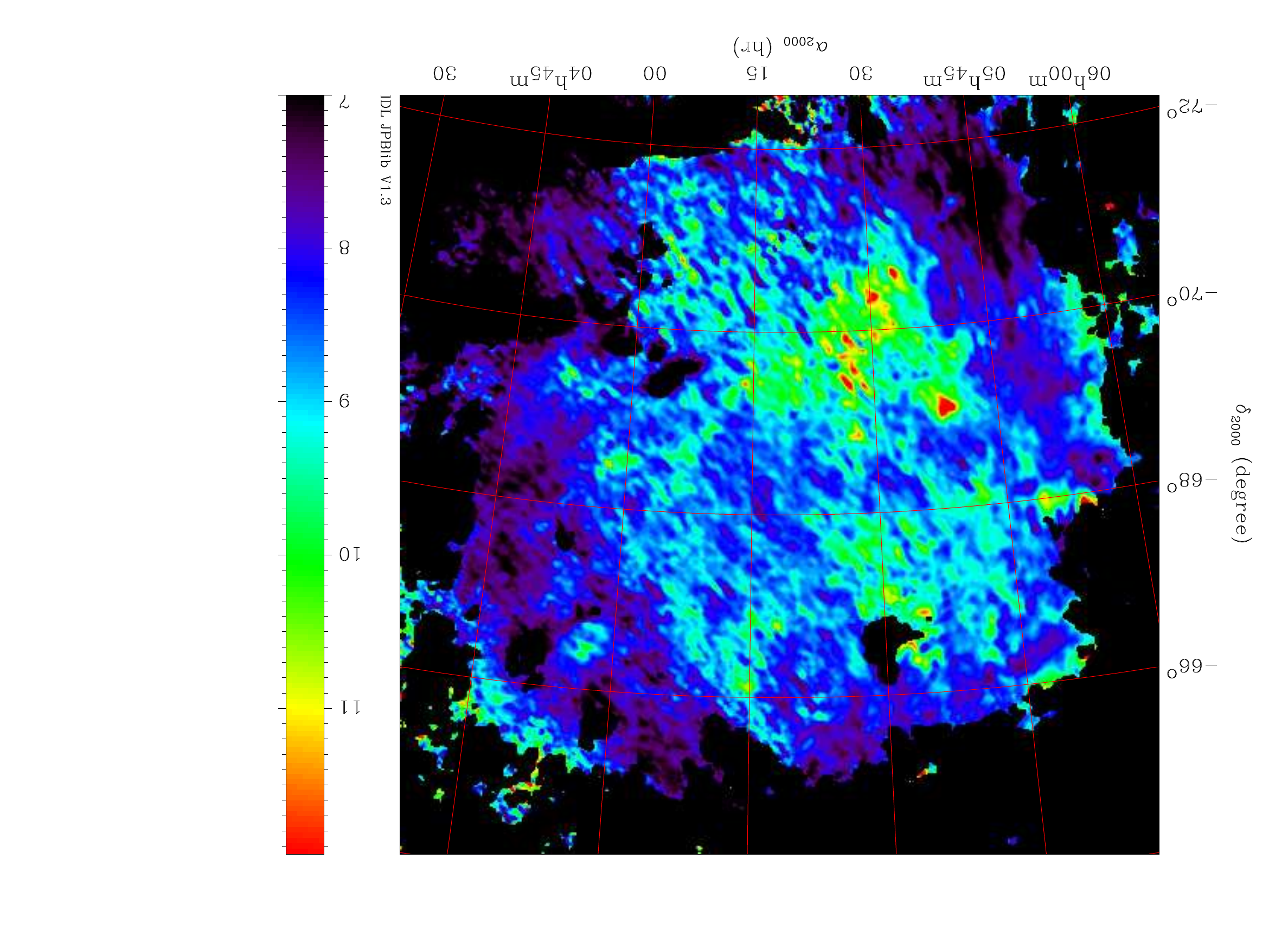}
\includegraphics[width=9cm,angle=180]{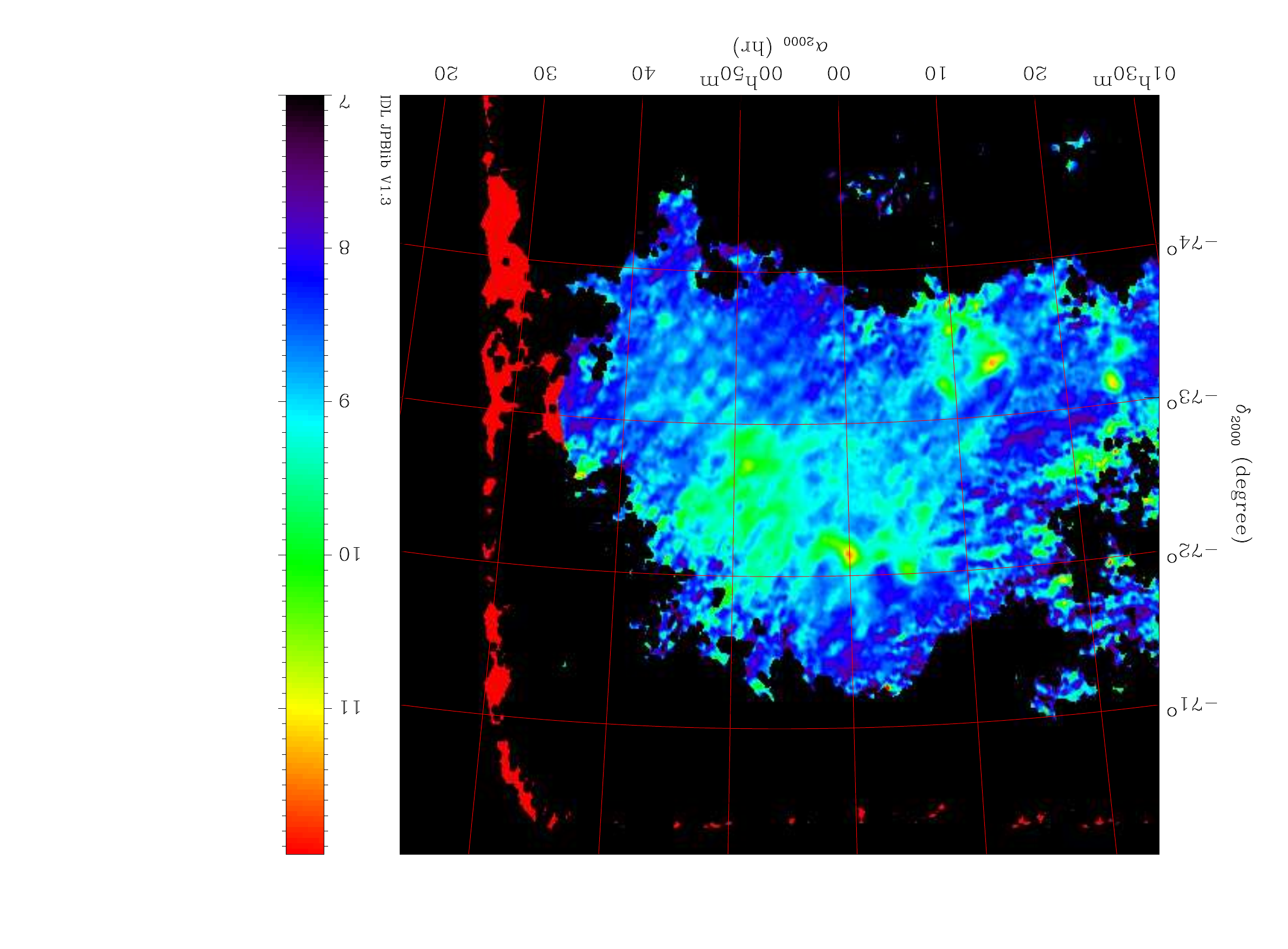}
\caption{
Upper panels: Dust temperature maps for the LMC (left) and SMC (right)
computed from the foreground subtracted maps using the
{\iris}\,$100\mic$, {\hfi}\,{\HFIonefreq} and {\HFItwofreq}\,{\GHz}
maps, using a fixed $\beta_{LMC}=\betalmc$ and
$\beta_{SMC}=\betasmc$. Lower panels: relative uncertainties on the
dust temperature at the same resolution, expressed as percentages.
\label{fig_Tmap}}
\end{center}
\end{figure*}

We then performed fits of the FIR emission using the fixed $\beta$
values, with the $\beta$ values given above for the two galaxies
(method referred to as ``fixed $\beta$'' and marked with a $^\dag$ in
Table\,\ref{tab:Tbeta_values}).  Although the median reduced $\chi^2$
was slightly higher than for the ``free $\beta$'' method, the temperature
maps showed fewer spurious values, in particular in low brightness
regions. This resulted in a more coherent distribution of the temperature
values than in the ``free $\beta$'' case. Since we later used the
temperature maps to investigate the spectral distribution of the dust
optical depth, and the dust temperature is a source of uncertainty, we
adopted the ``fixed $\beta$'' method maps in what follows. The
corresponding temperature maps for the LMC and SMC are shown in
Fig.\,\ref{fig_Tmap}.

The calculations were also carried out for $\beta=1.8$, which is the
average Galactic value \citep{planck2011-7.0,planck2011-7.12}, in
order to be able to derive dust emissivity values under the same
assumption as in the MW. We also experimented using the ``standard''
$\beta$=2 value.

\subsection{Angular distribution of dust temperature}

\subsubsection{LMC}

\begin{figure}
\begin{center}
\includegraphics[width=10cm,angle=180]{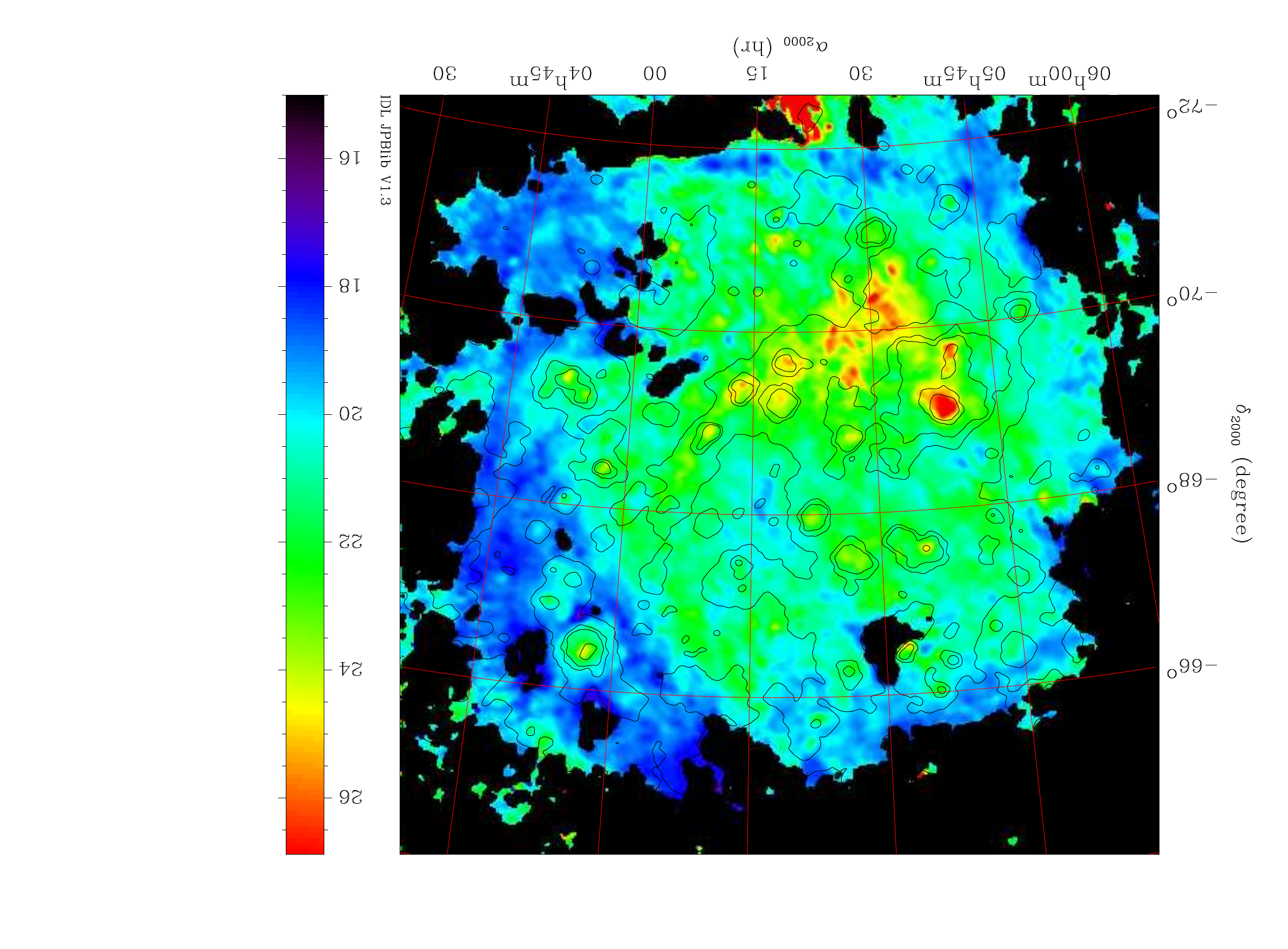}
\includegraphics[width=10cm,angle=180]{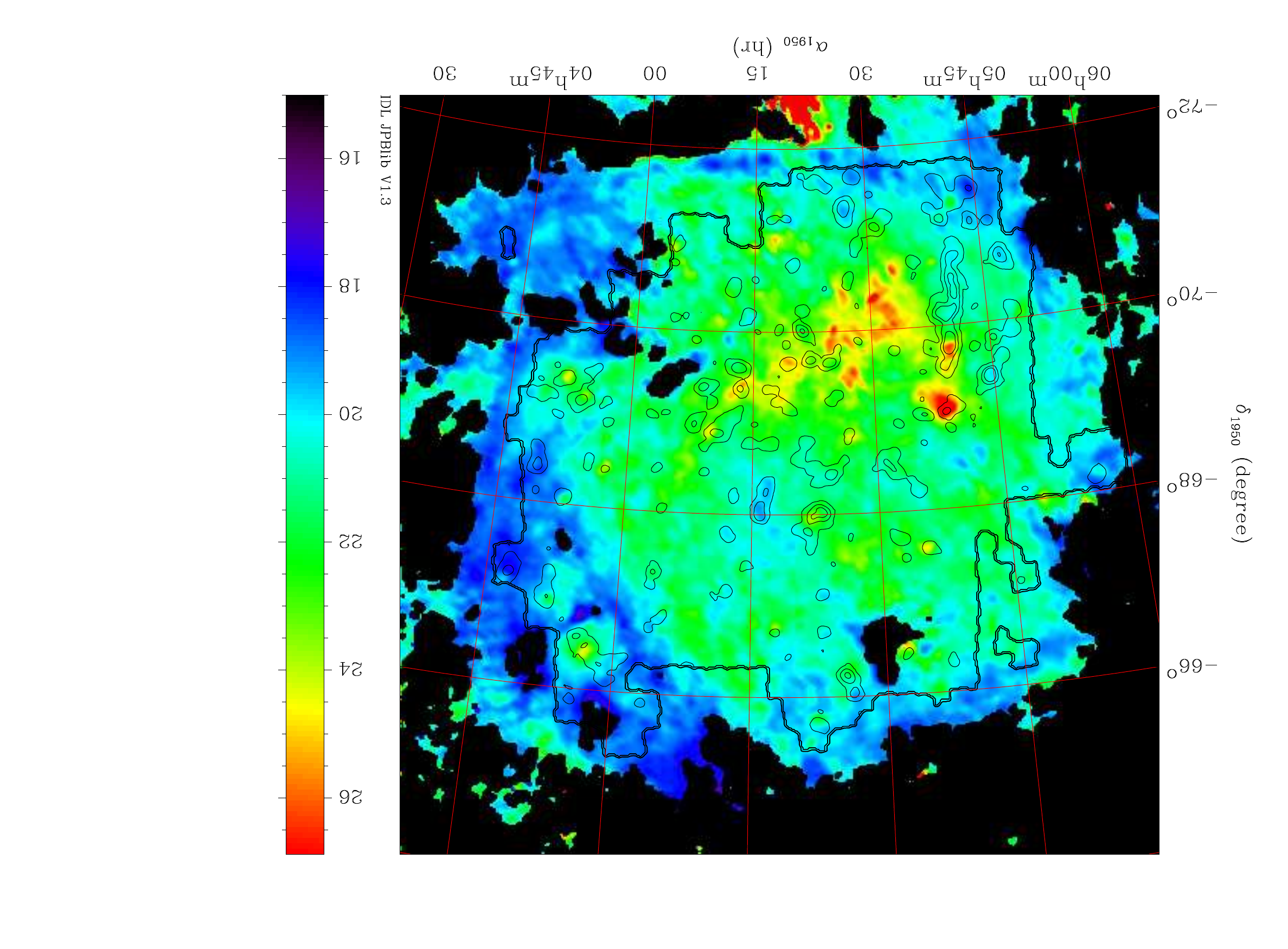}
\caption{
Comparison between the dust temperature map of the LMC with $\halpha$
(top) and CO emission (bottom).  The CO contours are at 0.5, 2, 4 and 10
K\kms. $\halpha$ contours are at 1, 10, 50,  100, 500 and 1000 Rayleigh.
The thick line shows the edge of the available CO surveys.
\label{fig_T_Ha_CO_LMCmap}}
\end{center}
\end{figure}

The temperature map derived here for the LMC shows a similar
distribution to the one derived from \iras and {\spitzer}\,data by
\cite{Bernard2008}. The highest temperatures are observed toward
30-Dor and are of the order of 24 K. The large scale distribution
shows the existence of an inner warm arm, which follows the
distribution of massive star formation as traced by known \ion{H}{ii}
regions. Figure\,\ref{fig_T_Ha_CO_LMCmap} shows that the warm dust at
$\rm \Td>20\,K$ in the LMC is reasonably well correlated with
$\halpha$ emission, indicating that it is heated by the
increased radiation field induced by massive star formation. Two
regions departing from the correlation are visible to the SW of the
30--Dor at $\rm \radeux$=05h40m, $\rm \decdeux$=$-70\degr30'$ and $\rm
\radeux$=$05h40m$, $\rm \decdeux$=$-72\degr30'$. The first region was
already identified in \cite{Bernard2008} who showed that this low
column density region was unexpectedly warm given that no star
formation is taking place in this area. They proposed that this could
be an artifact of the \iris $100\mic$ used in the analysis, possibly
due to \iras gain variations upon passage on the bright 30--Dor
source. However, the analysis carried out here and the comparison between
temperature maps obtained with and without MW background subtraction
favors a biased temperature due to low surface brightness.
This points to the necessity of correcting for
temperature bias at low brightness. The second region is close to the
edge of the map and is probably also affected by low surface
brightness.

The {\Planck} data however reveals cold regions in the outer regions
of the LMC, which were not seen using the {\spitzer}\,data. This is
essentially due to the large coverage of the {\Planck} data with
respect to the limited region, which was covered by the {\spitzer}
\,data. The south of the LMC exhibits a string of cold regions with
$\rm \Td<20\,K$. These regions correspond without exception to known
molecular clouds when they fall in the region covered by the CO
survey. They also correspond to peaks of the dust optical depth as
derived in the following sections. Similarly, a set of cold regions
exist at the nortwest periphery of the LMC.  The comparison with the
distribution of CO clouds is shown in
Fig.\,\ref{fig_T_Ha_CO_LMCmap}. As already noticed in
\cite{Bernard2008}, there is no systematic correlation between cold
dust and the presence of molecular material, at least in the inner
regions of the LMC. This was confirmed by the statistical
characteristics of the IR properties of LMC molecular clouds
established by \cite{Paradis2010}, which showed no systematic trend
for molecular regions to be colder than their surrounding neutral
material. However, toward the outer regions of the LMC, molecular
clouds appear systematically to show a decrease in the dust
temperature. This difference may be due to the absence of star
formation activity in the outer regions and/or to less mixing along
the line of sight.

\cite{Gordon2010} derived a dust temperature map from the \herschel
data (HERITAGE program) for a fraction of the
LMC observed during the Science Demonstration Phase (SDP). They used
the $100\mic$ to $350\mic$ \iras and \herschel bands to constrain the
temperature. They found that the temperatures derived this way only
differ from those derived by \cite{Bernard2008} by up to 10\%. The SDP
field covered an elongated region across the LMC roughly oriented
north--south. This strip crossed the warm inner arm, which is also
clearly seen in their map. However, we stress that, in their study, a
gradient was removed along the strip based on the values at the
edges. The fact that we detect significantly colder than average dust
in the outer cold arm of the LMC underlines the need to take those
variations into account when subtracting background emission in the
\herschel data.

\subsubsection{SMC}

\begin{figure}[ht]
\begin{center}
\includegraphics[width=10cm,angle=180]{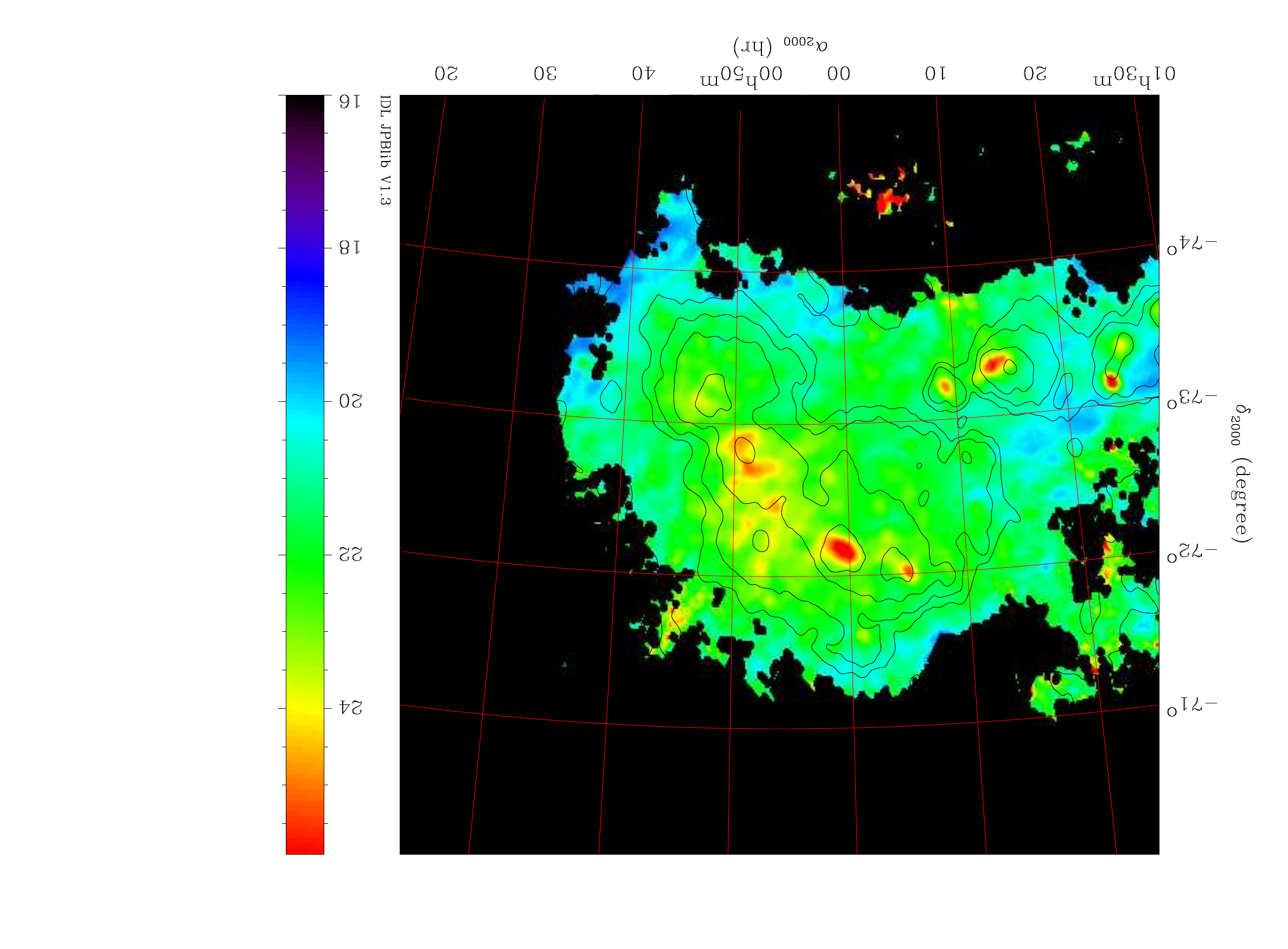}
\includegraphics[width=10cm,angle=180]{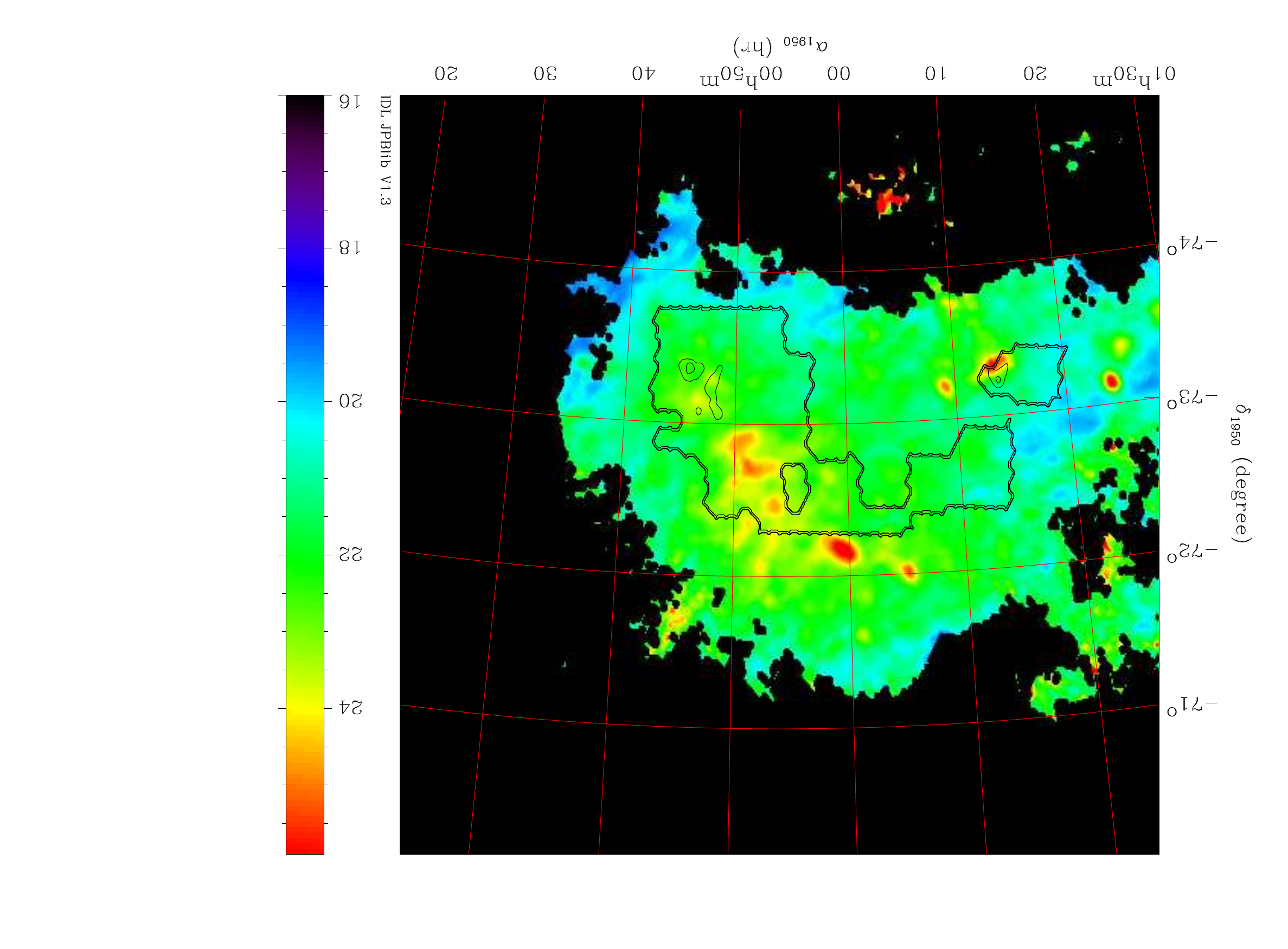}
\caption{
Comparison between the dust temperature map of the SMC with $\halpha$
(top) and CO emission (bottom).  The CO contours are at 0.5, 1 and 1.5
K\kms. $\halpha$ contours are at 1, 5, 30 and 100 Rayleigh. The thick
line shows the edge of the available CO surveys.
\label{fig_T_Ha_CO_SMCmap}}
\end{center}
\end{figure}

The temperature map derived for the SMC shows more moderate
temperature variations than those observed in the LMC but peaks
corresponding to well known HII regions are clearly identified, and
the overall dust temperature correlates with $\halpha$ emission, as
shown in Fig.\,\ref{fig_T_Ha_CO_SMCmap}.  In particular, the massive
star forming region SMC-N66 \citep{Henize1956} ($\rm \radeux$=$\rm
\raNsixsix$, $\rm \decdeux$=$\rm \decNsixsix$) corresponds to the
highest temperature ($\sim 20$ K) in the SMC. Other well known star
forming regions like SMC-N83/84 ($\rm \raNheightthree$, $\rm
\decNheightthree$), N81 ($\rm \raNheightone$, $\rm \decNheightone$),
N88/89/90 (in the wing: $\rm \raNheightheight$, $\rm
\decNheightheight$) and DEM S54 (at the center of the main bar: $\rm
\raSfivefour$, $\rm \decSfivefour$) appear as temperature peaks
compared to the surroundings.  In contrast, the infrared-bright star
forming region N76 (located at the northeast of N66: $\rm
\raNsevensix$, $\rm \decNsevensix$) and southwest star--forming
complex do not stand out.  This could be due to the presence of an
extended warm component that we observe in the main bar and that seems
spatially related to the diffuse $\halpha$ emission.

\subsection{Optical depth determination}
\label{sec_tau}

Optical depth is derived using:
\begin{equation}
\tau(\nu)=\frac{I_\nu}{B_\nu(T_d)},
\end{equation}
where $\rm B_\nu$ is the Planck function. We used resolution matched maps of $\Td$ and $\rm I_\nu$
and derived $\tau$ maps at the various resolutions of the data used here.
The uncertainty on $\tau$ ($\Delta \tau$) is computed accordingly as:
\begin{equation}
\Delta \tau(\nu)=\tau \left( \frac{\sigII^2}{I_\nu^2} + \left( \frac{\delta B_\nu(\Td)}{\delta T} \frac{\Delta \Td}{B_\nu(\Td)}\right)^2\right)^{1/2}.
\end{equation}

\begin{figure}[ht]
\begin{center}
\includegraphics[width=10cm,angle=180]{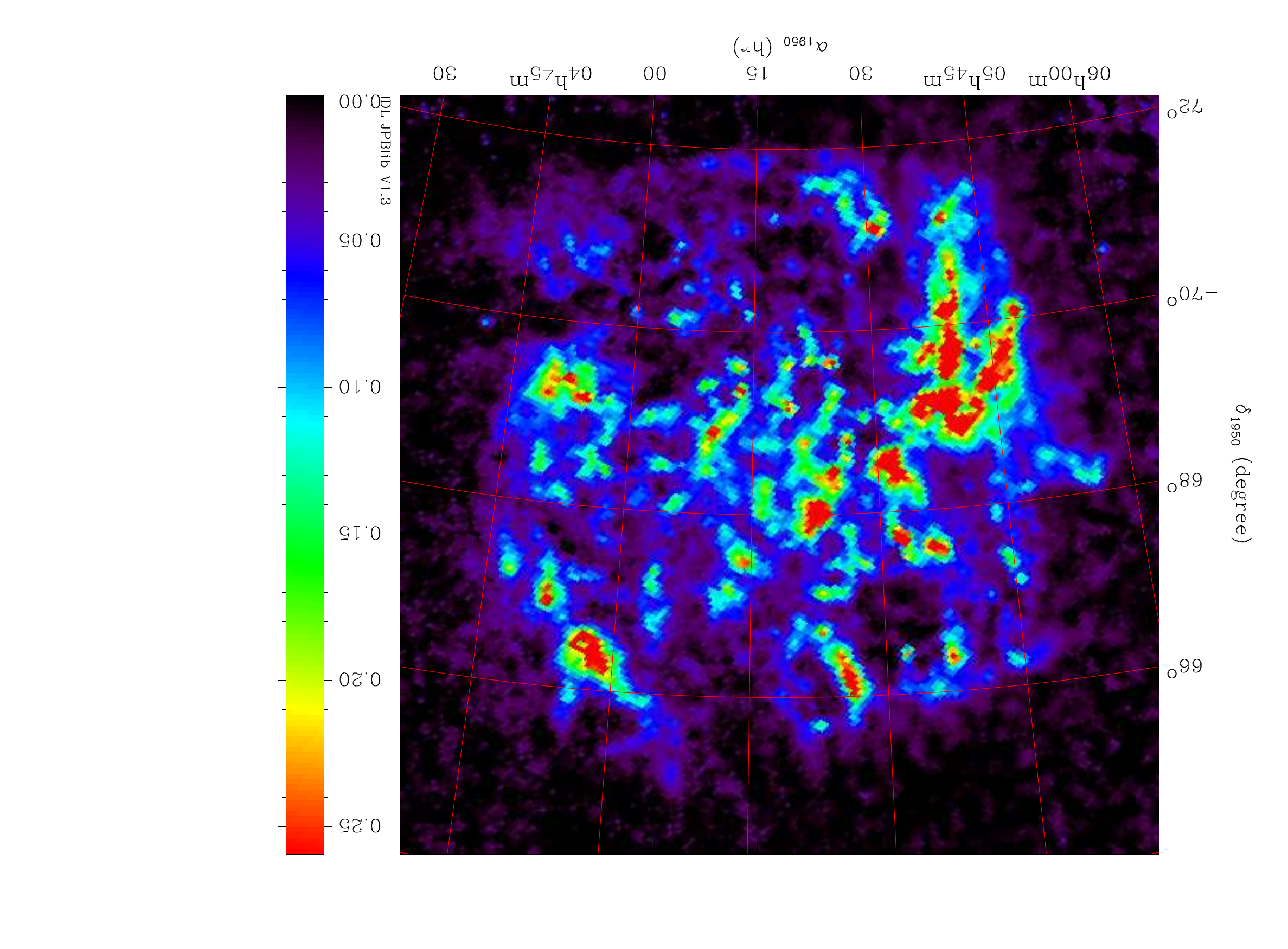}
\includegraphics[width=10cm,angle=180]{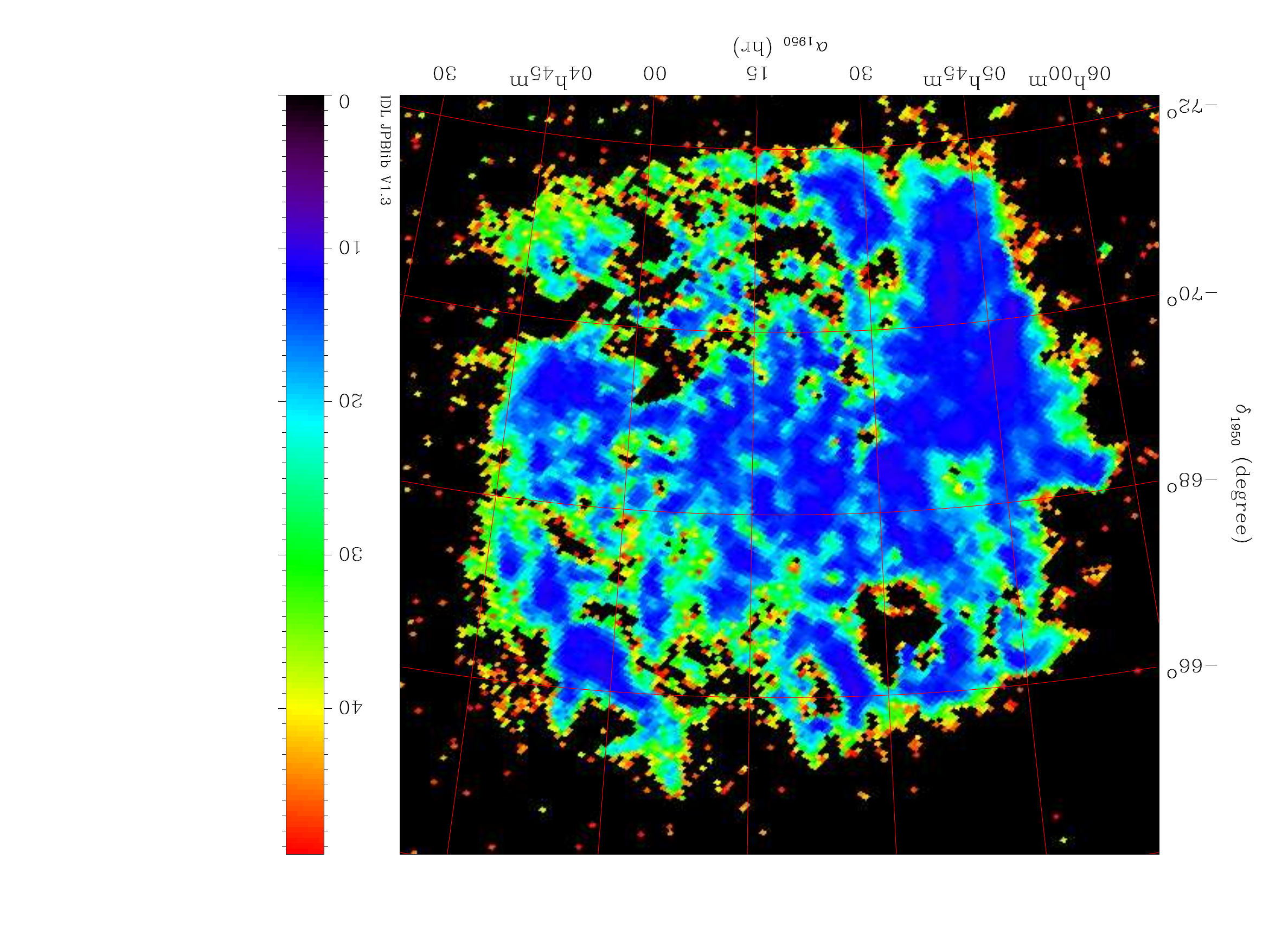}
\caption{\label{fig_taumap_lmc}
Upper panel: Map of the dust optical depths of the LMC at {\hfi}
{\HFIfourfreq}\,{\GHz}. Units are $\rm 10^4\times\tau$.  Lower panel: Map
of the dust optical depth relative uncertainty of the LMC at {\hfi}
{\HFIfourfreq}\,{\GHz} in percent.
Black pixels in the maps are masked and have
relative uncertainties larger than 50\%.
}
\end{center}
\end{figure}

\begin{figure}[ht]
\begin{center}
\includegraphics[width=10cm,angle=180]{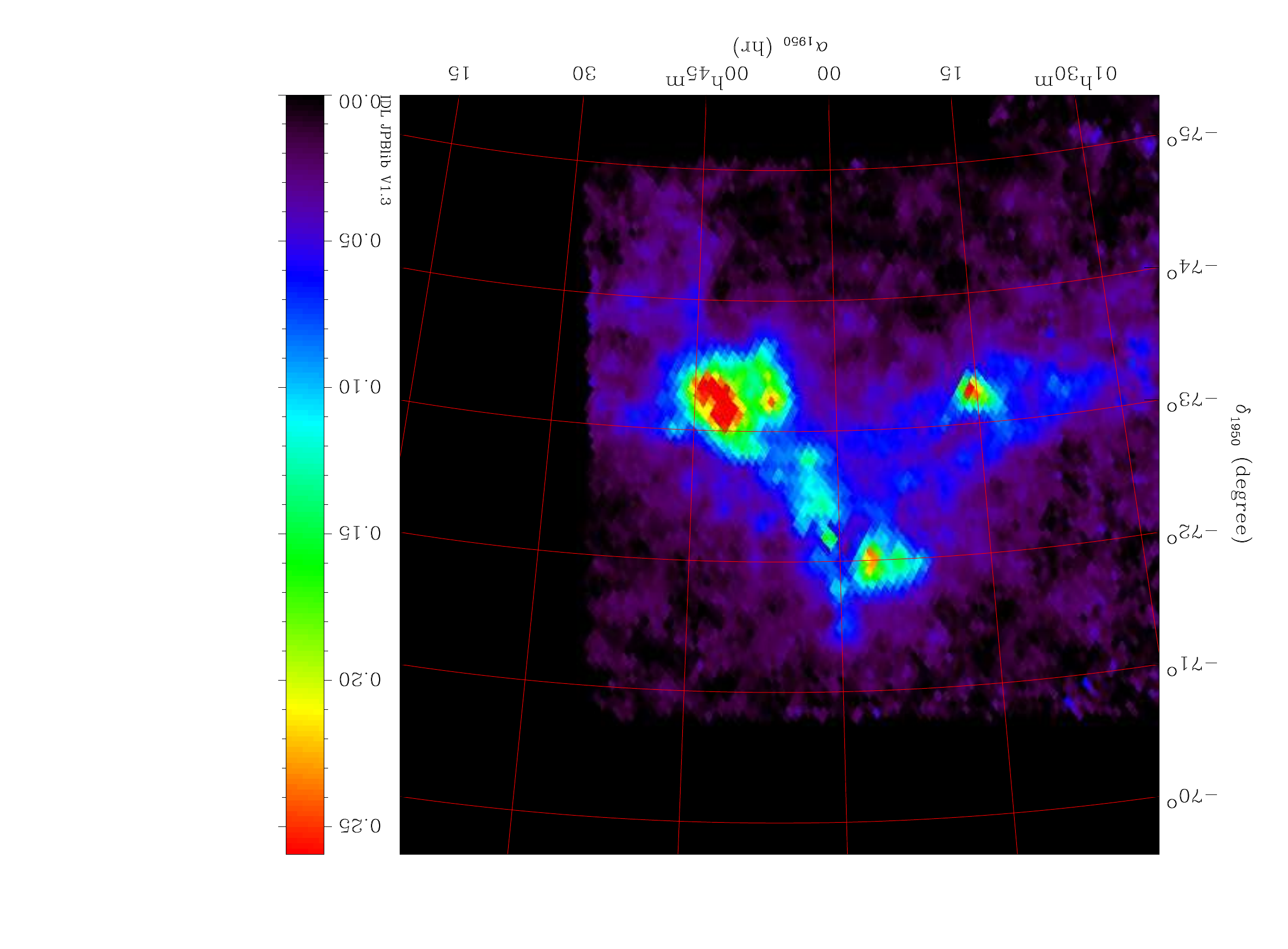}
\includegraphics[width=10cm,angle=180]{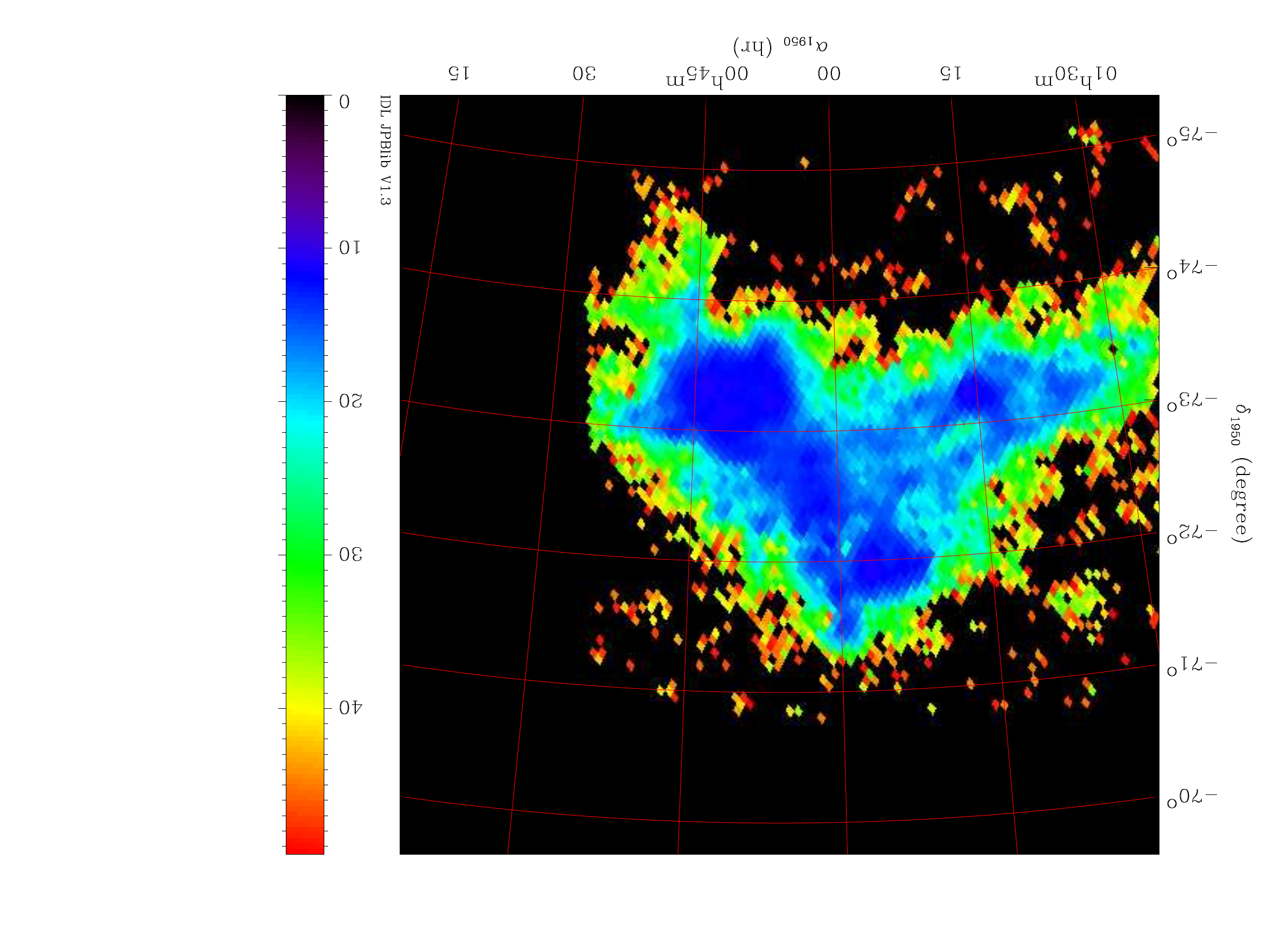}
\caption{
Upper panel: Map of the dust optical depths of the SMC at {\hfi} {\HFIfourfreq}\,{\GHz}. Units
are $\rm 10^4\times\tau$.
Lower panel: Map of the dust optical depth relative uncertainty at {\hfi} {\HFIfourfreq}\,{\GHz} in percent.
Black pixels in the maps are masked and have
relative uncertainties larger than 50\%.
\label{fig_taumap_smc}}
\end{center}
\end{figure}

The optical depth and optical depth uncertainty maps derived at \HFIonefreq\,\GHz \,are
shown in Fig.\,\ref{fig_taumap_lmc} and Fig.\,\ref{fig_taumap_smc} for
the LMC and the SMC respectively.

\section{Discussion}
\label{sec_discussion}

\subsection{The millimetre excess}

\begin{figure}[ht]
\begin{center}
\includegraphics[width=10cm,angle=180]{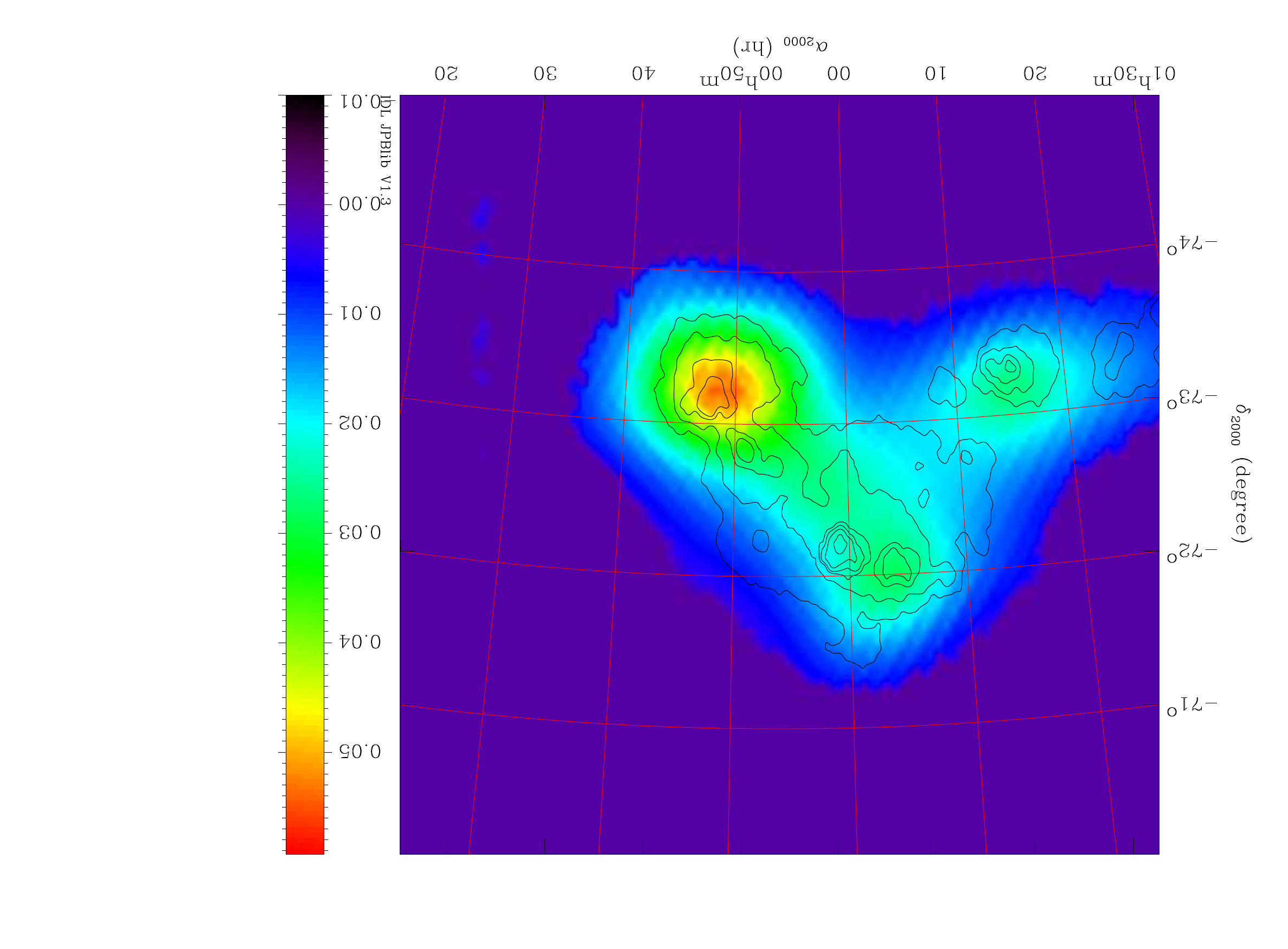}
\caption{
Map of the millimetre excess in the SMC at 3\,mm computed at the resolution
of the {\lfi}-{\LFIthreefreq}\,GHz channel, after free-free subtraction. The contours show the
$\halpha$ distribution.
\label{fig_excessmap}}
\end{center}
\end{figure}

\cite{Bot2010} (see Fig.\,\ref{fig_beforeaftersed}) found that
the integrated SED of the LMC and more noticeably of the SMC showed
excess millimetre emission with respect to a dust and
free-free/synchrotron model based respectively on the FIR and the
radio data available. They investigated several causes for this
excess. They noticed that the excess had precisely the colors of CMB
fluctuations.  They performed simulations by placing the LMC and SMC
structure at various positions over a CMB simulated sky, and concluded
that a CMB origin was unlikely, but not excluded. Other proposed
origins for the excess included the presence of very cold dust,
spinning dust or large modifications of the optical properties of
thermal dust.

Very cold dust ($\rm \Td \simeq 5-7\,K$) has been advocated to explain the
flattening of the millimetre emission observed in more distant low
metallicity galaxies \citep[e.g.][]{Galliano2005}. However, this usually
led to very large masses, and the existence of such very cold dust
remains controversial and difficult to understand in low metallicity
systems where the stronger star formation rate and the lower dust
abundances prevents efficient screening from UV photons.

\cite{Bot2010} applied spinning dust models to fit the SMC and LMC
SEDs and found a plausible match. However, since the observed excess peaked
at a significantly higher frequency than observed for spinning dust in
other regions \citep[e.g.][]{planck2011-7.2}, their fit required
extreme density and excitation conditions for the small dust
particles.

Using the Two Level System model by \cite{Meny2007} for the long
wavelength emission of amorphous solids also proved plausible for the
LMC, but a convincing fit could not be found for the SMC, essentially
because the model could not reproduce the shape of the excess.

The study carried out here, which takes advantage of the {\Planck}
measurements to constrain the CMB foreground fluctuations towards the
two galaxies, shows that part of the excess observed toward the SMC
cannot be accounted for by the fluctuations of the background CMB,
as discussed in Sec.\,\ref{sec:cmb_sub}, but the intensity of the excess 
has been greatly reduced compared to the one found by \cite{Bot2010}. 

To give more insight into this excess, we built a map to trace the
spatial distribution of the excess emission in the SMC. To do so, we
applied the fitting procedure performed for the integrated SEDs of the
Magellanic Cloud, to each point of the SMC at the angular resolution
of the {\lfi} smallest frequency channel. The SEDs were build at each
point using the CMB and foreground--subtracted data. To model the
SEDs, we assumed that, at each point, the SEDs are dominated by
free-free and dust emission up to the longest wavelengths covered by
{\Planck}, following what is observed for the integrated emission of
the SMC (see Fig.\,\ref{fig_beforeaftersed}).  The free-free component
at each wavelength was extrapolated from the $\halpha$ emission (as in
Sect.\,\ref{sec_seds}), assuming that the extinction in the SMC is
negligible. The thermal dust emission was then fitted to the data
points at $\rm \lambref< 550\mic$, using the \cite{Draine2007} dust
model. The millimeter excess was then defined as the difference
between the data and the dust-and-free-free model.  The resulting
spatial distribution of the excess emission at \HFIsixfreq\,GHz
(3\,mm) is shown in Fig.\,\ref{fig_excessmap}. This shows that the
peak of the excess is located at the southwest tip of the bar. This
region also corresponds to the maximum of the optical depth derived
from the FIR and shown in Fig.\,\ref{fig_taumap_smc} and the overall
excess spatial distribution is consistent with being proportional to
the dust column density.

\subsection{FIR dust emissivity}

\begin{figure}[ht]
\begin{center}
\includegraphics[height=6cm,angle=0]{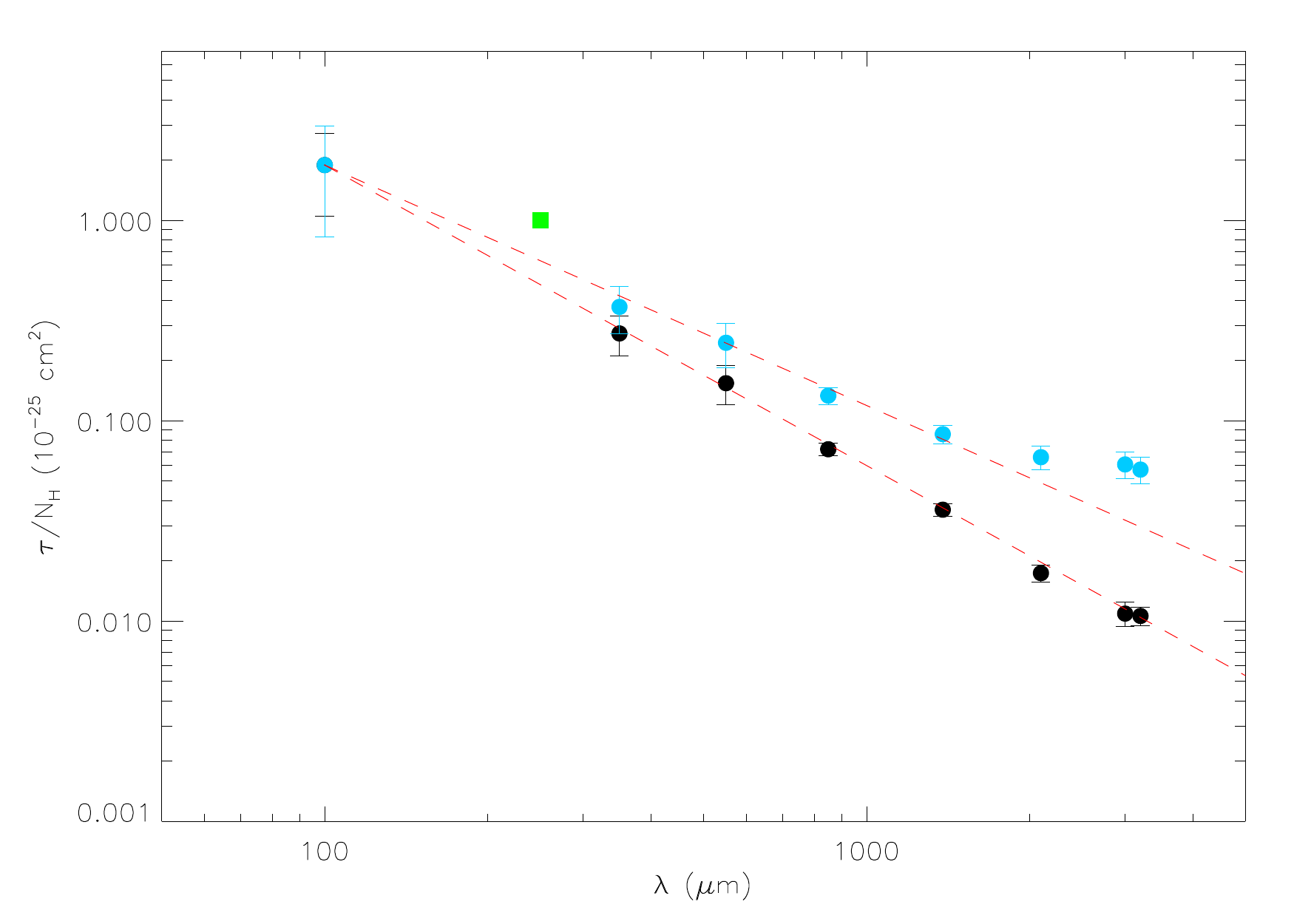}
\caption{
Average dust optical depth of the LMC (black) and SMC (blue) obtained
from the CMB, foreground and free-free subtracted SEDs.  The average
is taken in the same regions as for Fig.\,\ref{fig_beforeaftersed}.
The SMC SED was normalized (multiplied by 6.31) to that of the LMC at $100\mic$.
The square symbol (green) shows the reference value for the solar neighborhood
by \cite{Boulanger1996}.
The dashed lines show $\rm \tau \propto \nu^{\betalmc}$ and $\rm \tau
\propto \nu^{\betasmc}$ normalized at $100\mic$.  Error bars are $\pm3$-$\sigma$.
\label{fig:tauwav}}
\end{center}
\end{figure}

The wavelength dependence of the average dust optical depth in the LMC
and SMC is shown in Fig.\,\ref{fig:tauwav}. The spectral index of the
emissivity in the FIR is consistent with $\beta=\betalmc$ and
$\beta=\betasmc$ for the LMC and the SMC respectively. This value for
the LMC is consistent with findings by \cite{Gordon2010} using the
\herschel data.  We see no hint for for a change of the spectral index
with wavelength for the LMC.  In contrast, the SMC SED clearly
flattens at $\lambda>800\mic$ to reach extremely flat $\beta$ values
around $\lambda=3$ mm. The dust emissivities as interpolated using the
power laws shown in Fig.\,\ref{fig:tauwav} can be compared to the
reference value for the solar neighborhood of $\rm
\tau/\NH=10^{-25}\,cm^2$ at $250\mic$ \citep{Boulanger1996}.
This comparison indicates lower than solar dust abundances for the two
galaxies, by about 1/2.4 and 1/13 solar respectively. This is in rough
agreement with metallicity for the LMC and significantly lower than
metallicity for the SMC.

\cite{Gordon2010} found hints of such excess emission around $500\mic$
in the LMC {\herschel} data, but this remained within the
current calibration uncertainties of the {\herschel} Spectral and Photometric
Imaging Receiver ({\spire}) instrument. We also observe that the
$500\mic$ optical depth of the LMC is slightly higher (by $8.4\%$)
than the power law extrapolation shown in Fig.\,\ref{fig:tauwav},
but this is only marginally larger than the 1$\sigma$ uncertainty
at this wavelength. In any case, data points at longer wavelengths
do not support the existence of excess emission in addition to
a $\beta=\betalmc$ power law for the LMC.

\subsection{Possible interpretation of the excess}

\begin{figure*}[ht]
\begin{center}
\includegraphics[height=13cm,angle=0]{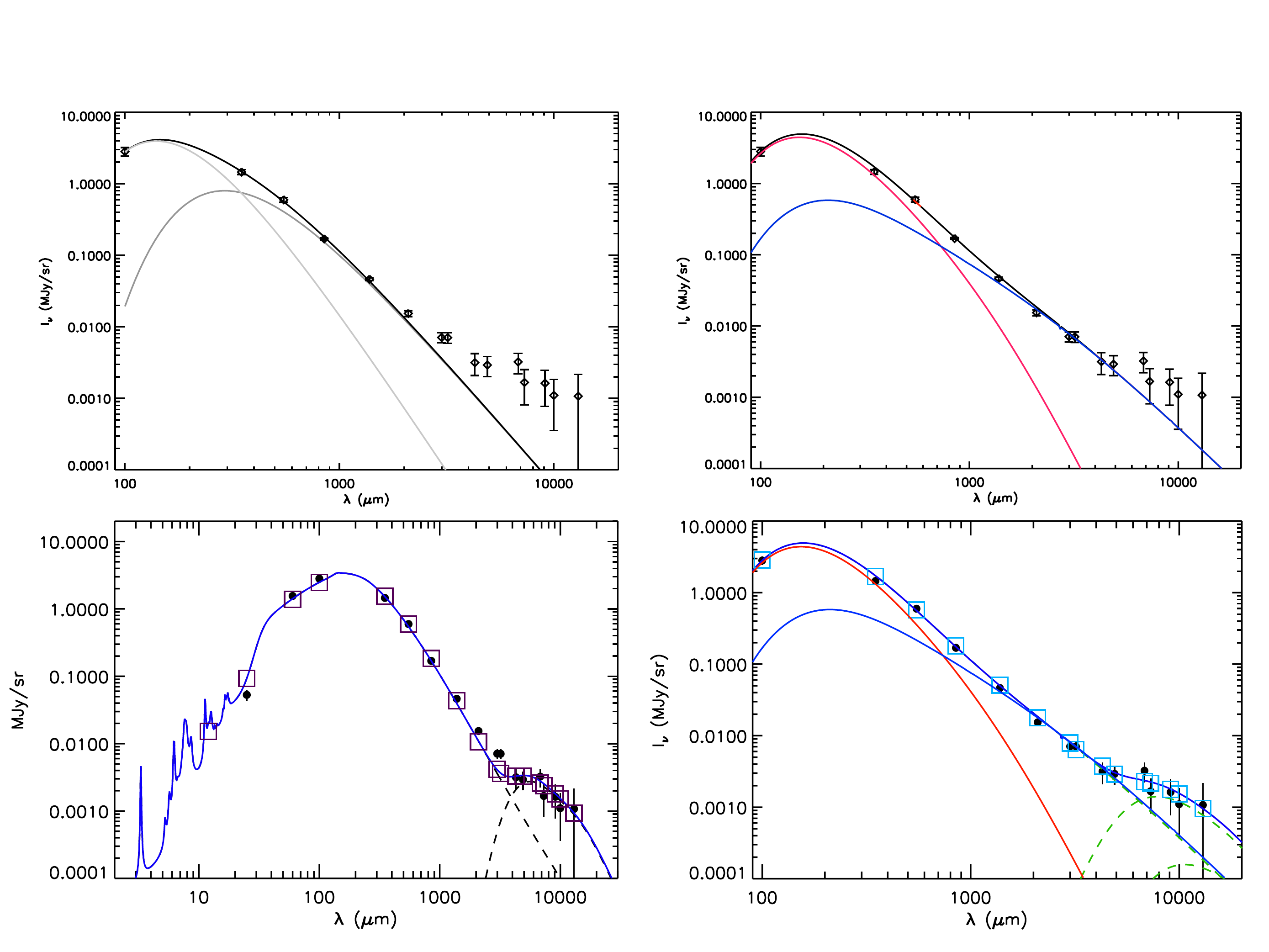}
\caption{
Fit of the SMC SED using the Finkbeiner model (upper left),
the TLS model (upper right), the spinning dust model (lower left) and
a combination of TLS and spinning dust models (lower right). In the
lower panels, the squares represent the flux in each band predicted by
the best model (the blue line).
\label{fig:Allfits}}
\end{center}
\end{figure*}

In this section, we attempt to fit the SED of the SMC using
various models which have been proposed to explain excess submm
emission in addition to a single dust emissivity power-law for thermal
dust.  Figure\,\ref{fig:Allfits} shows several fits to the FIR-Submm
SED of the SMC.

The first fit uses the \cite{Finkbeiner99} model $\sharp7$ (see their
Table 3), which was designed to explain the flatter than expected
emission spectrum of our Galaxy as observed by $\firas$. Here we use
this model to assess the possibility that the millimeter excess is due
to very cold dust. The model was fitted using the same $\beta$ values
as in \cite{Finkbeiner99} for the two dust components ($\rm
\beta_{warm}=2.6$, $\rm \beta_{cold}=1.5$).  It also assumes the same
type of relationship between the cold and warm dust temperatures,
which in the model reflects the fact that both dust species are
subjected to the same radiation field. However, we allow the
IR/optical opacity ratio ($\rm q_{cold}/q_{warm}$) to vary, so that
the ratio between the warm and cold dust temperatures is free to vary.
We also leave the cold dust component abundance ($\rm f_{cold}$) and
the warm dust temperature ($\rm T_{warm}$) as free parameters.  The
best fit values are found for $\rm T_{warm}=16.4\,K$ $\rm
q_{cold}/q_{warm}=168.2$, $\rm f_{cold}=8.7\,10^{-3}$. For
these parameters, the temperature of the cold component is $\rm
T_{cold}=5.9\,K$.  This cold temperature is needed to reproduce part
of the millimetre excess but is much colder than that obtained by
\cite{Finkbeiner99} for the Galaxy ($\rm T_{cold}=9.6\,K$). This is
obtained at the expense of strongly increasing the IR/optical opacity
ratio for the cold component ($\rm q_{cold}/q_{warm}$ increased by a
factor 15), which controls the $\rm T_{cold}/T_{warm}$ ratio.  The
mass fraction of the cold component derived here, $\rm f_{cold}$, is
about 4 times lower than that derived for the MW in
\cite{Finkbeiner99}, which compensates for the increased $\rm
q_{cold}/q_{warm}$.  It is apparent that, despite invoking a much
larger IR/optical opacity ratio for the cold particles, such a model
has difficulties producing the submm excess above about $\lambda=2$ mm.

The second fit employs the Two-Level-System (TLS) model developed by
\cite{Meny2007}. The fit was obtained by minimizing $\chi^2$ in the
range $\rm 100\mic < \lambda < 5\,mm$ against the following 3
parameters: $\Td$ the dust temperature, $\lcor$ the correlation
length of defects in the material and A the density of TLS sites in
the material composing the grains. The values derived for these
parameters are $\Td$=18.9 K, $\lcor$=12.85 nm and A=7.678. The
reduced $\chi^2$ for these parameters is $\chi^2$=2.56.  Compared to
the best values found from \cite{ParadisThesis} for the MW ($\Td$=17.9
K, $\lcor$=12.85 nm, A=2.42), this indicates dust material with
more TLS sites (more mechanical defects) but a similar defect
correlation length compared with the dust dominating the MW
emission. Note that a grey-body fit of the same SED over the same
frequency range leads to $\Td$=18.9 K and $\beta$=0.74 and a reduced
$\chi^2$=6.12, showing that the TLS model reproduces the data better
than a single grey-body fit.

The third fit uses spinning dust. In that case, the thermal dust
emission was reproduced using the Draine \& Li (2007) dust model.  The
remaining excess is then fitted with the spinning dust model described
in \cite{Silsbee2010}. Spinning dust emission has been proved to be
sensitive to the neutral and ionized gas densities and to the size
distribution of dust grains. The radiation field and the size
distributions are taken to be the same as for the Draine \& Li (2007)
model. The relevant gas parameters are computed with CLOUDY
\citep{Ferland1998}: we take the parameters from the optically thin
zone of isochoric simulations.  In both cases, we take into account
the lower metallicity and dust grains abundance when compared to the
MW. The electric dipole moment distribution of grains is taken as in
\cite{Draine1998a}\footnote{\cite{Ysard2010} showed that it is in good
agreement with the anomalous emission extracted from the {\wmap}
data.}.  This fit is obtained by adding two components according to
the PDR fraction inferred from the thermal emission model: a diffuse
medium with $n_{\rm H}$=$\rm30\,cm^{-3}$ and 100\% of the PAH mass
expected from the IR modeling, and a denser medium with $n_{\rm
H}$=$\rm5000\,cm^{-3}$ and 82\% of the PAH mass expected.

The fourth one is a combination of the TLS model and the spinning dust
model presented above.

It is clear from Fig.\,\ref{fig:Allfits} that the model combining TLS
and spinning dust is the only one of the proposed models giving a
satisfactory fit to the millimetre excess observed in the SMC over the
whole spectral range. In addition, it alleviates the need for using
more PAH than allowed by the NIR emission to produce the required
level of spinning dust emission observed.  Alternative models,
however, cannot be excluded. For instance, very large grains ($\rm
a>30\mic$) that would be efficient radiators
\citep{Rowan-Robinson1992} could create emission in the millimetre
wavelengths without too much mass. Exploring this possibility would
require specific modeling. This is beyond the scope of the current
article but should be explored in future studies.

\section{Conclusions}
\label{sec_conclusions}

We assessed the existence and investigated the origin of millimetre excess emission in the
LMC and the SMC using the \Planck \,data.  The integrated SED of the
two galaxies before subtraction of the foreground (Milky Way) and
background (CMB fluctuations) emission are in good
agreement with previous determinations.

The background CMB contribution was first subtracted using an Internal
Linear Combination (ILC) method performed locally around the two
galaxies.  The uncertainty of this contribution was measured through a
detailed Monte-Carlo simulation.  We subtracted the
foreground emission from the Milky Way using a Galactic \ion{H}{i} template
and the proper dust emissivity derived in a region surrounding the two
galaxies and dominated by MW emission.  We also subtracted the
free-free contribution from ionized gas in the galaxies, using the
$\halpha$ emission, taking advantage of the low extinction in those
galaxies.  The remaining emission of both galaxies correlates with the
gas emission of the LMC and SMC.

We showed that the excess 
previously reported in the LMC can be fully explained by CMB fluctuations. 
For the SMC, subtracting the CMB fluctuations decreases
the intensity of the excess but a significant millimetre emission above the 
expected thermal dust, free-free and synchrotron emission remains.

We combined the {\Planck} and \iris data at $100 \mic$ to produce thermal
dust temperature and optical depth maps of the two galaxies. The LMC
temperature map shows the presence of a warm inner arm already
found with the {\spitzer} data, but also shows the existence of a
previously unidentified cold outer arm.  Several cold regions were
found along this arm, some of which are associated to known
molecular clouds.

We used the dust optical depth maps to constrain the thermal dust
emissivity spectral index ($\beta$). The average spectral index in
the FIR ($\lambda<500\mic$) is found to be consistent with $\beta=1.5$
and $\beta=1.2$ for the LMC and the SMC respectively.  This is
significantly flatter than what is observed in the Milky Way.  The
absolute values of the emissivities in the FIR, when compared to that
in our solar neighborhood, are compatible with D/G mass ratio of 1/2.4
and 1/13 for the LMC and SMC. This is compatible with the
metallicity for the LMC but significantly lower than metallicity for the SMC.
In the submm, the LMC SED remains consistent with
$\beta=1.5$, while the SED of the SMC flattens even more.

The spatial distribution of the mm excess in the SMC appears to follow
the general pattern of the gas distribution. It therefore appears
unlikely that the excess could originate from very cold dust. Indeed,
this is confirmed by attempts to fit the SMC emission SED with 2 dust
component models, which led to poor fits. Alternative models, such as
emission excess due to the amorphous nature of large grains are likely
to provide a natural explanation to the observed SED, although
spinning dust is needed to explain the SED above $\lambda=$3\,mm.

\begin{acknowledgements}
A description of the Planck Collaboration and a list of its members can be found at \url{http://www.rssd.esa.int/index.php?project=PLANCK&page=Planck_Collaboration}
\end{acknowledgements}

\bibliographystyle{aa}

\bibliography{Planck_lmcsmc2,Planck_bib}

\end{document}